\documentclass[twocolumn]{aastex631}
\usepackage{hyperref}
\usepackage{graphicx}
\usepackage{natbib}

\usepackage{xurl}
\usepackage[hyphens]{url}
\usepackage{booktabs}

\shorttitle{JWST-TST DREAMS: WASP-17{\rm b} NIRISS SOSS Transmission}
\shortauthors{Louie et al.}

\begin{document}

\title{JWST-TST DREAMS: A Precise Water Abundance for Hot Jupiter WASP-17b from the NIRISS SOSS Transmission Spectrum}

\correspondingauthor{Dana R. Louie}
\email{dana.louie.astro21@gmail.com}


\author[0000-0002-2457-272X]{Dana R. Louie}
\altaffiliation{GSFC Sellers Exoplanet Environments Collaboration}
\affiliation{Catholic University of America, Department of Physics, Washington, DC, 20064, USA}
\affiliation{Exoplanets and Stellar Astrophysics Laboratory (Code 667), NASA Goddard Space Flight Center, Greenbelt, MD 20771, USA}
\affiliation{Center for Research and Exploration in Space Science and Technology II, NASA/GSFC, Greenbelt, MD 20771, USA}

\author[0000-0003-0814-7923]{Elijah Mullens}
\affiliation{Department of Astronomy and Carl Sagan Institute, Cornell University, 122 Sciences Drive, Ithaca, NY 14853, USA}

\author[0000-0001-8703-7751]{Lili Alderson}
\affiliation{University of Bristol, HH Wills Physics Laboratory, Tyndall Avenue, Bristol, UK}

\author[0000-0002-5322-2315]{Ana Glidden}
\affiliation{Department of Earth, Atmospheric and Planetary Sciences, Massachusetts Institute of Technology, Cambridge, MA 02139, USA}
\affiliation{Kavli Institute for Astrophysics and Space Research, Massachusetts Institute of Technology, Cambridge, MA 02139, USA}

\author[0000-0002-8507-1304]{Nikole K. Lewis}
\affiliation{Department of Astronomy and Carl Sagan Institute, Cornell University, 122 Sciences Drive, Ithaca, NY 14853, USA}

\author[0000-0003-4328-3867]{Hannah R. Wakeford}
\affiliation{University of Bristol, HH Wills Physics Laboratory, Tyndall Avenue, Bristol, UK}

\author[0000-0003-1240-6844]{Natasha E. Batalha}
\affiliation{NASA Ames Research Center, MS 245-3, Moffett Field, CA 94035, USA}

\author[0000-0001-8020-7121]{Knicole D. Col\'{o}n}
\affiliation{Exoplanets and Stellar Astrophysics Laboratory (Code 667), NASA Goddard Space Flight Center, Greenbelt, MD 20771, USA}

\author[0000-0003-0854-3002]{Am\'{e}lie Gressier}
\affiliation{Space Telescope Science Institute, 3700 San Martin Drive, Baltimore, MD 21218, USA}

\author[0000-0002-2508-9211]{Douglas Long}
\affiliation{Space Telescope Science Institute, 3700 San Martin Drive, Baltimore, MD 21218, USA}

\author[0000-0002-3328-1203]{Michael Radica}
\affiliation{Institut Trottier de Recherche sur les Exoplanètes and Département de Physique, Université de Montréal, 1375 Avenue Thérèse-Lavoie-Roux, Montréal, QC, H2V 0B3, Canada}
\affiliation{Department of Astronomy \& Astrophysics, University of Chicago, 5640 South Ellis Avenue, Chicago, IL 60637, USA}

\author[0000-0001-9513-1449]{N\'{e}stor Espinoza}
\affiliation{Space Telescope Science Institute, 3700 San Martin Drive, Baltimore, MD 21218, USA}
\affiliation{William H. Miller III Department of Physics and Astronomy, Johns Hopkins University, Baltimore, MD 21218, USA}

\author[0000-0002-8515-7204]{Jayesh Goyal}
\affiliation{School of Earth and Planetary Sciences (SEPS), National Institute of Science Education and Research (NISER), HBNI, Odisha, India}

\author[0000-0003-4816-3469]{Ryan J. MacDonald}
\altaffiliation{NHFP Sagan Fellow}
\affiliation{Department of Astronomy, University of Michigan, 1085 S. University Ave., Ann Arbor, MI 48109, USA}

\author[0000-0002-2739-1465]{Erin M. May}
\affiliation{Johns Hopkins University Applied Physics Laboratory, 11100 Johns Hopkins Rd, Laurel, MD 20723, USA}

\author[0000-0002-6892-6948]{Sara Seager}
\affil{Department of Earth, Atmospheric, and Planetary Sciences, Massachusetts Institute of Technology, Cambridge, MA 02139, USA}
\affiliation{Kavli Institute for Astrophysics and Space Research, Massachusetts Institute of Technology, Cambridge, MA 02139, USA}
\affil{Department of Aeronautics and Astronautics, Massachusetts Institute of Technology, Cambridge, MA 02139, USA}

\author[0000-0002-7352-7941]{Kevin B. Stevenson}
\affiliation{Johns Hopkins University Applied Physics Laboratory, 11100 Johns Hopkins Rd, Laurel, MD 20723, USA}

\author[0000-0003-3305-6281]{Jeff A. Valenti}
\affiliation{Space Telescope Science Institute, 3700 San Martin Drive, Baltimore, MD 21218, USA}

\author[0000-0002-0832-710X]{Natalie H. Allen}
\affiliation{William H. Miller III Department of Physics and Astronomy, Johns Hopkins University, Baltimore, MD 21218, USA}

\author[0000-0003-4835-0619]{Caleb I. Ca\~{n}as}
\affiliation{Exoplanets and Stellar Astrophysics Laboratory (Code 667), NASA Goddard Space Flight Center, Greenbelt, MD 20771, USA}

\author[0000-0002-8211-6538]{Ryan C. Challener}
\affiliation{Department of Astronomy and Carl Sagan Institute, Cornell University, 122 Sciences Drive, Ithaca, NY 14853, USA}

\author[0000-0001-5878-618X]{David Grant}
\affiliation{University of Bristol, HH Wills Physics Laboratory, Tyndall Avenue, Bristol, UK}

\author[0000-0001-5732-8531]{Jingcheng Huang}
\affiliation{Department of Earth, Atmospheric and Planetary Sciences, Massachusetts Institute of Technology, Cambridge, MA 02139, USA}

\author[0000-0003-0525-9647]{Zifan Lin}
\affiliation{Department of Earth, Atmospheric and Planetary Sciences, Massachusetts Institute of Technology, Cambridge, MA 02139, USA}

\author[0000-0002-2643-6836]{Daniel Valentine}
\affiliation{University of Bristol, HH Wills Physics Laboratory, Tyndall Avenue, Bristol, UK}

\author{Mark Clampin}
\affiliation{NASA Headquarters, 300 E Street SW, Washington, DC 20546, USA}

\author[0000-0002-3191-8151]{Marshall Perrin}
\affiliation{Space Telescope Science Institute, 3700 San Martin Drive, Baltimore, MD 21218, USA}

\author{Laurent Pueyo}
\affiliation{Space Telescope Science Institute, 3700 San Martin Drive, Baltimore, MD 21218, USA}



\author[0000-0001-7827-7825]{Roeland P. van der Marel}
\affiliation{Space Telescope Science Institute, 3700 San Martin Drive, Baltimore, MD 21218, USA}
\affiliation{William H. Miller III Department of Physics and Astronomy, Johns Hopkins University, Baltimore, MD 21218, USA}

\author{C. Matt Mountain}
\affiliation{Association of Universities for Research in Astronomy, 1331 Pennsylvania Avenue NW Suite 1475, Washington, DC 20004, USA}



\begin{abstract}

Water has proven to be ubiquitously detected in near-infrared (NIR) transmission spectroscopy observations of hot Jupiter atmospheres, including WASP-17b. However, previous analyses of WASP-17b's atmosphere based upon Hubble Space Telescope (HST) and Spitzer data could not constrain the water abundance, finding that sub-solar, super-solar and bimodal posterior distributions were all statistically valid. In this work, we observe one transit of the hot Jupiter WASP-17b using JWST's Near Infrared Imager and Slitless Spectrograph Single Object Slitless Spectroscopy (NIRISS SOSS) mode. We analyze our data using three independent data analysis pipelines, finding excellent agreement between results. Our transmission spectrum shows multiple H$_2$O absorption features and a flatter slope towards the optical than seen in previous HST observations. We analyze our spectrum using both \texttt{PICASO+Virga} forward models and free retrievals. \texttt{POSEIDON} retrievals provide a well-constrained super-solar $\log$(H$_2$O) abundance (-2.96$^{+0.31}_{-0.24}$), breaking the degeneracy from the previous HST/Spitzer analysis. We verify our \texttt{POSEIDON} results with \texttt{petitRADTRANS} retrievals. Additionally, we constrain the abundance of $\log$(H$^-$), -10.19$^{+0.30}_{-0.23}$, finding that our model including H$^-$ is preferred over our model without H$^-$ to 5.1 $\sigma$. Furthermore, we constrain the $\log$(K) abundance (-8.07$^{+0.58}_{-0.52}$) in WASP-17b's atmosphere for the first time using space-based observations. Our abundance constraints demonstrate the power of NIRISS SOSS's increased resolution, precision, and wavelength range to improve upon previous NIR space-based results. This work is part of a series of studies by our JWST Telescope Scientist Team (JWST-TST), in which we use Guaranteed Time Observations to perform Deep Reconnaissance of Exoplanet Atmospheres through Multi-instrument Spectroscopy (DREAMS).

\end{abstract}



\section{Introduction} \label{sec:intro}

Transiting exoplanet observations using the JWST Near Infrared Imager and Slitless Spectrograph Single Object Slitless Spectroscopy (NIRISS SOSS) mode\footnote{\url{https://jwst-docs.stsci.edu/jwst-near-infrared-imager-and-slitless-spectrograph/niriss-observing-modes/niriss-single-object-slitless-spectroscopy}} continue to fulfill pre-launch expectations \citep[e.g.,][]{Beichman2014,Greene2016,Howe2017,Batalha2017,Louie2018}. Since July 2022 when NASA publicized JWST's Early Release Observations \citep[ERO,][]{Pontoppidan2022}, NIRISS SOSS has observed planets ranging in size from terrestrial to gas giant. Many results are still forthcoming, but publications so far include precise constraints on both water and potassium in exoplanet atmospheres \citep[e.g.,][]{feinstein2023,taylor2023,Fournier-Tondreau2024}. 

The Canadian Space Agency specifically designed NIRISS SOSS for transiting exoplanet time series observations (TSOs) at medium resolution R$\sim$700 \citep{albert2023,doyon2023,radica2022,darveau-bernier2022}. To enable observations of bright exoplanet host stars without reaching saturation levels, the GR700XD grism incorporates a weak defocussing lens that spreads light across $\sim$23 pixels in the spatial direction. The NIRISS SOSS composite order 1 and order 2 bandpass spans wavelengths between 0.6 to 2.8$\mu$m, a high signal-to-noise (S/N) regime in the exoplanet host star spectral energy distribution. The wavelength range includes strong water (H$_2$O), methane (CH$_4$), carbon monoxide (CO), and carbon dioxide (CO$_2$) bands, as well as the 0.767 $\mu$m potassium resonance doublet K\,I and the 1.083 $\mu$m helium triplet line He\,I \citep{Seager2000,Brown2001}.

Following the ERO observations of WASP-96b and HAT-P-18b, the first NIRISS SOSS exoplanet observations were those of WASP-39b as part of the Transiting Exoplanet Community Early Release Science (ERS) program\footnote{\url{https://ers-transit.github.io/index.html}} \citep{Stevenson2016,Bean2018}. Multiple groups published analyses of these three Saturn-mass gas giant exoplanet observations. 

\cite{feinstein2023} reported the first NIRISS SOSS results for WASP-39b. In their analysis, they detected H$_2$O to greater than $30\sigma$ and the potassium doublet at $6.8\sigma$, as well as a greater than $8\sigma$ preference for non-gray, inhomogeneous clouds. Notably, they showed that the precision and broad wavelength coverage of NIRISS SOSS spectra enabled the breaking of degeneracies between cloud abundances and atmospheric metallicity. The \cite{holmberg2023} case study of WASP-39b agreed with these results.

\cite{fu2022} published the first NIRISS SOSS results for HAT-P-18b, finding an unambiguous H$_2$O detection, an escaping metastable He tail at 1.083 $\mu$m, a sub-Rayleigh haze scattering slope, and CH$_4$ depletion indicative of disequilibrium processes. \cite{Fournier-Tondreau2024} independently analyzed HAT-P-18b, reporting detections of H$_2$O (12.5$\sigma$), CO$_2$ (7.3$\sigma$), a cloud deck (7.4$\sigma$), and unocculted star-spots (5.8$\sigma$), with hints of Na (2.7$\sigma$) and the absence of CH$_4$ (log CH$_4$ $<$ -6 to 2$\sigma$).

\cite{radica2023} and \cite{taylor2023} published complementary analyses on WASP-96b, highlighting both NIRISS SOSS data reduction best practices, as well as the power of the instrument in constraining multiple molecular and atomic species. They constrained the volume mixing ratios of H$_2$O, CO$_2$, and potassium, notably providing the first abundance constraint on potassium in the planet's atmosphere. In their forward model case study of WASP-96b, \cite{holmberg2023} noted H$_2$O as the most prominent spectral feature for the planet atmosphere.

Additional results from the past year have examined NIRISS SOSS observations of smaller planets. \cite{Radica2024} reported muted spectral features on the 4.72$R_\earth$ hot Neptune desert planet LTT 9779b. \cite{Benneke2024} observed the 2.2$R_\earth$ sub-Neptune TOI-270d using both NIRISS SOSS and the Near Infrared Spectrograph (NIRSpec) G395H, finding CH$_4$, CO$_2$, and H$_2$O in a metal-rich atmosphere. \cite{lim2023} observed 2 transits of TRAPPIST-1b using NIRISS SOSS. Their results confirmed the absence of a cloud-free, hydrogen-rich atmosphere. They also found strong evidence of contamination from unocculted stellar heterogeneities.

Here, we expand upon previous NIRISS SOSS transiting exoplanet results by reporting the findings from our analysis of the hot Jupiter WASP-17b. The 1.932$R_{\rm Jup}$, 0.477$M_{\rm Jup}$ gas giant WASP-17b revolves around its 1.583$R_{\odot}$ F6-type host star on a 3.735 d retrograde orbit \citep{anderson2010,anderson2011,southworth2012,Alderson2022}. The accompanying low density (0.06$\rho_{\rm Jup}$) and hot equilibrium temperature, $T_{\rm eq}$, 1755\,K result in a $\sim$1609 km atmospheric scale height. Combined with a bright ($m_J = 10.509$), quiet host star \citep{sing2016,Khalafinejad2018}, these qualities make WASP-17b an exceptional target for atmospheric characterization studies. Additionally, theoretical studies including WASP-17b predict large variations in day-to-night and morning-to-evening cloud and atmospheric properties for this planet \citep{kataria2016,Zamyatina2023}.

Early WASP-17b space-based transmission spectroscopy using the Hubble Space Telescope (HST) Wide Field Camera 3 (WFC3) and Space Telescope Imaging Spectrograph (STIS), as well as \textit{Spitzer} Infrared Array Camera (IRAC), found evidence of H$_2$O and sodium Na\,I absorption features \citep{mandell2013,sing2016,wakeford2016}. Ground-based high resolution observations also reported evidence of sodium Na\,I absorption \citep{Wood2011,Zhou2012,Khalafinejad2018}, as well as the detection of the wings of the potassium K\,I resonance doublet \citep{sedaghati2016}. Dedicated theoretical and retrieval analyses based upon these early observations yielded inconclusive results concerning the existence of clouds or Rayleigh scattering aerosols in the atmosphere. Additionally, H$_2$O abundances spanned sub-solar, to solar, to super-solar values \citep{barstow2017,pinhas2019,fisherHeng2018,welbanks2019}.

Recently, \cite{Alderson2022} reevaluated previous HST STIS and \textit{Spitzer} data alongside newly obtained HST WFC3/G102 and G141 data taken in spatial scanning mode to complete a comprehensive 0.3--5 $\mu$m reanalysis of the WASP-17b transmission spectrum. They found H$_2$O absorption at greater than $7\sigma$ and CO$_2$ absorption at greater than $3\sigma$, but no conclusive evidence for either Na\,I or K\,I. Additionally, across an assortment of retrieval models---employing both free and equilibrium chemistries---they found their data favored a bimodal solution, with both high- or low-metallicity modes plausible. \cite{grant_miritransit_2023} recently complemented the \cite{Alderson2022} analysis by adding Mid-Infrared Instrument Low Resolution Spectrometer (MIRI LRS) transit observations, which span 5--12 $\mu$m. Their spectrum showed an opacity source at 8.6$\mu$m, which they identified as being due to SiO$_2$(s) (quartz) clouds comprised of small $\sim$0.01 $\mu$m particles extending high up in the atmosphere.

Here, we present our observational analysis and interpretation of the NIRISS SOSS transmission spectrum for WASP-17b (GTO~1353, PI: Lewis). As part of our Deep Reconnaissance of Exoplanet Atmospheres through Multi-instrument Spectroscopy (DREAMS), our team has also observed WASP-17b in transit using MIRI LRS \citep{grant_miritransit_2023} and NIRSpec G395H. We observed WASP-17b in eclipse using NIRISS SOSS \citep{Gressier2024_arxiv}, MIRI LRS \citep{Valentine2024}, and NIRSpec G395H. Papers on the full transmission (Lewis et al., in prep) and emission (Wakeford et al., in prep) spectra are forthcoming. In this work, we perform a retrieval analysis on NIRISS SOSS data alone, and we also fit forward models and perform retrievals on  NIRISS SOSS data combined with our team's MIRI LRS data, as well as previous observations from HST and Spitzer.


This paper is part of a series to be presented by the JWST Telescope Scientist Team (JWST-TST), which uses Guaranteed Time Observer (GTO) time awarded by NASA in 2003 (PI M. Mountain) for studies in three different subject areas: 
(a) Transiting Exoplanet Spectroscopy (lead: N. Lewis); (b) Exoplanet and Debris Disk Coronagraphic Imaging (lead: M. Perrin); and (c) Local Group Proper Motion Science (lead: R. van der Marel). A common theme of these investigations is the desire to pursue and demonstrate science for the astronomical community at the limits of what is made possible by the exquisite optics and stability of JWST. An up-to-date listing of papers published by our team across these three areas is maintained on our public JWST-TST website.\footnote{https://www.stsci.edu/$\sim$marel/jwsttelsciteam.html} The present paper is part of our work on Transiting Exoplanet Spectroscopy, which focuses on detailed explorations of three transiting exoplanets representative of key exoplanet classes: Hot Jupiters (WASP-17b, GTO~1353), Warm Neptunes (HAT-P-26b, GTO~1312), and Temperate Terrestrials (TRAPPIST-1e, GTO~1331). 

We organize our paper as follows. Section \ref{sec:obs} describes our observations. In Section \ref{sec:DataAnalysis} we explain our data analysis techniques for our three independent pipelines, including the presentation and comparison of our three resulting transmission spectra (\S\ref{subsec:intercomparison}) as well as a discussion of limb darkening treatments (\S\ref{subsec:limbdarkeningtreatment}). We describe and present our \texttt{PICASO+Virga} forward modeling analysis in Section \ref{sec:Picaso}, followed by our \texttt{POSEIDON} and \texttt{petitRADTRANS} \citep[\texttt{pRT}, ][]{Molliere2019} retrieval analyses in Section \ref{sec:Retrievals}. We discuss our results in Section \ref{sec:discuss} and conclude in Section \ref{sec:conclusions}.

\section{Observations} \label{sec:obs}

WASP-17b was observed in transit using NIRISS SOSS as part of JWST Cycle 1 GTO program 1353 (PI: Lewis), making use of the GR700XD grism with the clear filter and collecting data with the SUBSTRIP256 \citep[256 x 2048 pixels,][]{doyon2023,albert2023}. We observed for 12.72 hrs on 20 March 2023, beginning at UTC 09:44:14, and concluding at UTC 21:06:25. Our TSOs comprised a total of 720 integrations, with 8 groups per 49.47 sec integration. We observed the entire 4.4 hr transit, as well as $\sim$8.1 hrs of baseline outside of transit. Following our transit observations, we added a second optional exposure of WASP-17 using the GR700XD grism with the F277W filter, which eliminates wavelengths $\lesssim$2.6 $\mu$m, thus revealing locations of field star contaminants and isolating the region of SOSS orders 1 and 2 spectral overlap \citep{albert2023,radica2023}.\footnote{See NIRISS SOSS Recommended Observing Strategies at \href{https://jwst-docs.stsci.edu/jwst-near-infrared-imager-and-slitless-spectrograph/niriss-observing-strategies/niriss-soss-recommended-strategies\#NIRISSSOSSRecommendedStrategies-AddinganoptionalF277Wexposuretoyourobservingprogram}{https://jwst-docs.stsci.edu}.} Our F277W exposure consisted of 10 integrations with 8 groups per integration.

We previously attempted the same observation on 18 February 2023, between UTC 12:35:38 and UTC 23:57:35. However, the observation failed since an incorrect guide star was acquired, resulting in the order 1 trace falling partially off the SUBSTRIP256 array, while the order 2 trace was offset. The 20 March 2023 observations took place following JWST Telescope Time Review Board (TTRB) approval of our Webb Operation Problem Report (WOPR).

\section{NIRISS SOSS Data Analysis} \label{sec:DataAnalysis}

To ensure confidence in our results, we analyzed the WASP-17b NIRISS SOSS transit observations using 3 independent pipelines: \texttt{Ahsoka}, \texttt{transitspectroscopy/juliet}, and \texttt{supreme-SPOON}. We describe the basic detector-level and spectroscopic data reduction procedures (through spectral extraction) for the 3 pipelines in Sections \ref{subsec:AhsokaReduction}, \ref{subsec:transitspectroscopyReduction}, and \ref{subsec:supremeSpoonReduction}, respectively. In Section \ref{subsec:AhsokaLightCurve} we describe our methods to produce both white light curves and spectroscopic light curves using the \texttt{Ahsoka} pipeline, while in Sections \ref{subsec:julietLightCurve} and \ref{subsec:supremeSPOONLightCurve} we do the same for \texttt{transitspectroscopy/juliet} and \texttt{supreme-SPOON}, respectively. In Section \ref{subsec:intercomparison} we present our final WASP-17b transmission spectra from the three pipelines, and we highlight some key similarities and differences between them. Finally, in Section \ref{subsec:limbdarkeningtreatment} we report our findings regarding the effect of various limb darkening treatments on the resultant transmission spectra.

\subsection{\texttt{Ahsoka} Data Reduction} \label{subsec:AhsokaReduction}

We introduce a new data analysis pipeline called \texttt{Ahsoka},\footnote{\url{https://github.com/Witchblade101/ahsoka}} which we use to perform an independent data reduction of the WASP-17b transmission spectrum. We created the \texttt{Ahsoka} pipeline by assembling portions of the \texttt{jwst calibration} \citep{bushouse2023}, \texttt{supreme-SPOON} \citep{feinstein2023,radica2023}, \texttt{nirHiss} \citep{feinstein2023}, and \texttt{Eureka!} \citep{bell2022} pipelines. We employed \texttt{jwst calibration} pipeline version 1.8.2 in our analysis. We describe the steps of  \texttt{Ahsoka}'s 6 stages in the subsections below. We include one section describing our use of the F277W observations to identify field star contaminants ($\S$\ref{subsubsec:AhsokaF277Wusage}). \texttt{Ahsoka} Stages 4, 5, and 6 coincide with those same stages of \texttt{Eureka!}, so we describe those latter 3 stages in Section \ref{subsec:AhsokaLightCurve}--describing how we applied the \texttt{Eureka!} pipeline to \textit{light curve fitting}, and refer the reader to \texttt{Eureka!}'s documentation 
\citep{bell2022} for further details.

\subsubsection{\texttt{Ahsoka} Stage 1: Detector-level processing}\label{subsubsec:Ahsokastage1}

We begin our analysis using the \textit{uncal.fits} files downloaded from the Barbara A. Mikulski Archive for Space Telescopes (MAST), and perform the following \texttt{jwst} pipeline stage 1 detector-level\footnote{\texttt{calwebb\_detector1}, see \url{https://jwst-pipeline.readthedocs.io/en/stable/jwst/pipeline/calwebb_detector1.html}} steps: \texttt{dq\_init, saturation, superbias,} and \texttt{refpix.} 

Following these initial pipeline steps, we apply the \texttt{supreme-SPOON} background subtraction and 1/$f$ noise removal techniques at the group level \citep{radica2023}. The 1/$f$ noise \citep[see, e.g.,][]{rauscher2014,schlawin2020} is introduced during detector readout. Because 1/$f$ noise is one of the last noise sources affecting JWST data, ideally 1/$f$ noise should be one of the first noise sources removed, which necessitates its removal at the group level \citep{albert2023,radica2023}. The reason to perform background subtraction at the group level \textit{prior} to 1/$f$ noise subtraction is to remove any contaminating Zodiacal light from the images. However, following 1/$f$ noise removal, the group-level background is then added back into the observation frames, so that additional steps in the \texttt{Ahsoka} pipeline are applied to the as-observed astrophysical images. Refer to \cite{radica2023} for additional details. 

\begin{figure*}
    \includegraphics[width=\textwidth]{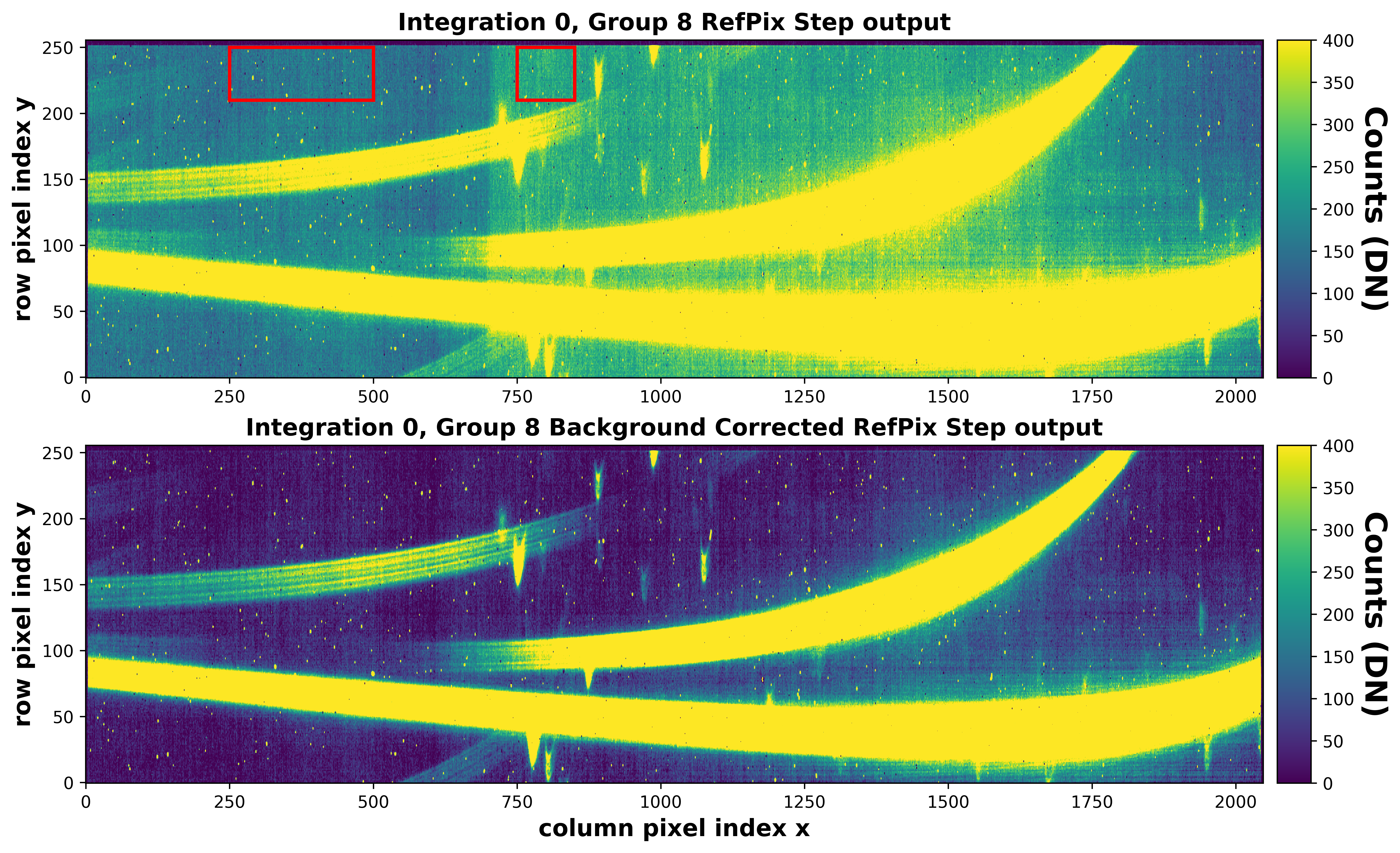}
    \caption{Comparison of one frame (integration 0, group 8) before and after \texttt{supreme-SPOON} group-level background subtraction applied in \texttt{Ahsoka} Stage 1. The order 1 spectral trace lies in the bottom part of the frame; The order 2 spectral trace lies in the middle; and the order 3 spectral trace is at the upper left. 
    \underline{\textbf{Top:}} Frame shown after the \texttt{jwst} pipeline \texttt{refpix} step. The change in background level intensity near column $\sim$700 is clearly evident as a ``step function.'' Order 0 field star contaminants appear as chevron shaped bright splotches on the right side of the ``step.'' Additionally, we show with red boxes the regions we use on either side of column 700 to scale the STScI background model to each side of this ``step.'' We chose these regions because they are relatively free of contaminants. \underline{\textbf{Bottom:}} The same frame shown after the application of the \texttt{supreme-SPOON} group-level background subtraction. The ``step function'' near column 700 is no longer evident. The next step to be applied after group-level background subtraction is 1/$f$ noise removal. The 3 spectral traces and order 0 field star contaminants are all masked during 1/$f$ noise removal.
    }
    \label{fig:beforeBGsubtraction}
\end{figure*}

\texttt{Ahsoka} employs a slightly modified version of the \texttt{supreme-SPOON} background subtraction algorithm to scale the STScI background model\footnote{ See SOSS Background Observations at \href{https://jwst-docs.stsci.edu/jwst-near-infrared-imager-and-slitless-spectrograph/niriss-observing-strategies/niriss-soss-recommended-strategies\#NIRISSSOSSRecommendedStrategies-SOSSBackgroundObservations} {https://jwst-docs.stsci.edu}.} to group-level median frames of our observations. The 8 median frames (one for each group) are created from the out-of-transit integrations for each of the 8 groups \citep{radica2023}. The background level intensity changes abruptly near column pixel index 700. As discussed by \cite{albert2023}, we apply separate scalings to each side of this break in the background intensity. On the left side of the break, we scale the STScI background model to the following small region in the upper left side of our median frames: x $\in$ [250,500], y $\in$ [210,250]. On the right side of the break, we use x $\in$ [750,850], y $\in$ [210,250]. We chose these regions because they are free of contamination from field stars and the three spectral orders. \texttt{Ahsoka}'s implementation of background scaling to either side of the apparent step function near column 700 differs slightly from the implementation in the \texttt{supreme-SPOON} pipeline. Figure \ref{fig:beforeBGsubtraction} shows an example frame before and after background subtraction, where the scaling regions are outlined using red boxes.

Following background subtraction, we apply the \texttt{supreme-SPOON} group-level 1/$f$ noise subtraction algorithm, masking the field star contaminants and spectral traces, which, like the Zodiacal light background, could bias 1/$f$ noise removal. Next, the noise-weighted average of each column in the group-level median frames is then computed, and this is subtracted from each column of the raw image frames for each group. The final step in the 1/$f$ noise subtraction algorithm is to add the background back into the image frames.  

The \texttt{Ahsoka} detector-level reduction concludes with the following \texttt{jwst} pipeline steps: \texttt{linearity, jump, ramp\_fitting,} and \texttt{gain\_scale.} 

\subsubsection{\texttt{Ahsoka} Stage 2: Spectroscopic processing}\label{subsubsec:Ahsokastage2}

We next apply stage 2 spectroscopic processing\footnote{\texttt{calwebb\_spec2}, see \url{https://jwst-pipeline.readthedocs.io/en/latest/jwst/pipeline/calwebb_spec2.html}} to our stage 1 outputs, beginning with the following \texttt{jwst} pipeline steps: \texttt{assign\_wcs, srctype,} and \texttt{ flat\_field.} We then employ our slightly modified version of the  \texttt{supreme-SPOON} background subtraction algorithm, using the same method (to include scaling regions) as with the detector-level application, except that only one median frame is constructed during stage 2. 

We conclude Stage 2 with the \texttt{supreme-SPOON BadPix} custom cleaning step to flag and correct outlying/hot pixels \citep{radica2023}. The \texttt{BadPix} step first creates a median frame using the out-of-transit integrations from the Stage 2 background subtraction output. Then, each pixel of the median frame is compared to the 5 surrounding pixels on either side of it within the same column. Any pixel with a NaN or negative value, or that differs from surrounding pixels by more than 5$\sigma$, is flagged. Flagged pixels are then replaced by the median value of surrounding pixels. Additionally, a mask is created to record the locations of the flagged pixels on the NIRISS SOSS subarray. Finally, the outlying/hot pixels (indicated by the mask) in each integration frame (those output from the Stage 2 background subtraction step) are replaced by the corresponding pixel values on the corrected median frame, which is scaled to the transit white light curve.

\subsubsection{\texttt{Ahsoka} Stage 3: Spectral Extraction}\label{subsubsec:Ahsokastage3}

In \texttt{Ahsoka} stage 3, we apply the box extraction algorithm developed as part of the \texttt{nirHiss} pipeline \citep{feinstein2023} to the \texttt{BadPix} step output frames, thus generating a time series of 1D stellar spectra for NIRISS SOSS orders 1 and 2. The NIRISS SOSS wavelength solution has been shown to shift up to a few pixels between visits due to slight variations in the position of the pupil wheel, which maneuvers the GR700XD grism into the optical path. We therefore use the STScI-developed \texttt{PASTASOSS} package\footnote{https://github.com/spacetelescope/pastasoss} to determine the order 1 and order 2 spectral trace positions and wavelength solutions. \cite{baines2023wavelength} developed the \texttt{PASTASOSS} package using data from commissioning and calibration programs to map the variation in the locations of the order 1 and 2 traces as a function of the pupil wheel position. They  applied two independent polynomial regression models (one for each spectral order) to derive the spectral trace positions and wavelength solutions to sub-pixel level accuracy \citep{baines2023wavelength,baines2023traces}.

We varied spectral extraction widths between 24 and 38 pixels, ultimately choosing widths of 36 pixels for order 1 and 35 pixels for order 2. We chose these extraction widths because they minimized the median absolute deviations (MAD) for the out-of-transit normalized flux values of the raw white light curves for each order. While the \texttt{PASTASOSS} spectral trace positions for each column are computed in floating point values, the pixels used by the \texttt{nirHiss} box extraction algorithm are integer values. To find the pixels used for spectral extraction in a given column, the \texttt{nirHiss} box extraction algorithm adds and subtracts half the spectral width to the spectral trace position for that column, and then truncates the resulting values to integers. This computation yields the lowest and highest pixel indices which are included when summing the flux across the spectral width.

\begin{figure*}
    \includegraphics[width=\textwidth]{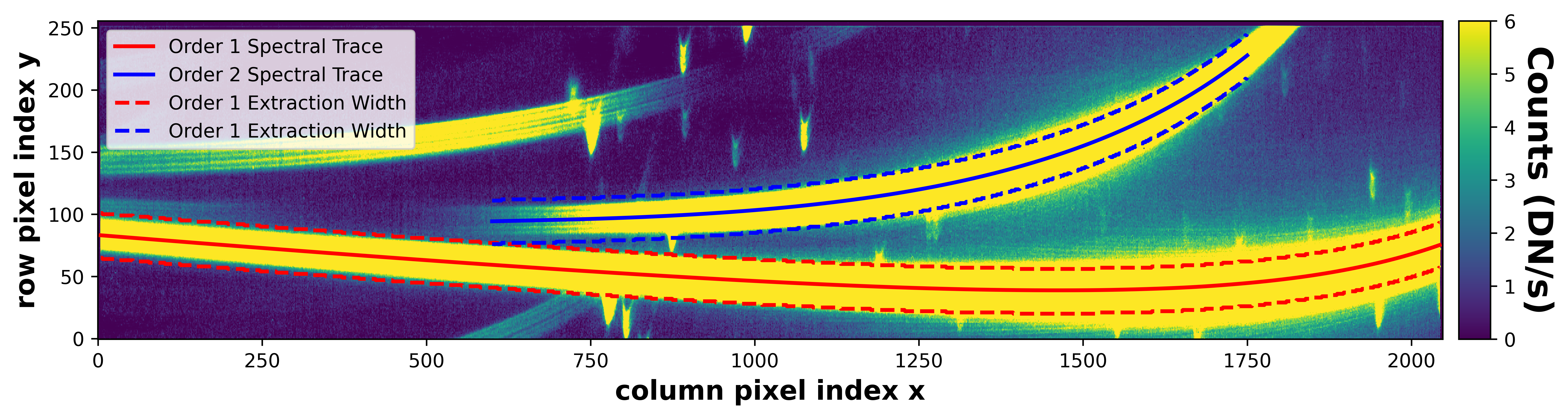}
    \caption{\texttt{Ahsoka} spectral traces and spectral widths. We employ the STScI-developed \texttt{PASTASOSS} package to determine the spectral traces and wavelength solutions to sub-pixel level accuracy \citep{baines2023wavelength,baines2023traces}. Our spectral extraction widths were chosen to minimize the raw white light curve median absolute deviations (MAD) for out-of-transit normalized flux values for each order (See Figure \ref{fig:rawWhiteLight}). 
    }    \label{fig:spectralTracesWidthsOnImageFrame}
\end{figure*}

\begin{figure}[t!]
   \includegraphics[width=0.48\textwidth]{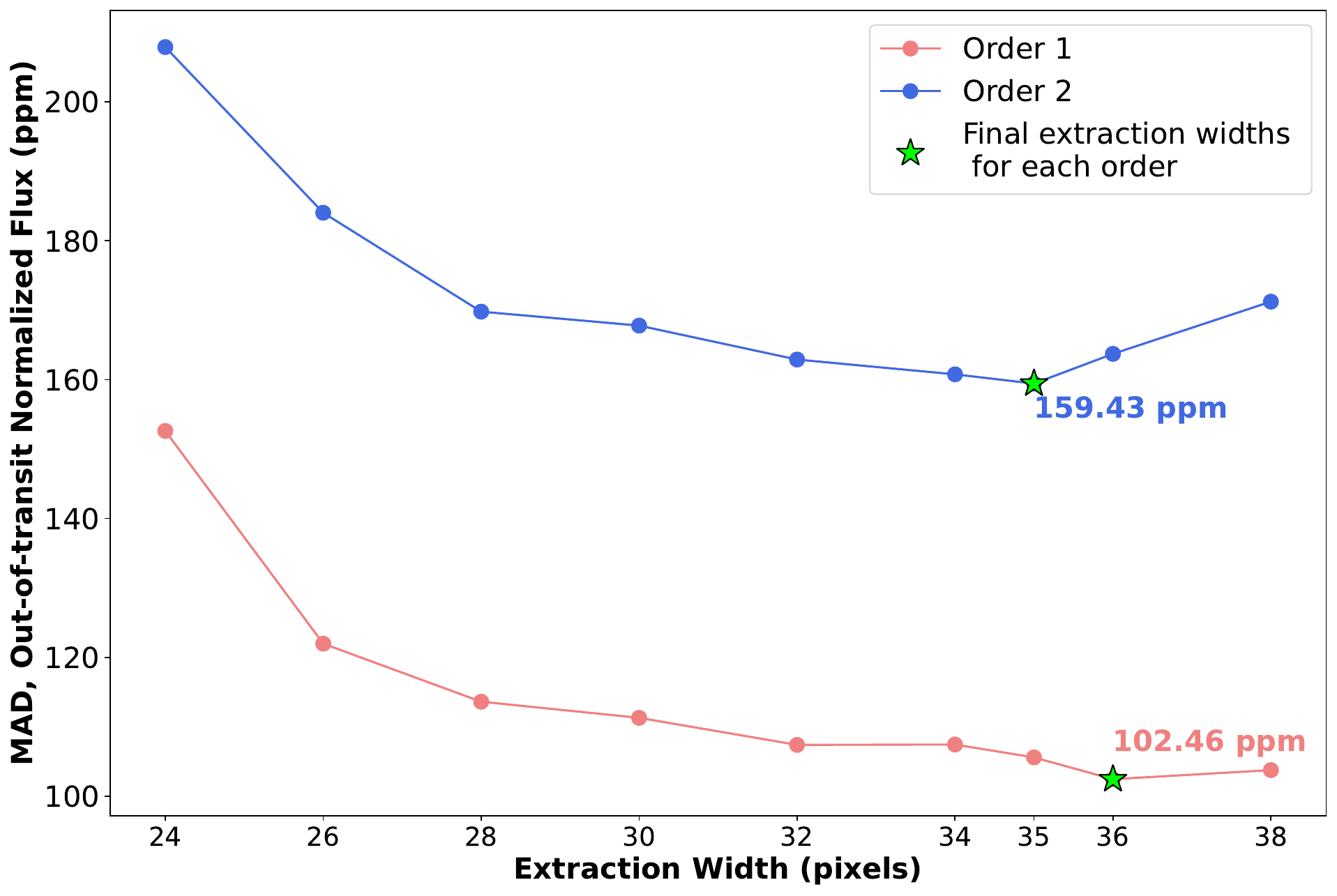}
    \caption{ Median absolute deviation (MAD) values corresponding to various extraction widths for \texttt{Ahsoka} order 1 (red) and order 2 (blue). For each extraction width, we computed a MAD value based upon the raw white light curve out-of-transit (OOT) normalized flux values. We defined the OOT points as those from integration indices 0 through 229, combined with integration indices 600 to 719. Our final spectral extraction widths (green stars) are those that minimize MAD in each order. The MAD values for our final extraction widths are labeled next to the green stars on the plot. See text for additional information.
    }
    \label{fig:rawWhiteLight}
\end{figure}

White light curves are created by summing the flux across all wavelengths (all columns) within the spectral extraction width for \textit{each} integration. These total flux values can then be plotted against integration number (or time) to produce a transit light curve. The flux values are then normalized by the median out-of-transit flux. Here, we defined the out-of-transit integrations as those from integration indices 0 through 229, combined with integration indices 600 to 719. Figure \ref{fig:spectralTracesWidthsOnImageFrame} depicts the spectral traces and spectral widths upon the image frame for integration 0, as output from the stage 2 \texttt{BadPix} step. Figure \ref{fig:rawWhiteLight} shows MAD values corresponding to various extraction widths for order 1 and order 2, with MAD values labeled for our final extraction widths.

We export the extracted stellar spectra for both orders and all integrations to \texttt{HDF5} files in an \texttt{Xarray} format\footnote{See \url{https://kevin218.github.io/Astraeus/}} that are compatible with \texttt{Eureka!} stage 4. The extracted order 1 and order 2 spectra overlap in a region beginning at $\sim$0.83 $\mu$m. The order 1 signal within this overlap region is much higher than that of order 2. We therefore only export order 2 spectra for values less than 0.95 $\mu$m, thus preserving a small region of spectral overlap for comparative purposes. We apply \texttt{Eureka!} to generate and fit white light and spectroscopic light curves, as described in Section \ref{subsec:AhsokaLightCurve}.

\subsubsection{\texttt{Ahsoka}: Use of F277W exposure}\label{subsubsec:AhsokaF277Wusage}

\begin{figure*}
    \includegraphics[width=\textwidth]{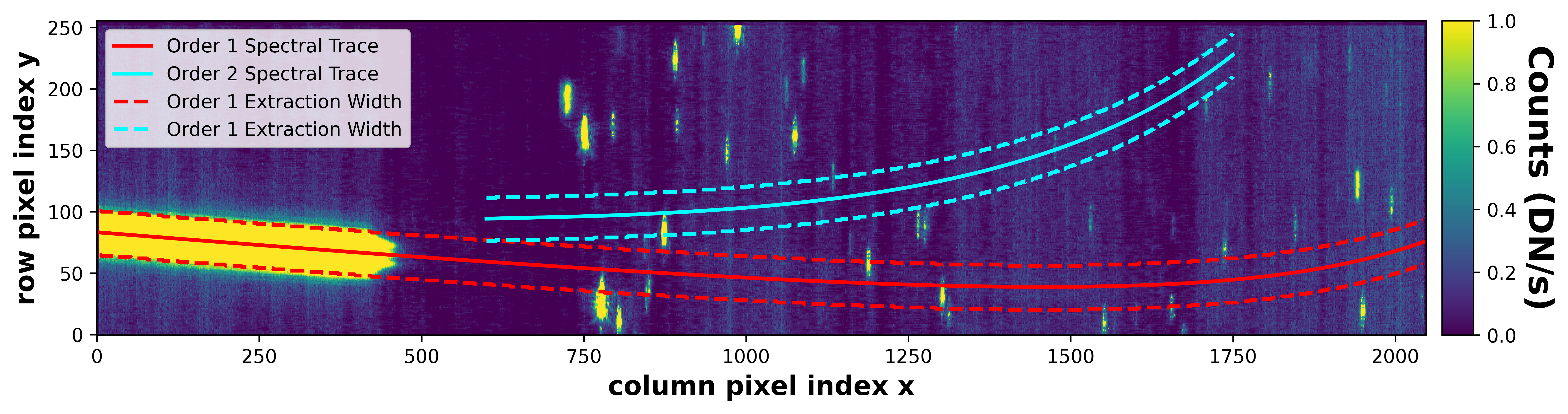}    
    \caption{WASP-17b NIRISS SOSS F277W exposure following application of \texttt{Ahsoka} stages 1 and 2. The \texttt{PASTASOSS}-derived spectral traces and optimized spectral extraction widths are shown for reference. The F277W exposure reveals bright chevron-shaped order 0 field star contaminants, which dilute the transmission spectrum at those wavelengths corresponding to regions where they overlap the order 1 and order 2 spectral extraction regions. Through examination of the spatial cross sections of the order 0 field star contaminants, we determined the most significant contaminant lies between columns 768 and 791 of order 1. See Figure \ref{fig:ExtractedStellarSpectra} for details. } \label{fig:F277WSubarrayWSpectralTraces}
\end{figure*}

\texttt{Ahsoka} makes use of the F277W exposure to identify the location of field star contaminants. We apply the same \texttt{Ahsoka} stage 1 and 2 steps outlined previously to the F277W \textit{uncal.fits} file. Figure \ref{fig:F277WSubarrayWSpectralTraces} shows the median frame constructed from all integrations of our WASP-17b F277W observations. We overlay the \texttt{PASTASOSS}-derived spectral traces and optimized spectral extraction widths for reference. Order 0 field star contaminants appear as bright chevron-shapes, which are somewhat narrow in the wavelength direction, but spread out spatially due to the cross-dispersing prism \citep{albert2023}. Order 0 field star contaminants will dilute the transmission spectrum at any wavelengths corresponding to regions where they overlap with extracted spectra. As explained by \cite{albert2023}, order 0 field star contaminants only appear at column pixel indices higher than $\sim$700.

\begin{figure*}
    \includegraphics[width=\textwidth]{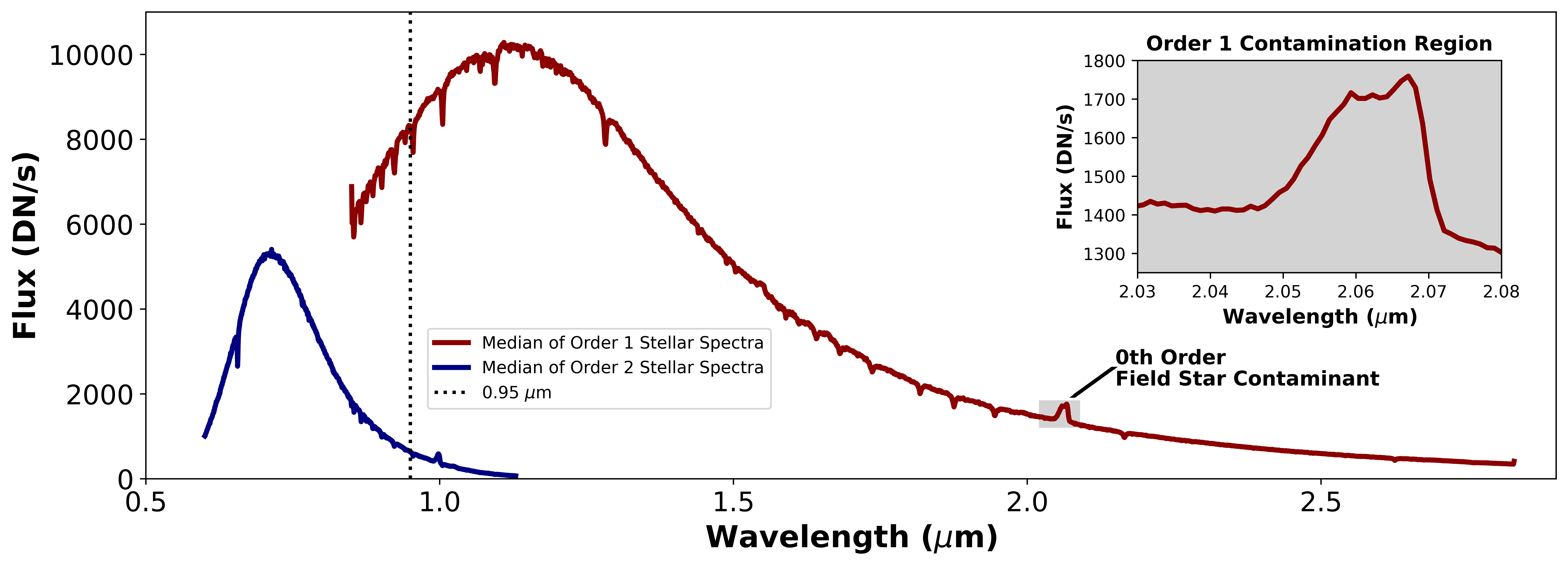}
    
    \caption{ Examination of order 0 field star contaminant. Here, we construct the median of our time series of extracted 1D stellar spectra. The order 0 field star contaminant visible in Figure \ref{fig:F277WSubarrayWSpectralTraces} between columns 768 (2.050 \textmu m) and 791 (2.073 \textmu m) manifests as excess flux in order 1 over the corresponding wavelength range. We depict the contamination region using gray shading. Field star contaminants dilute the transmission spectrum at wavelengths where the contamination occurs. A field star contaminant is also visible near 1.0 \textmu m in order 2, but we do not extract order 2 wavelengths beyond 0.95 \textmu m (indicated by dotted line), since the signal-to-noise for order 1 is much higher in this wavelength regime. 
    }
    \label{fig:ExtractedStellarSpectra}
\end{figure*}

Visual inspection reveals $\sim$4 order 0 field star contaminants of potential concern for order 1 and one for order 2. We cross-reference the location of the order 0 contaminants with the \texttt{PASTASOSS} wavelength solution to identify those wavelengths that may be contaminated by field stars in our transmission spectrum. We examined the spatial cross sections of the major order 0 field star contaminants and determined that the most significant contaminant for order 1 lies from column 768 to column 791, corresponding to wavelengths from 2.050 $\mu$m (column 791) to 2.073 $\mu$m (column 768). The most significant contaminant for order 2 lies near 1.0 $\mu$m, but this contaminant is not of concern since it is beyond 0.95 $\mu$m and therefore we do not use it in our analysis (See $\S$\ref{subsubsec:Ahsokastage3}). In Figure \ref{fig:ExtractedStellarSpectra} we plot the median of our time series of extracted 1D order 1 and order 2 stellar spectra. The order 0 field star contaminant just beyond 2.0 $\mu$m is clearly visible in the median stellar spectrum. The field star contaminant near 1.0 $\mu$m in order 2 is also visible. 

At this time, \texttt{Ahsoka} does not correct order 0 field star contaminants. The transit depths at the impacted wavelengths are masked in our interpretation analyses. Additionally, we note that the \texttt{supreme-SPOON} analysis procedure ($\S$ \ref{subsec:supremeSpoonReduction}) does correct for field star contaminants. As described in Section \ref{subsec:supremeSpoonReduction}, the \texttt{supreme-SPOON} pipeline also attempted to correct the dispersed contaminant visible in Figure \ref{fig:beforeBGsubtraction} near order 1 between columns 550 and 800.

\subsection{\texttt{transitspectroscopy} Data Reduction} \label{subsec:transitspectroscopyReduction}
We independently reduced the data from WASP-17b's transit observation using the \texttt{transitspectroscopy} pipeline \citep{espinoza_nestor_2022}\footnote{\url{https://github.com/nespinoza/transitspectroscopy}}, following the process described in \cite{Gressier2024_arxiv}. This pipeline uses stage 1 \textit{rateints.fits} files from the \texttt{jwst} pipeline, applying default settings for various corrections including group scale, data quality initialization, saturation detection, superbias subtraction, reference pixel correction, linearity correction, dark current subtraction, jump detection, ramp fitting, and gain scale. \texttt{transitspectroscopy} is then used to determine order 1 and order 2 trace positions, subtract the zodiacal background, correct for 1/$f$ noise, extract the stellar spectrum, and generate light curves. 

We trace the positions of NIRISS SOSS order 1 and order 2 using the \texttt{transitspectroscopy.trace\_spectrum} routine, which determines the trace center by maximizing cross-correlation with each detector column. We employ a double Gaussian input function with parameters derived from prior observations. The tracing spans $x$-pixels from 4 to 2043 for order 1 and 750 to 1750 for order 2. Subsequently, we smooth trace positions using spline functions. 

To remove the background, we scale the model background provided in the STScI JDox User Documentation to match our observations. We select a small portion of median integrations from the NIRISS SOSS subarray [x\,=\,210:250, y\,=\,500:800], encompassing the ``pick-off mirror'' region without order 0 contaminants \citep{albert2023}. By computing the ratio between the pixels in this portion and the background model, we determine a scaling factor. We adjust the background model by this scaling factor, and then subtract the adjusted background from all integrations to ensure consistent background correction. 

To correct for 1/$f$ noise in each integration frame, we generate an out-of-transit median frame after correcting for the zodiacal background. Then, we evaluate the 1/$f$ noise for each column of the residual frames by computing the median for pixels within a range of 20 to 35 pixels from the trace center. This median represents the estimated 1/$f$ noise in the column and is subsequently subtracted from each column across all integrations.

The stellar spectral extraction is conducted on the background and 1/$f$ noise-corrected frames using the \texttt{transitspectroscopy.spectroscopy. getSimpleSpectrum} routine. This routine employs a 15-pixel radius aperture extraction, totaling 30 pixels, centered around the trace positions, utilizing a box-extraction method to calculate the summed flux within the defined aperture. To address potential outliers in the time series of 1D stellar spectra for order 1 and order 2, we apply a correction. Outliers exceeding a 5$\sigma$ threshold are replaced using a 1D median version of the spectra. We did not correct for order 0 contamination and clipped the wavelength bins impacted by the contaminants after the extraction of the spectrum. The \texttt{transitspectroscopy} pipeline identified the locations of the impacted wavelengths using the F277W filter, similar to the technique implemented by the Ahsoka pipeline ($\S$ \ref{subsubsec:AhsokaF277Wusage}).

\subsection{\texttt{supreme-SPOON} Data Reduction} \label{subsec:supremeSpoonReduction}

In parallel, we perform a reduction using the \texttt{supreme-SPOON} pipeline \citep{feinstein2023, radica2023, coulombe2023, lim2023, Radica2024, Radica2024_exotedrf, Radica2024b, Piaulet2024}, which performs the end-to-end reduction of NIRISS SOSS TSOs. We closely follow the steps laid out in \citet{radica2023} for Stages 1 -- 3. During the correction of column-correlated 1/$f$ noise, we mask all undispersed order 0 contaminants of background field stars, as well as the dispersed contaminant below the target order 1 between columns 550 and 800. We perform a ``piecewise" background subtraction \citep[e.g.,][]{lim2023, Fournier-Tondreau2024}, whereby we separately scale the STScI background model either side of the background ``step" near column 700, using scaling values of 0.75590 and 0.79568, respectively red-wards and blue-wards of the step.

We perform the spectral extraction using the ATOCA algorithm \citep{darveau-bernier2022, radica2022} to explicitly model the self-contamination of the first two diffraction orders of the target spectra on the detector, though this is expected to be minimal. We use an extraction aperture of 32 pixels, as we find this minimizes the scatter in the white light curves.

After the spectral extraction, we correct for the contamination from undispersed background sources following the methodology presented in \citet{radica2023}. We account for the dilution caused by six contaminants in order 1 and one in order 2. However, we are unable to correct for the effects of the dispersed contaminant intersecting the target order 1, visible between columns 550 and 800. The reasons for this are two-fold: firstly, there is simply not enough of the contaminant trace visible on the detector to accurately estimate the spectrum of the contaminant star, and thus extrapolate the resulting contamination on the target trace \citep[e.g.,][]{radica2023}. Furthermore, simulations using a custom SOSS contamination tool\footnote{\url{http://maestria.astro.umontreal.ca/niriss/SOSS_cont/SOSScontam.php}}, show that this dispersed contaminant is actually the very bluest end of an order 1 trace ($\lambda<0.75$\,µm). We do not see this segment of order 1 in the target trace as it is truncated by the detector edge at $\sim$0.85\,µm. As such, reference throughput is relatively poor in this region, all but precluding the necessary flux calibration of the contaminant.

\subsection{\texttt{Ahsoka} Light Curve Fitting}\label{subsec:AhsokaLightCurve}

We employ \texttt{Eureka!} version 0.10, stage 4, to extract the time series of 1D stellar spectra from \texttt{Ahsoka} and generate both white light and spectroscopic light curves. We fit both our white light (\S \ref{subsubsec:AhsokaWLCs}) and spectroscopic light curves (\S \ref{subsubsec:AhsokaSpectroscopicLCs}) with \texttt{Eureka!} stage 5, making use of the \texttt{emcee}  \citep{Foreman-Mackey2013} affine invariant Markov Chain Monte Carlo (MCMC) ensemble sampler \citep{GoodmanWeare2010} package to find the best fit parameters. For each of our \texttt{emcee} fits, we run 1000 steps, discarding the first 500 as burn-in, using 200 walkers. We model transit light curves with the  \texttt{batman} package \citep{kreidberg2015} and systematics with a polynomial model comprised of constant and linear coefficients. We also fit a white noise multiplicative parameter to the expected noise for the light curves extracted from the time series of 1D stellar spectra. In Sections \ref{subsubsec:AhsokaWLCs} and \ref{subsubsec:AhsokaSpectroscopicLCs}, we describe our application of these tools to the white light curves and spectroscopic light curves, respectively. 

\texttt{Eureka!} stage 5 provides an assortment of diagnostic tools by which we confirmed MCMC convergence and the quality of our fits. We examined Allan deviation (RMS error over various time intervals) to check for correlated noise, normalized residual distributions to confirm our residuals were Gaussian, \texttt{emcee} fitting chains showing fit parameter values versus number of steps, and corner plots of our fits. Our fit parameters converged to median values well before reaching the specified number of burn-in steps. 

\texttt{Eureka!} stage 5 outputs include median values and $1\sigma$ error values for planet-to-star radius ratio ($R_{\rm p}/R_{\star}$).  The lower (upper) error values are computed by finding the differences between the 16th (84th) percentile and median values of $R_{\rm p}/R_{\star}$. \texttt{Eureka!} stage 6 computes the transmission spectrum (\S\ref{subsec:intercomparison}) using these stage 5 output values. Specifically, median transit depths are found by squaring the median values of $R_{\rm p}/R_{\star}$. The lower (upper) errors on transit depth are the differences between the median transit depths and the transit depths computed for the 16th (84th) percentile values of ($R_{\rm p}/R_{\star})^2$.

\subsubsection{\texttt{Ahsoka} White Light Curves}\label{subsubsec:AhsokaWLCs}

\begin{figure*}
    \includegraphics[width=0.5\textwidth]{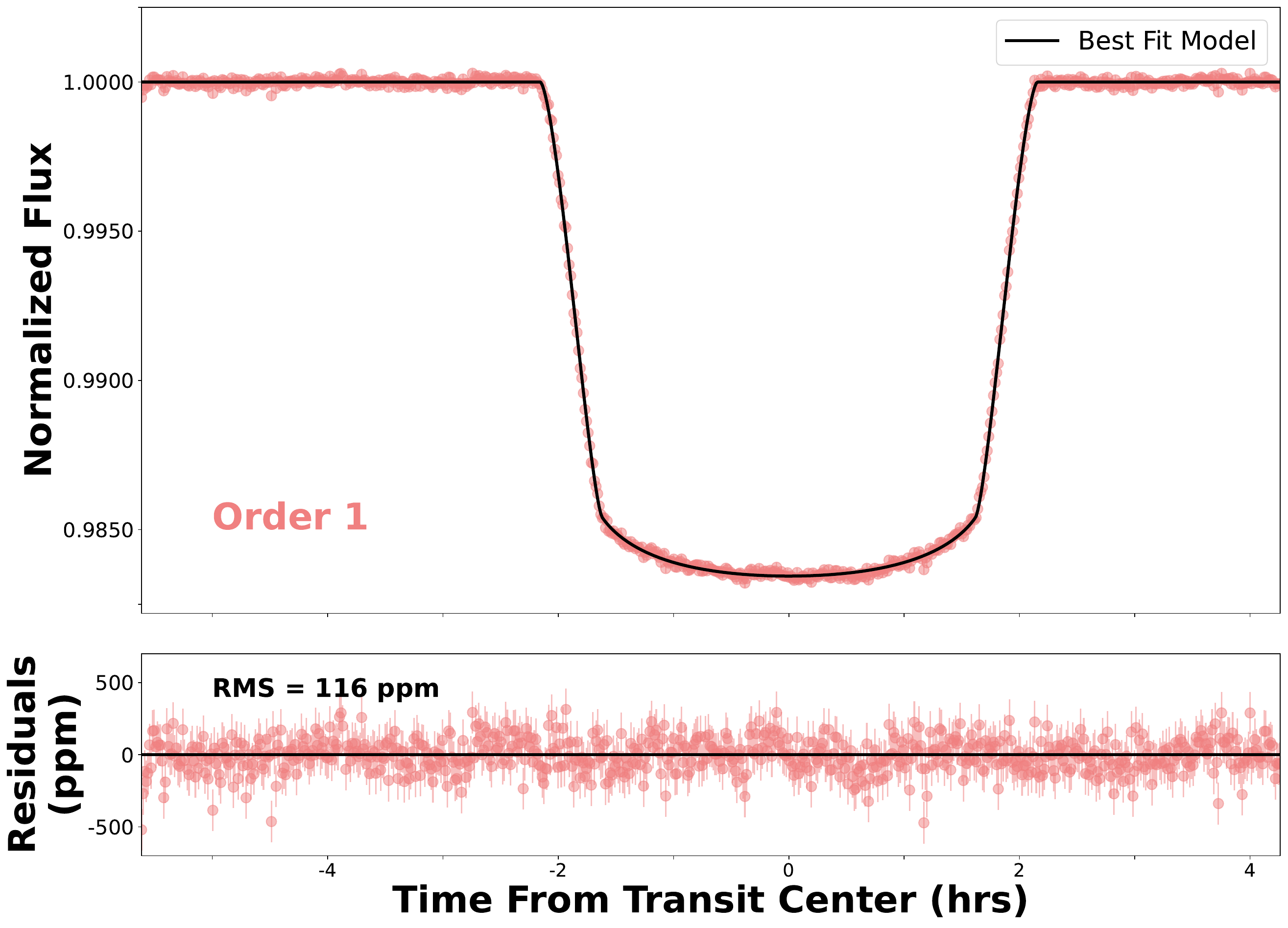}
    \includegraphics[width=0.5\textwidth]{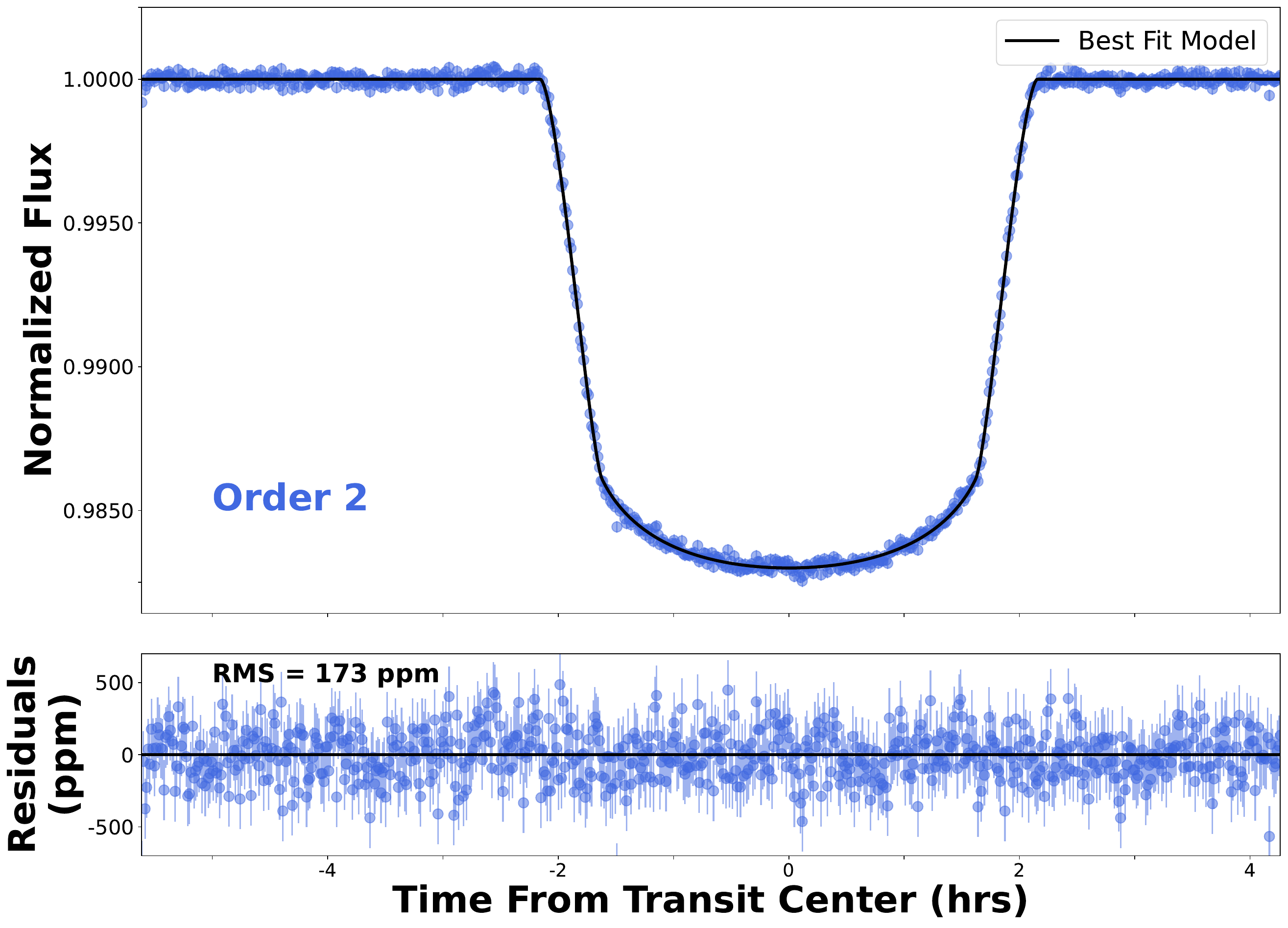}
    \includegraphics[width=0.5\textwidth]{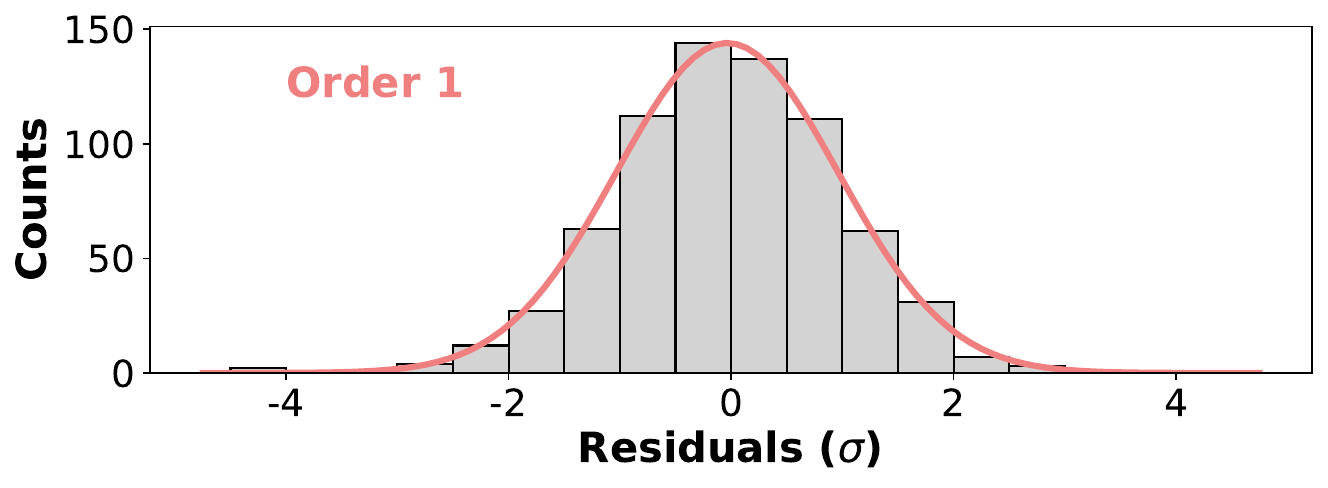}
    \includegraphics[width=0.5\textwidth]{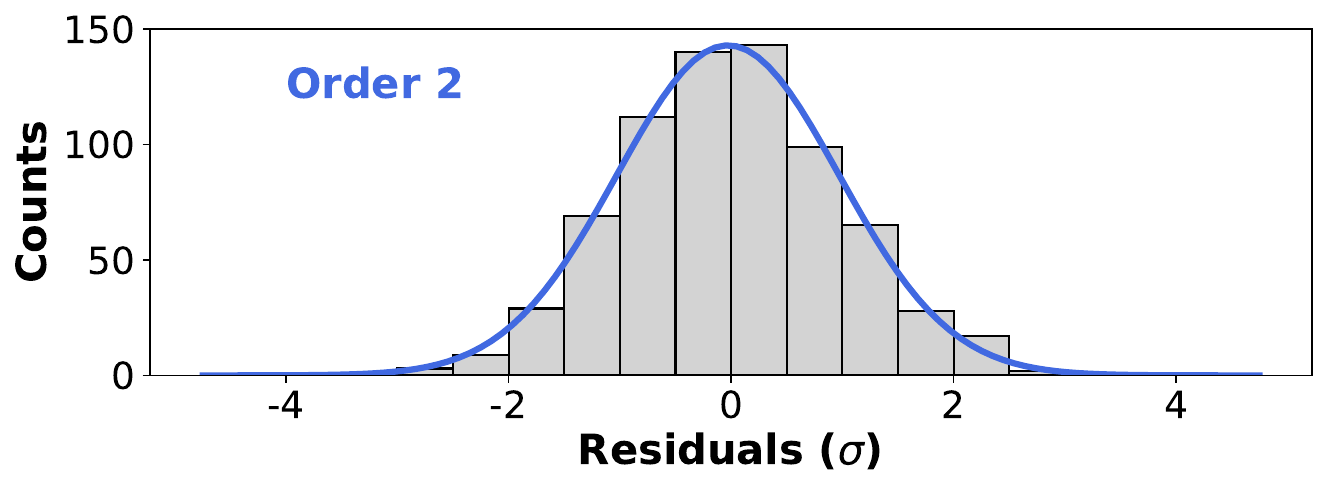}
    \caption{\texttt{Eureka!} \citep{bell2022} stage 5 white light curve \texttt{emcee} \citep{Foreman-Mackey2013} fits to the \texttt{Ahsoka} data reduction for order 1 (\underline{\textbf{left}}) and order 2 (\underline{\textbf{right}}). The \underline{\textbf{top}} panels depict the data corrected with our polynomial systematics model (colored points), and overplotted with the best fit transit model (black curves). The \underline{\textbf{center}} panels show residuals, with RMS scatter indicated in black text. In the top and center plots, mean error bars are 145 ppm for order 1 and 209 ppm for order 2. The \underline{\textbf{bottom}} panels show histogram distributions of residuals in terms of number of standard deviations $\sigma$. We overplot Gaussian curves matching the values of mean and standard deviation for the residuals of each order. The alignment of our distributions with a Gaussian trend indicates the systematics  are well handled. See Table \ref{table:wlc_fitting} for our white light curve prior and best fit parameter values.
    }
    \label{fig:AhsokaFitWLCs}
\end{figure*}

We constructed white light curves for each spectral order by summing the flux across all wavelengths for each of our 720 1D stellar spectra. We then fit the order 1 and order 2 white light curves separately, using the prior parameters and distributions listed in Table \ref{table:wlc_fitting}. Comparing our results (Table \ref{table:wlc_fitting}) between the two spectral orders, our fit values for mid-transit time $t_0$ agree to within 10.0 seconds, while values for the semi-major axis ($a/R_{\star}$) and inclination $i$ agree within the 1-$\sigma$ error bars. $R_{\rm p}/R_{\star}$ and quadratic limb darkening coefficients $u_{1}$ and $u_{2}$ are wavelength dependent, and thus expected to differ between the two spectral orders. Figure \ref{fig:AhsokaFitWLCs} depicts our fits for both order 1 and order 2 white light curves.

\begin{table*}[htpb]
  	\caption{White Light Curve Fitting Parameter Information. }         
  	\label{table:wlc_fitting}
        \begin{flushleft}
  	\begin{tabular}{ l c c c }        
  	\hline\hline
   &\multicolumn{3}{c}{\hspace{28 mm} Fixed Values and \texttt{Ahsoka} Fit Values} \\ 
    \cmidrule(l){3-4}
    Parameters\tablenotemark{a} & Prior &  Order 1 & Order 2 \\
    \hline
    $P$ (days) & fixed \tablenotemark{b}  & 3.73548546 & 3.73548546 \\
    $t_0$ (BMJD$_{\rm TDB}$) & $\mathcal{N}(60023.697, 0.05)$ & $60023.69742169^{+1.871 \times 10^{-5}}_{-1.874 \times 10^{-5}}$ & $60023.69730635^{+2.930 \times 10^{-5}}_{-2.968 \times 10^{-5}}$ \\
    $a/R_{\star}$ & $\mathcal{N}(7.025, 1.46)$ \tablenotemark{c} & $7.1330^{+0.0156}_{-0.0163}$ & $7.1387^{+0.0259}_{-0.0250}$ \\
    $i$ (degrees) & $\mathcal{N}(86.9, 7.0)$ \tablenotemark{b} & $87.2793^{+0.0612 }_{-0.0614}$ & $87.3090^{+0.1039 }_{-0.0965}$ \\
    $e$ & fixed & 0 & 0 \\
    $\omega$ (degrees) & fixed & 90 & 90 \\
    $R_{\rm p}/R_{\star}$ & $\mathcal{N}(0.1255, 0.085)$ \tablenotemark{d}  &  $0.124006^{+0.000103}_{-0.000100}$ & $0.123254^{+0.000198}_{-0.000193}$ \\
    $u_{1}$ & $\mathcal{U}(0, 1)$ \tablenotemark{e}  & $0.1424^{+0.0139}_{-0.0134}$ & $0.2355^{+0.0197}_{-0.0195}$ \\
    $u_{2}$ & $\mathcal{U}(0, 1)$ \tablenotemark{e}  & $0.1632^{+0.0250}_{-0.0264}$ & $ 0.2192^{+0.0385}_{-0.0395}$ \\
    \\ [-2.5 ex]
    \hline \\ [-2.5 ex]
    $c_{0}$ & $\mathcal{N}(1, 0.05)$  & $1.006047^{+5.488 \times 10^{-6} }_{-5.668 \times 10^{-6} }$ & $1.006032^{+8.397 \times 10^{-6} }_{-8.386 \times 10^{-6} }$ \\
    $c_{1}$ & $\mathcal{N}(0, 0.01)$  & $0.0000106^{+3.789 \times 10^{-5} }_{-3.812 \times 10^{-5} }$ & $0.0000519^{+5.44 \times 10^{-5} }_{-5.55 \times 10^{-5} }$ \\
    \\ [-2.5 ex]
    \hline
    \end{tabular}
    \tablenotetext{a}{Parameter definitions: orbital period, $P$; time of transit center, $t_0$, where BMJD$_{\rm TDB} =$ BJD$_{\rm TDB}$ - 2400000.5; semi-major axis in units of stellar radii, $a/R_{\star}$; inclination, $i$; eccentricity, $e$; argument of periastron, $\omega$; planet radius in units of stellar radii, $R_{\rm p}/R_{\star}$; quadratic limb darkening coefficients $u_{1}$ and $u_{2}$; systematics polynomial coefficients $c_{0}$ (constant) and $c_{1}$ (linear). }
    \tablenotetext{b}{\cite{Alderson2022} median values used for $P$ and $i$, with 10$\times$ wider 1-$\sigma$ prior value used in \texttt{Ahsoka} fit for inclination.}
    \tablenotetext{c}{\cite{sedaghati2016} median value used for $a/R_{\star}$, with 10$\times$ wider 1-$\sigma$ prior value.}
    \tablenotetext{d} {\cite{southworth2012} median value used for $R_{\rm p}/R_{\star}$, with wider 1-$\sigma$ value.}
    \tablenotetext{e}{Quadratic limb darkening coefficient prior values u$_{1}$ and u$_{2}$ were computed using \texttt{ExoTiC-LD} \citep{david_grant_2022_7437681}, followed by fitting on a uniform distribution. Prior value of u$_{1}$ for order 1 was 0.1331, while that for order 2 was 0.2435. Prior value of u$_{2}$ for order 1 was 0.1938, while that for order 2 was 0.2192. }
    \end{flushleft}
\end{table*}

\subsubsection{\texttt{Ahsoka} Spectroscopic Light Curves}\label{subsubsec:AhsokaSpectroscopicLCs}

We analyzed the spectroscopic light curves for each spectral order at the pixel-level (one light curve fit per detector column), fitting for $R_{\rm p}/R_{\star}$ as well as $u_{1}$ and $u_{2}$ using the prior distributions shown in Table \ref{table:wlc_fitting}. We used the \texttt{ExoTiC-LD} package \citep{david_grant_2022_7437681,Grant_2024JOSS} to compute prior values for the limb darkening coefficients based upon the wavelengths corresponding to each spectral channel. We fixed values for $a/R_{\star}$ and inclination to the white light curve fit values for each order listed in Table \ref{table:wlc_fitting}. Period $P$, eccentricity $e$, and argument of periastron $\omega$ were fixed to the same values used in the white light curve fits.

\begin{figure*}
    \includegraphics[width=\textwidth]{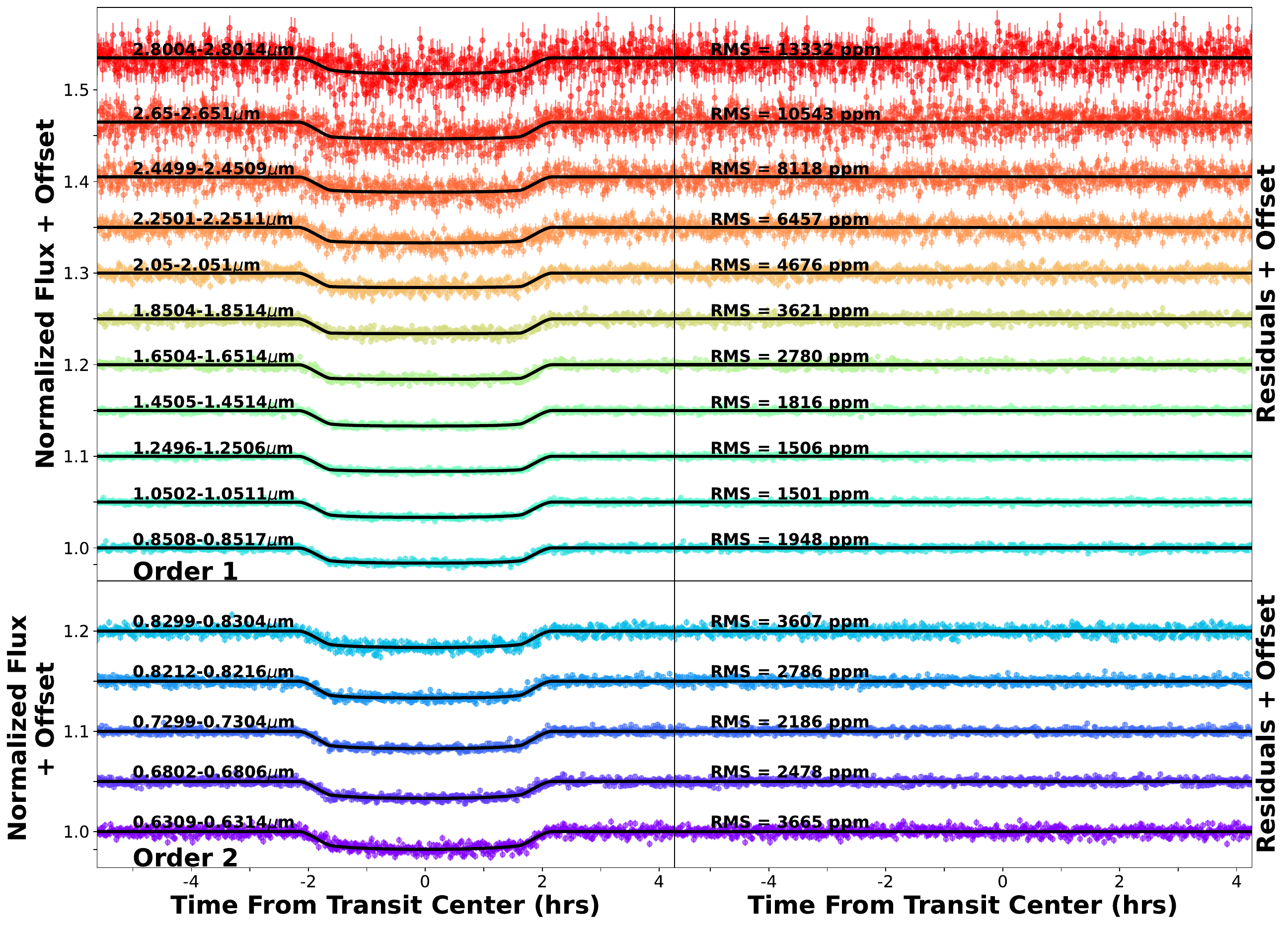}
    
    \caption{ Examples of our \texttt{Eureka!} \citep{bell2022} stage 5 spectroscopic light curve \texttt{emcee} \citep{Foreman-Mackey2013} pixel-level fits to the \texttt{Ahsoka} data reduction for orders 1 and 2. Plots on the left show the pixel-level data corrected with polynomial systematics models (colored points), and overplotted with the best fit transit models (black curves). The wavelengths for each spectral channel are shown above the corresponding transit light curve. The panels on the right show residuals for the corresponding data, with RMS scatter indicated in black text. Data and best fit transit curves are offset for each spectral channel for clarity. See Section \ref{subsubsec:AhsokaSpectroscopicLCs} for further details.}
    \label{fig:AhsokaFitPixelLevelLCs}
\end{figure*}

Our strategy to fit spectroscopic light curves at the pixel-level was motivated by commissioning studies \citep[see, e.g.,][]{espinoza2023}, which showed that the most accurate and precise results from JWST NIR detectors were obtained by analyzing data at the column level, and then binning parameters of interest (e.g., transit depths) during post processing. For the \texttt{Ahsoka} analysis, we fit 2021 spectral channels for order 1, and 705 for order 2. The spectral channels for order 1 consisted of wavelengths $0.85 \mu m \leq \lambda \leq 2.81\mu m$, while those for order 2 encompassed $0.63 \mu m \leq \lambda \leq 0.95\mu m$. We show sample pixel-level light curve fits in Figure \ref{fig:AhsokaFitPixelLevelLCs}.

During post processing, we binned our pixel-level results to R=100 for comparison with the \texttt{transitspectroscopy} and \texttt{supreme-SPOON} pipelines, as well as for interpretation using \texttt{PICASO} (\S\ref{sec:Picaso}), \texttt{petitRADTRANS} (\S\ref{subsec:pRTconfig}), and \texttt{POSEIDON} (\S\ref{subsec:POSEIDONconfig}).

\subsection{\texttt{juliet} Light Curve Fitting (\texttt{transitspectroscopy} Reduction)}\label{subsec:julietLightCurve}
We generate white light curves and pixel-level light curves from the time series of 1D stellar spectra using \texttt{transitspectroscopy}'s reduction. White light curves integrate across all wavelengths for each order. We perform light curve fitting with the \texttt{juliet} Python package \citep{Espinoza_2019}, employing nested sampling via \texttt{dynesty} \citep{Speagle_2020}. The period, eccentricity, and argument of periastron are fixed to the same values used in our \texttt{Ahsoka} analysis, as listed in Table \ref{table:wlc_fitting}. We fit for $R_{\rm p}/R_{\star}$, $t_0$, impact parameter ($b$), and $a/R_{\star}$. Limb-darkening coefficients ($q_1$, $q_2$) are determined using a quadratic law with a uniform prior between 0 and 1. The parameters $q_1$ and $q_2$ are related to $u_1$ and $u_2$ using the \cite{kipping2013} parameterization. Additionally, a mean-out-of-transit offset ($M_{\text{SOSS}}$) and a jitter parameter ($\sigma_{w, \text{SOSS}}$) for white noise are included in the fit. Systematic trends are addressed using a Gaussian Process (GP) with a Matèrn 3/2 kernel via the \texttt{celerite} package \citep{Foreman_Mackey_2017}, with log-uniform priors for the GP amplitude ($\sigma_{\mathrm{GP}_\mathrm{SOSS}}$) and length-scale ($\rho_{\mathrm{GP}_\mathrm{SOSS}}$) between 10$^{-5}$ and 10$^{3}$. We decided to fit the white light curves from Order 1 and Order 2 separately using \texttt{juliet}. This approach allows us to account for the unique characteristics and correlated noise in each order independently. We provide prior and fit values for the \texttt{transitspectroscopy/juliet} analysis in Appendix Table \ref{table:wlc_fitting_transitspectroscopy}.

For spectral light curve fitting at pixel resolution, we adopt a similar setup to the white light curve, but we fix the values of $t_0$, $b$, and $a/R_{\star}$ based on the combined best-fit results of order 1 and order 2 white light curve fits. This approach enhances the interpretability of the results by refining the orbital parameters with information from both white light curve analyses. Additionally, it allows us to focus on the wavelength-dependent variations and other parameters of interest without introducing additional degrees of freedom. The detrending model follows a similar approach to that of the white light curve. The values found from the white light curve fits and then used in the spectroscopic fits are as follows: $t_0 = 60023.697373797 \pm 4 \times 10^{-5}$ BMJD$_{\rm TDB}$, $b = 0.34 \pm 0.01$, and $a/R_{\star} = 7.12 \pm 0.03$. The limb-darkening coefficients $q_1$ and $q_2$ are fitted using a truncated normal distribution, with the mean set to the value estimated using the \citet{Morello_2020} \texttt{
ExoTETHyS} package between 0 and 1, and standard deviation set to 0.1. The transmission spectrum is obtained by binning down the pixel-level transit depths, resulting in a resolution of 100. The pixel-level transit depths and corresponding transit depth errors reported for the \texttt{transitspectroscopy/juliet} analysis represent the median and the variance of the posterior distribution of $R_p/R_*$ for each wavelength bin.

\subsection{\texttt{supreme-SPOON} Light Curve Fitting }\label{subsec:supremeSPOONLightCurve}

We fit the \texttt{supreme-SPOON} light curves using the \texttt{juliet} package \citep{Espinoza_2019}. Firstly, we jointly fit the order 1 and 2 white light curves, such that the fit shared all orbital parameters ($P$, $t_0$, $a/R_*$, $i$) between both orders, but individually fit the transit depth, two parameters of the quadratic limb darkening law following the \citet{kipping2013} parameterization, and an additive error inflation term individually to each order. We found the light curves to be well-behaved, and adequately fit without any additional systematics model. Wide uninformative priors were used on all parameters, except for the period, which was fixed to 3.73548546 days \citep{Alderson2022}. 

We fit the spectrophotometric light curves at the pixel-level, that is one light curve per detector column. We fixed the planet's orbital parameters (i.e., $t_0$, $i$, $a/R_*$) to the values listed in Table \ref{table:wlc_fitting}, except for limb-darkening, for which we used Gaussian priors centered around the predictions of ExoTiC-LD \citep{david_grant_2022_7437681}, with widths of 0.2 \citep{patelEspinoza2022}. As with the white light curve fits, we do not include any additional systematics model. The final transit depths used in the analysis that follows are the medians of the resulting posterior distributions, and the transit depth errors represent the 16$\rm ^{th}$--84$\rm ^{th}$ percentile range of the sampled $R_p/R_*$ values for each wavelength bin, after marginalizing over all other fitted parameters.

\subsection{Transmission Spectra Intercomparison} \label{subsec:intercomparison}

\begin{figure*}
    \includegraphics[width=\textwidth]{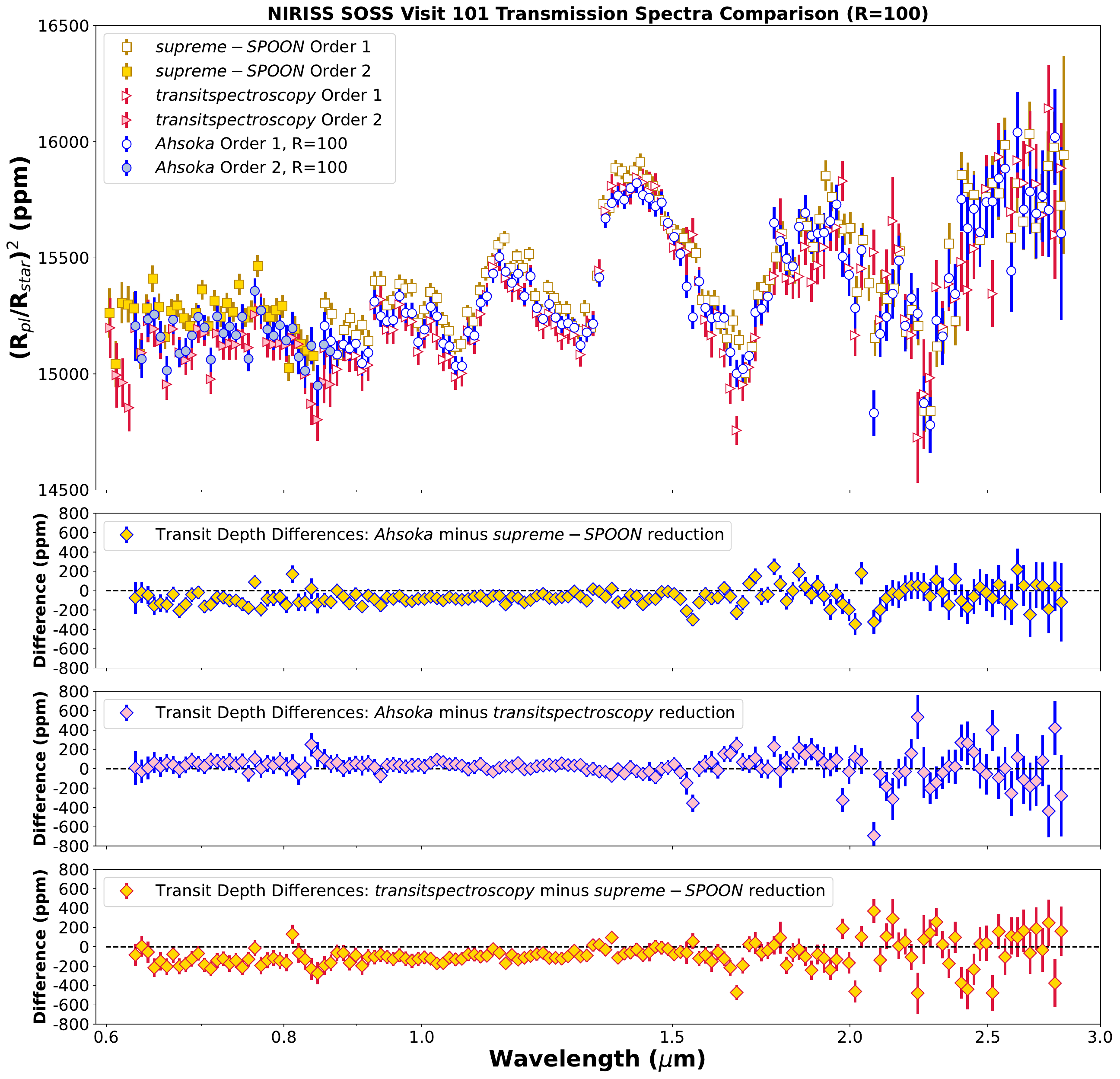}
    
    \caption{WASP-17b NIRISS SOSS transmission spectra from our three independent pipelines, presented at R=100. \textbf{Top:} Order 1 and order 2 reductions from each pipeline. \textbf{Bottom Three Panels:} We plot the differences in transit depths between pipelines. In the wavelength region where we have transit depths for both order 1 and order 2, we compare only the values for order 1. Additionally, we make no comparisons for points where we do not have transit depth values for all three pipelines. (For example, the \texttt{Ahsoka} analysis does not include data shortward of $0.63 \, \mu$m, so we make no comparisons at those shorter wavelengths.) See text for a discussion of similarities and differences between the results of the three pipelines. 
    }
    \label{fig:R100TransmissionSpectraComparison}
\end{figure*}

We present the WASP-17b NIRISS SOSS transmission spectra from our three independent pipelines in Figure \ref{fig:R100TransmissionSpectraComparison}. The average (median) uncertainties across the order 1 spectra produced by our three pipelines are 78 (55) ppm for \texttt{Ahsoka}, 88 (69) ppm for \texttt{transitspectroscopy}, and 63 (46) ppm for \texttt{supreme-SPOON}. The average (median) precision across the order 2 spectra of our three pipelines are 67 (60) ppm for \texttt{Ahsoka}, 79 (73) ppm for \texttt{transitspectroscopy}, and 57 (52) ppm for \texttt{supreme-SPOON}. In the text that follows, we describe the similarities and differences between the three pipelines, and point to the reasons for any apparent differences in our resulting transmission spectra.

Our team standardized many options in our data analysis to better enable comparison between our pipelines, and thus verification of our data analysis results. All pipelines analyzed the data at pixel-level resolution \citep[i.e., column on the detector, see][]{espinoza2023}. The resulting pixel-level transit depths and errors were then binned to R = 100. We established prior values from the literature to use for system parameters during our initial white light curve fits (see Tables \ref{table:wlc_fitting} and \ref{table:wlc_fitting_transitspectroscopy}), although the \texttt{Ahsoka} and \texttt{transitspectroscopy} pipelines then used their white light curve fit values as priors in their spectroscopic fits. All pipelines used quadratic limb darkening coefficients, and all pipelines fit for limb darkening values (see also \S\ref{subsec:limbdarkeningtreatment}). 

The \texttt{Ahsoka} and \texttt{supreme-SPOON} pipelines share many similarities in their stage 1 (detector-level) and stage 2 (spectroscopic) processing, since \texttt{Ahsoka} makes use of the \texttt{supreme-SPOON} background subtraction, 1/f noise removal, and \texttt{BadPix} algorithms. However, during background subtraction, the two pipelines differ in their exact implementation of STScI background model scaling to either side of the ``step'' near column 700 (\S\ref{subsubsec:Ahsokastage1}, \S\ref{subsec:supremeSpoonReduction}). While the \texttt{Ahsoka} and \texttt{supreme-SPOON} pipelines begin with \textit{uncal.fits} files, the \texttt{transitspectroscopy} pipeline begins analysis with \textit{rateints.fits} files downloaded from MAST. During spectral extraction, both \texttt{Ahsoka} and  \texttt{transitspectroscopy} rely upon a simple box extraction, although the pipelines differ in the computation of spectral traces and their choice of extraction widths. These minor differences should primarily affect S/N across the transmission spectra. The \texttt{supreme-SPOON} pipeline makes use of the \texttt{ATOCA} algorithm for spectral extraction, which models contamination between the two spectral orders, which is expected to be minimal \citep{darveau-bernier2022,radica2022}. In this work, the only pipeline that corrects for field star contaminants is \texttt{supreme-SPOON}. 

During light curve fitting, both the \texttt{transitspectroscopy} and \texttt{supreme-SPOON} pipelines employed the \texttt{juliet} package, using \texttt{dynesty} nested sampling. \texttt{Ahsoka} made use of \texttt{Eureka!} with the \texttt{emcee} MCMC package. The three pipelines differed in their treatment of systematics, with \texttt{Ahsoka} using a polynomial with constant and linear terms, \texttt{transitspectroscopy} a GP model with a Matèrn 3/2 kernel, and \texttt{supreme-SPOON} finding that the light curves were well behaved so they did not require any additional systematics model. 

Referencing the top panel of Figure \ref{fig:R100TransmissionSpectraComparison}, all three transmission spectra show spectral water features clearly distinguishable by eye. Additionally, the order 2 spectra from each pipeline exhibit deeper transit depths near the $\sim 0.77 \, \mu$m potassium resonance doublet. 

In the lower three panels of Figure \ref{fig:R100TransmissionSpectraComparison} we see that the three independent analyses agree to within the 1$\sigma$ errors across a large portion of the spectrum. However, some larger differences are apparent just beyond $0.8 \, \mu$m, $1.5 \, \mu$m, and $2.0 \, \mu$m. Below, we examine each of these differences in turn.

The region beyond $0.8 \, \mu$m coincides with the region where we transition from order 2 to order 1 in our composite transmission spectrum. The transit depths from the \texttt{transitspectroscopy} pipeline are consistently lower in this regime. Although the  \texttt{transitspectroscopy} depths are more than 1$\sigma$ lower than the other two reductions in this region of the spectrum, the discrepancy does not impact our spectral interpretation.

The regions near $1.5 \, \mu$m and $2.0 \, \mu$m correspond to areas affected by field star contamination. Whereas \texttt{supreme-SPOON} applied a correction to treat field star contaminants, \texttt{transitspectroscopy} and \texttt{Ahsoka} did not. In Figure \ref{fig:ExtractedStellarSpectra}, we can see a small blip of excess flux near $1.5 \, \mu$m. The figure highlights the contaminant near $2.0 \, \mu$m, which is the area of worst contamination from field stars. (See also Figure \ref{fig:F277WSubarrayWSpectralTraces}.) 

Field star contaminants cause lower transit depths within contaminated regions. Note that the lowest transit depth beyond $2.0 \, \mu$m is removed from both \texttt{Ahsoka} and \texttt{transitspectroscopy}, and is not used in our retrieval analysis. 

\subsection{Effect of Limb Darkening Treatment on Transit Depths} \label{subsec:limbdarkeningtreatment}

Our final transmission spectra presented in Section \ref{subsec:intercomparison} were produced by fitting spectroscopic light curves at the pixel-level, and then binning the pixel-level results to R=100. We emphasize that all pipelines \textit{fit} for quadratic limb darkening coefficients (LDCs) in their final analyses. Our pixel-level fits were based upon JWST commissioning results showing the optimal method to work with near infrared (NIR) detector data is to perform analysis at the instrument's spectral sampling (column-to-column) level, followed by binning during post-processing. Our decision to \textit{fit} LDCs was driven by previous work showing that fitting LDCs using transit light curves can prevent systematic effects introduced by fixing theoretically derived values, which would in turn affect parameters of interest such as transit depths \citep[e.g.,][]{espinoza2015, espinoza2016, patelEspinoza2022}.

Initially, our \texttt{Ahsoka} analysis opted to \textit{fix} quadratic LDCs to the \texttt{ExoTiC-LD} computed values at the pixel-level, whereas the two other pipelines did not. In that original analysis, we found that the resulting \texttt{Ahsoka} transit depths were noticeably higher than those output by the other pipelines at wavelengths beyond $\sim2\,\mu$m. This motivated us to further explore the effects of limb darkening treatments on the \texttt{Ahsoka} transmission spectra. 

Here, we present three versions of the \texttt{Ahsoka} pipeline transmission spectra, all employing \texttt{Eureka!} stage 5 for light curve fitting, and differing only in their treatment of quadratic limb darkening. In each case, we computed quadratic LDCs using \texttt{ExoTiC-LD}. A major distinction is that we \textit{fixed} LDCs in our original treatment, but \textit{fit} for limb darkening using \texttt{ExoTiC-LD} coefficients as priors in the two latter treatments. We summarize our three treatments below.

\begin{enumerate}
    \item Original \texttt{Ahsoka} analysis: quadratic LDCs computed by \texttt{ExoTiC-LD} \textit{fixed} at pixel-level; results binned to R=100.
    \item Intermediate analysis: pixel-level data binned \textit{directly} to R=100; \texttt{ExoTiC-LD} coefficients computed for R=100 bins, which were then used as priors in fits that included limb darkening.
    \item Final \texttt{Ahsoka} analysis: \texttt{ExoTiC-LD} coefficients computed at pixel-level and used as priors in pixel-level fits; results binned to R=100.
\end{enumerate}

\begin{figure*}
    \includegraphics[width=\textwidth]{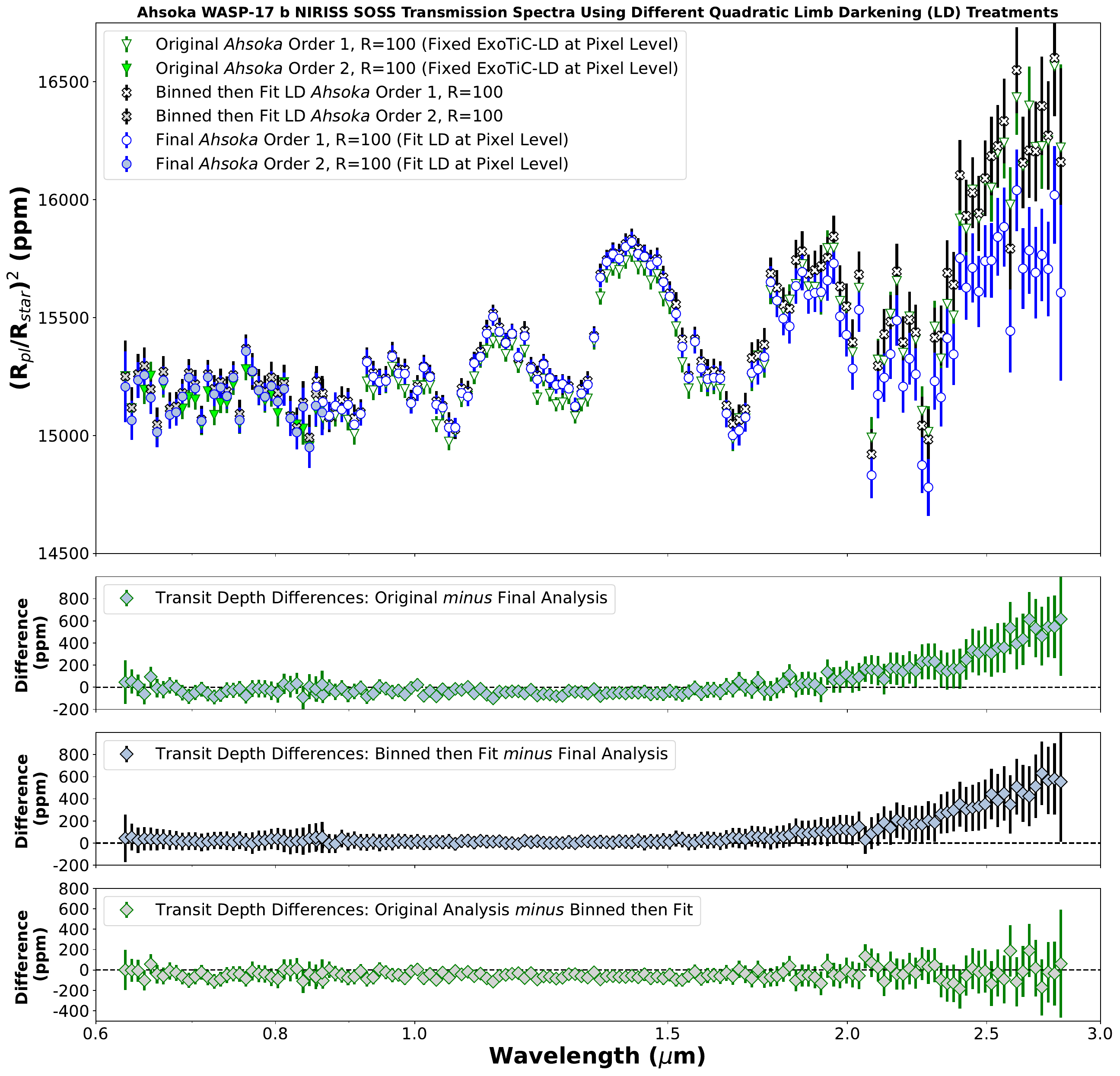}
    
    \caption{Examination of three different Limb Darkening (LD) treatments on \texttt{Ahsoka} WASP-17b NIRISS SOSS R=100 transmission spectrum. The \textbf{top} panel presents the three versions of the transmission spectrum. The \textbf{second (middle top)} and \textbf{third (middle bottom)} panels show the differences between two earlier LD treatments and our final \texttt{Ahsoka} analysis, where we \textit{fit} LD at pixel-level, and then binned our results to R=100. The \textbf{second} panel illustrates this difference with our original analysis, where we \textit{fixed} LD at pixel-level, then binned to R=100. The \textbf{third} panel displays this difference with an analysis where the pixel-level data were binned to R=100 and then fit for LD. The two middle panels reveal that limb darkening treatments have a progressively higher impact at longer wavelengths. The \textbf{bottom} panel compares our two earlier LD treatments, where we see a general agreement to within 1$\sigma$ across most of the NIRISS SOSS bandpass. Note that all spectra are affected by field star contamination in the region just beyond 2\,$\mu$m (see Figure \ref{fig:ExtractedStellarSpectra}), resulting in a region of lower transit depths. See Section \ref{subsec:limbdarkeningtreatment} for detailed discussion. }
    
    \label{fig:R100AhsokaSpectra_3LDTreatments}
\end{figure*}

We present the transmission spectra derived from our three limb darkening treatments in Figure \ref{fig:R100AhsokaSpectra_3LDTreatments}. We report the following findings based upon comparisons between the three treatments.

\begin{itemize}
    \item \textit{Fixing} versus \textit{fitting} LDCs for pixel-level fits. At shorter wavelengths, binned R=100 values largely agree within the 1$\sigma$ error bars, although the analysis using \textit{fixed} LDCs has slightly lower transit depths at these wavelengths. Additionally, the analysis using \textit{fixed} LDCs results in progressively larger transit depths at longer wavelengths, beginning to diverge beyond the 1$\sigma$ error bars at wavelengths greater than $\sim2\,\mu$m, and reaching  differences $\sim600$ ppm at the longest wavelengths in the NIRISS SOSS bandpass. See \textbf{second} panel in Figure \ref{fig:R100AhsokaSpectra_3LDTreatments}.
    \item Two options for LD fits: binning data to R=100, then \textit{fitting} (prebinning), versus \textit{fitting} at pixel-level, then binning to R=100 (postbinning). Of the three subpanel comparisons, these two treatments show the best agreement at \textit{shorter} wavelengths, being virtually indistinguishable by eye in some portions of the transmission spectrum itself (\textbf{top} panel). However, the transit depths progressively diverge from the 1$\sigma$ error bars beyond $\sim2\,\mu$m, and reach differences $\sim600$ ppm at the longest wavelengths in the NIRISS SOSS bandpass See \textbf{third} panel in Figure \ref{fig:R100AhsokaSpectra_3LDTreatments}.
    \item \textit{Fixing} LDCs at pixel-level, then binning results to R=100, versus binning data to R=100, then \textit{fitting} (prebinning) LDCs. These LD treatments agree to within 1$\sigma$ across most of the NIRISS SOSS bandpass, although \textit{fixing} LDCs at pixel-level yields consistently lower transit depths at shorter wavelengths. See \textbf{bottom} panel in Figure \ref{fig:R100AhsokaSpectra_3LDTreatments}.
\end{itemize}

Other recent JWST work has examined the question of whether to \textit{fix} or \textit{fit} LDCs, and also whether to analyze data at the pixel-level, then bin (postbinning), or whether to bin first and then fit light curves (prebinning). \cite{holmberg2023} examined the effects of \textit{fixing} versus \textit{fitting} LDCs for NIRISS SOSS observations of the gas giant exoplanets WASP-96b and WASP-39b. They found that \textit{fitting} LDCs can significantly increase the uncertainties in the resultant transmission spectrum for planets such as WASP-96b, which has a relatively high impact parameter. \cite{May2023} compared the effects of prebinning versus postbinning on NIRSpec G395H data for the rocky exoplanet GJ 1132b. To minimize uncertainties in their final transmission spectrum for this small planet, \cite{May2023} ultimately chose to prebin their NIRSpec G395H data, and then \textit{fixed} LDCs during \texttt{Eureka!} light curve fitting.

Our findings here call for careful consideration in the application of limb darkening treatments to analysis of JWST NIR transmission spectroscopy data. In the case of NIRISS SOSS, different treatments produce disagreements in median WASP-17b transit depths up to $\sim600$ ppm---well outside the 1$\sigma$ error bars---at the longest wavelengths. When comparing JWST transmission spectra produced by multiple pipelines, variations in limb darkening treatments should be examined as potential sources of any differences in resultant transit depths.

\section{Atmospheric Forward Modeling with \texttt{PICASO}} \label{sec:Picaso}

To analyze the transmission spectrum of NIRISS SOSS, we largely follow the model fitting procedure introduced in \citet{grant_miritransit_2023}. Specifically, we employ the identical grid of radiative-convective thermochemical equilibrium (RCTE) models developed to analyze the MIRI LRS spectrum, and we employ the identical procedure to fit for the presence of clouds. We briefly describe the procedure below and also highlight one minor modification, which allows us to jointly fit for HST STIS, NIRISS SOSS and MIRI LRS.  

The climate grid introduced in \citet{grant_miritransit_2023} was computed using the open-source model \texttt{PICASO} v3.1 \citep{batalha2019exoplanet, mukherjee2023picaso}. Note, that since the release of v3.1 modifications have been made to the climate code, however, they only affect models with high internal temperatures $>$300 K, and not those used in this analysis. The grid of models is computed as a function of interior temperature (200 K \& 300 K), atmospheric metallicity (9 values logarithmically spaced between 1--100$\times$Solar), C/O ratio (5 values between 0.25--2$\times$Solar), and the heat redistribution factor (0.5, 0.6, 0.7, 0.8). In the methodology used in \texttt{PICASO}, a heat redistribution factor of 0.5 represents the case of fully efficient heat redistribution \citep{mukherjee2023picaso}. We compute chemical equilibrium using the procedure developed in  \citet{gordon1994computer}, \citet{fegleylodders1994}, \citet{lodders99}, \citet{lodders02}, \citet{LoddersFegley2002}, \citet{visscher06}, and \citet{channon10}, using elemental abundances from \citet{Lodders2010} (where Solar C/O=0.458). Note, because solar elemental abundances are not consistent between model setups, we report our results moving forward in units of absolute C/O, as opposed to relative to Solar. 

Because clouds have already been confirmed via the analysis of \citet{grant_miritransit_2023}, it was necessary to add aerosols in the fitting procedure. Because of the difficulties in converging self-consistent models in climate and aerosol \citep[e.g.][]{Diamondback} we follow the methodology of \citet{grant_miritransit_2023} and use the cloud-free RCTE pressure-temperature profiles and compute approximate cloudy  profiles using the open source cloud code \texttt{Virga} \citep{Batalha2020, Rooney2022}. \texttt{Virga} is based on the methodology of \citet{Ackerman-Marley2001}, which has been widely used to model both exoplanets \citep{wakeford2016high}, and Brown Dwarfs \citep{Burningham2021}. \texttt{Virga} leverages two parameters---vertical turbulent diffusion (K$_{zz}$) and sedimentation (f$_\mathrm{sed}$)---to compute the mean effective particle radius as a function of altitude, and cloud optical depth as a function of both altitude and wavelength. Available optical properties of SiO$_2$ do not vary significantly at the NIRISS SOSS wavelengths, thus we use the optical properties for SiO$_2$(s) $\alpha$-crystal computed at 928 K by \citet{zeidler2013optical} because they were shown to better represent the spectral features in the MIRI LRS data \citep{grant_miritransit_2023}.

\begin{figure*}
    \includegraphics[width=0.95\textwidth]{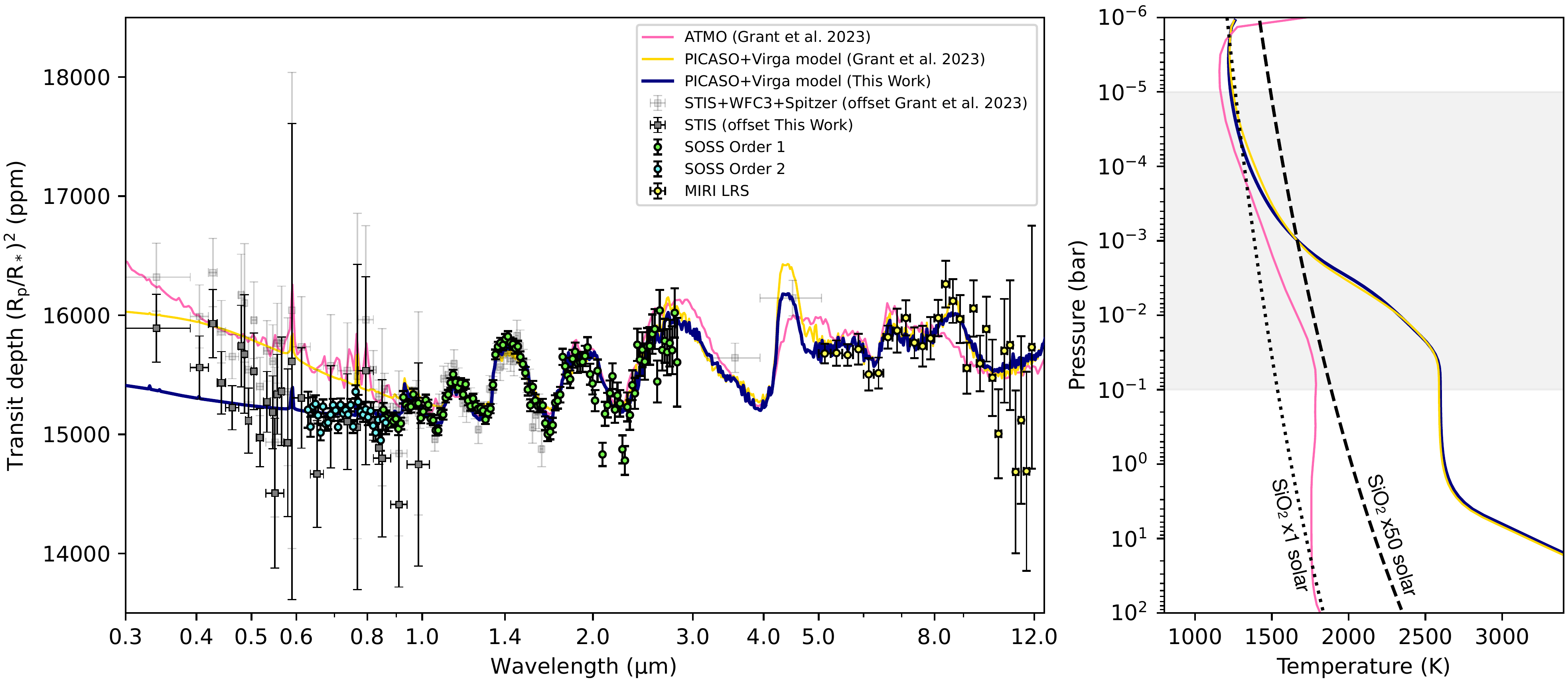}
    \caption{WASP-17b transmission spectrum with the best-fit \texttt{PICASO+Virga} forward model and P-T profile, compared to forward model fits from \cite{grant_miritransit_2023}. \underline{\textbf{Left:}} forward models presented by \cite{grant_miritransit_2023} fit to HST+Spitzer+MIRI spectra are shown in pink (\texttt{ATMO}) and yellow (\texttt{PICASO+Virga}) with the fit offsets for the HST data (light grey). The best-fit \texttt{PICASO+Virga} forward model applied to this work (STIS+SOSS+MIRI) is shown in dark blue with the offsets applied to the HST STIS data (dark grey). Both \texttt{PICASO+Virga} models show similar properties in the IR but diverge in the optical where the STIS data are allowed to ``float'' (with a fit offset) to match the higher precision JWST data. \underline{\textbf{Right:}} pressure-temperature profiles for each of the models along with the condensation curve for SiO$_2$[s] at 1$\times$ and 50$\times$ solar metallicity. The shaded region denotes the pressures probed in transmission over the full wavelength range.  }
    \label{fig:forward_model}
\end{figure*}

\begin{deluxetable*}{l c c c}
    \renewcommand{\arraystretch}{1.1}
    \tabletypesize{\footnotesize}
    \tablecolumns{5} 
    \tablecaption{Best-fit atmospheric forward model parameters in comparison to \citet{grant_miritransit_2023}.}
    \tablehead{ & \multicolumn{1}{c}{\textsc{Atmo} best-fit} & \multicolumn{2}{c}{\textsc{PICASO+Virga} 1$\sigma$ range}}
    \startdata
    Parameter\phantom{gap}  & \phantom{gap}Grant et al. (2023)\phantom{gap} & \phantom{gap}Grant et al. (2023)\phantom{gap} & \phantom{gap}This Work\phantom{gap} \\
    \hline
    Metallicity & 1 & 30--100 & 40--60 \\
    C/O &  0.7 & 0.4--0.7 &  0.4--0.7 \\
    Heat redistribution  & 0.5 & 0.6--0.7 & 0.5--0.55\\
    Internal temperature [K]  & fixed 100 & 200--300 & 205--240 \\
    Gray cloud factor & 0.5 & $\cdots$ & $\cdots$\\
    Rayleigh scattering haze factor & 10 & $\cdots$ & $\cdots$ \\
    $\log K_{zz}$ [cm$^2$/s] & $\cdots$ & 8.7--9.8 & 7.8--8.6 \\
    $\log f_{\rm{sed}}$ & $\cdots$ & -0.7--0.3 & -0.4-- -0.2 \\
    $\chi^2$/N & 1.33 & 0.98 & 0.88 \\
    \enddata
    \label{tab:forward_model_params}
    \tablecomments{Parameter definitions: metallicity is given in units of Solar metallicity. C/O ratio is given in absolute units. Gray cloud factor is in units of H$_2$ Rayleigh scattering cross-section at 350 nm. Haze factor is in units of nominal Rayleigh scattering. A heat redistribution value of 0.5 represents efficient redistribution and a value of 1.0 means no redistribution. $K_{zz}$ is the vertical turbulent diffusion parameter and $f_{\rm{sed}}$ is the sedimentation efficiency parameter. Units for logarithmic parameters refer to the argument.}
\end{deluxetable*}

Using the RCTE models+\texttt{Virga} setup, we use a MultiNest fitting routine to fit simultaneously for our four grid parameters (interior temperature, M/H, and C/O, and heat redistribution) and two cloud parameters (K$_{zz}$, and f$_\mathrm{sed}$). In addition to these six, we include three parameters to account for: 1) scaling of the 10 bar pressure radius, which scales the data to NIRISS SOSS baseline, 2) an instrumental offset for STIS, and 3) an instrumental offset for MIRI LRS. A positive instrumental offset of $\sim$ 100-200 ppm between MIRI LRS and extant observations was noted in \citet{grant_miritransit_2023}. The instrumental offset is subtracted from the data, i.e. a positive offset results in a decreased transit depth. In total we fit for 9 parameters using the open source nested sampling code, Ultranest \citep{Ultranest}. 

We fit our cloudy spectra to the \texttt{Ahsoka} SOSS data along with the HST STIS \citep{Alderson2022} and MIRI LRS \citep{grant_miritransit_2023} data, fixing the SOSS position and an offset for STIS and MIRI separately with the free parameters of the grid detailed above. Our best fit model has a super-solar metallicity ($\sim$55$\times$ solar) and solar C/O, requiring an internal temperature of 200\,K and redistribution factor of 0.5 (see Table\,\ref{tab:forward_model_params}). 
Figure\,\ref{fig:forward_model} shows the best fit \texttt{PICASO+Virga} model from our grid along with the best fit \texttt{PICASO+Virga} and ATMO model presented by \citet{grant_miritransit_2023}. The largest differences can be seen in the optical wavelengths with a different offset required when just fitting for HST+MIRI compared to adding in our new SOSS data. However, even with a difference in the optical spectra, associated with aerosol opacity, we still require an abundant SiO$_2$[s] cloud opacity to fit the MIRI spectra.
Table\,\ref{tab:forward_model_params} details the best-fit values for the \texttt{PICASO+Virga} forward models from this work in comparison to those presented in \cite{grant_miritransit_2023} and shown in Figure\,\ref{fig:forward_model}.

\section{Atmospheric Retrieval Analysis} \label{sec:Retrievals}

\begin{figure*}
    \includegraphics[width=0.9\textwidth]{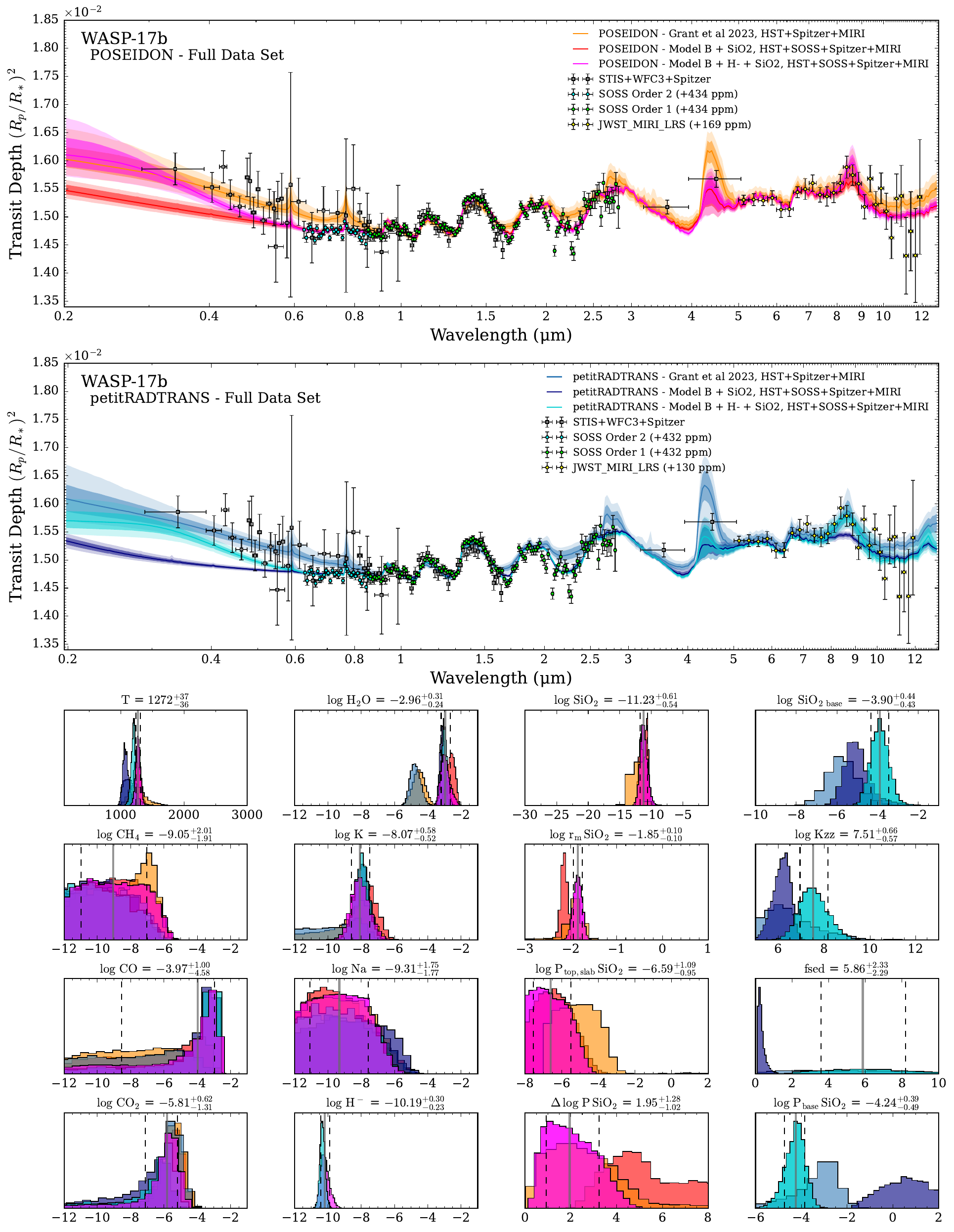}
    
    \caption{Atmospheric retrievals with \texttt{POSEIDON} and \texttt{petitRADTRANS} (\texttt{pRT}) on HST+Spitzer data \citep{Alderson2022}, SOSS order 1 and 2 data (\texttt{Ahsoka} reduction, this work), and MIRI LRS data \citep{grant_miritransit_2023}. \underline{\textbf{Top and middle:}} median retrieved spectra (solid lines), with 1$\sigma$ and 2$\sigma$ confidence intervals (dark and light shaded regions) for \texttt{POSEIDON} retrievals on the full dataset (top) and \texttt{pRT} retrieval on the full dataset (middle). Retrievals displayed are those using the chemical inventory of Model B (Appendix Table \ref{tab:model_set_up}), with and without H$^-$, and the retrievals from \cite{grant_miritransit_2023} for comparison. \underline{\textbf{Bottom panels:}} retrieved posteriors for isothermal atmospheric temperature and gas-phase volume mixing ratios (\textbf{columns one and two}) and cloud properties from \texttt{POSEIDON} (\textbf{column three}) and \texttt{pRT} (\textbf{column four}).   } 
    \label{fig:RetrievalResults}
\end{figure*}

\begin{figure*}
    \includegraphics[width=0.9\textwidth]{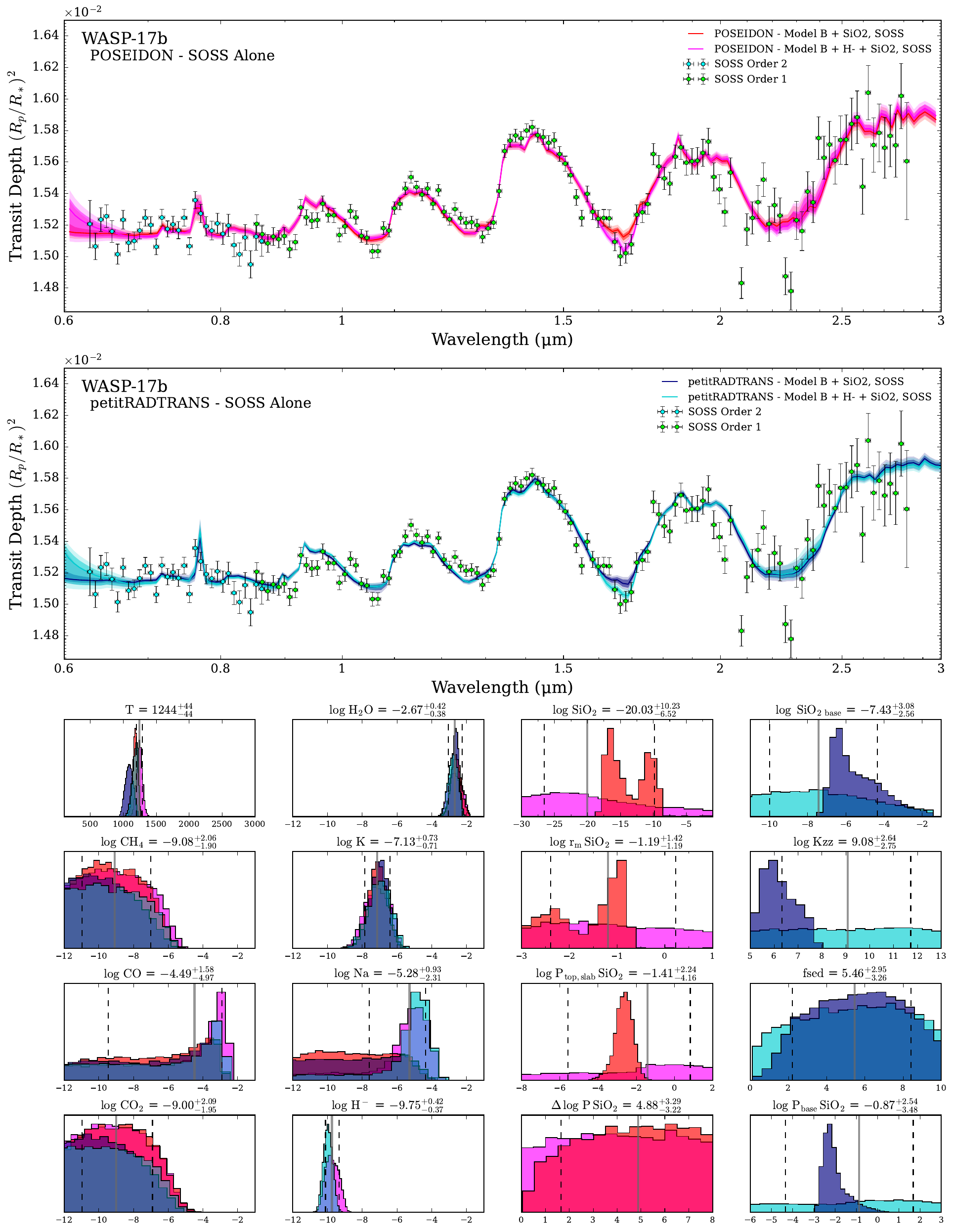}
    
    \caption{Atmospheric retrievals with \texttt{POSEIDON} and \texttt{petitRADTRANS} (\texttt{pRT}) on SOSS order 1 and 2 data (\texttt{Ahsoka} reduction from this work). \underline{\textbf{Top and middle:}} median retrieved spectra (solid lines), with 1$\sigma$ and 2$\sigma$ confidence intervals (dark and light shaded regions) for \texttt{POSEIDON} (top) and \texttt{pRT} (middle) retrievals. Both retrievals use the chemical inventory of Model B (Appendix Table \ref{tab:model_set_up}), with and without H$^-$. \underline{\textbf{Bottom panels:}} Retrieved posteriors for isothermal atmospheric temperature and gas-phase volume mixing ratios (\textbf{columns one and two}) and cloud properties from \texttt{POSEIDON} (\textbf{column three}) and \texttt{pRT} (\textbf{column four}).}   
    \label{fig:RetrievalResults_SOSSOnly}
\end{figure*}

We interpret WASP-17b's transmission spectrum using two independent atmospheric retrieval codes, \texttt{POSEIDON}\footnote{\url{https://github.com/MartianColonist/POSEIDON}} \citep{MacDonaldMadhusudhan2017, MacDonald2023} and \texttt{petitRADTRANS} (\texttt{pRT})\footnote{\url{https://gitlab.com/mauricemolli/petitRADTRANS}} \citep{Molliere2019}. We ran retrievals with the chemical inventories as follows:
\begin{itemize}
    \item Model A: H$_2$O, CH$_4$, CO$_2$, TiO, H$^-$
    \item Model B: H$_2$O, CH$_4$, CO$_2$, Na, K, CO
    \item Model C: H$_2$O, CH$_4$, CO$_2$, Na, K, CO, VO, TiO, H$^-$
    \item Model D: H$_2$O, CH$_4$, CO$_2$, H$^{-}$, Na, K, CO, VO, TiO, AlO, CaO, CrH, FeH, ScH, TiH, Li, HCN, NH$_3$, SO$_2$, H$_2$S
    \item Model D$^{*}$: H$_2$O, CH$_4$, CO$_2$, H$^{-}$, Na, K, CO, VO, TiO, AlO, CrH, FeH, Li, HCN, NH$_3$, SO$_2$
\end{itemize}

Models A, B, and C are the most statistically preferred models from the suite presented in \citet{Alderson2022}, in which they individually retrieved bimodal (A), sub-solar (B) and super-solar (C) H$_2$O abundance modes. We additionally run a fourth model, which we title model D, which represents an exploratory, expanded chemical inventory. Model D$^{*}$ was run with {\tt pRT} and is model D without CaO, ScH, TiH, and H$_2$S as the cross-sections were not available in the correct format, and they have negligible spectral impact. For our retrieval analysis we also include SiO$_2$ clouds due to their inference from MIRI LRS observations of WASP-17b from \citet{grant_miritransit_2023}.

Appendix Table \ref{tab:model_set_up} shows the chemical inventories of each model with pertinent line lists. Our retrievals jointly consider the information from JWST (NIRISS SOSS R = 100 \texttt{Ahsoka} spectrum from this work, and MIRI LRS 0.25 µm ExoTiC-MIRI reduction from \citealt{grant_miritransit_2023}), HST (STIS G430, STIS G750, WFC3 G102, and WFC3 G141 from \citealt{Alderson2022}), and {\it Spitzer} (IRAC1 and IRAC2 from \citealt{Alderson2022}) observations (Figure \ref{fig:RetrievalResults}). We then perform retrieval analyses on NIRISS SOSS (R = 100 \texttt{Ahsoka} spectrum from this work) data alone (Figure \ref{fig:RetrievalResults_SOSSOnly}).

\subsection{\texttt{POSEIDON} Retrieval Configuration and Results }\label{subsec:POSEIDONconfig}

We conducted free-chemistry retrievals with the inclusion of SiO$_2$ aerosols using the open-source atmospheric retrieval code \texttt{POSEIDON} \citep{MacDonaldMadhusudhan2017, MacDonald2023} on the full complement of WASP-17b transmission spectra datasets published to date. We follow the same methodology as that presented in \citet{Alderson2022} and \citet{grant_miritransit_2023}, which we repeat here.

We assume a one-dimensional H$_2$-He dominated atmosphere (with He/H$_2$ = 0.17) with the gas-phase containing the freely fit log mixing ratios of the chemical species described in models A, B, C, and D. The specific line lists used are detailed in the appendix of \citet{macdonald2022trident}, with particular references specified in Appendix Table \ref{tab:model_set_up}. Retrievals include continuum opacity from H$_2$ and He collision-induced absorption \citep{Karman2019}, H$^-$ bound-free absorption \citep[for models with H$^-$ included,][]{John1988}, and H$_2$ Rayleigh scattering \citep{Hohm1994}. We assume an isothermal pressure-temperature profile, following the modeling of \citet{Alderson2022}. The reference pressure was set at 10\,bar with model atmospheres covering 10$^{-8}$ - 100 bar with 100 layers uniformly distributed in log-pressure space. We include a Mie-scattering SiO$_2$(s) (crystalline) cloud model parameterized by the mean particle size, $r_{\rm{m}}$, the cloud-top pressure, $P_{\rm{top,slab}}$, the width of the cloud in log pressure space, $\Delta\log$\,$P$, and the constant-in-altitude log mixing ratio of the aerosol in the cloud. The effective aerosol extinction cross section is pulled from a precomputed database with mean particle sizes ranging from 0.001--10\,\textmu m and wavelengths spanning 0.2--30\,\textmu m at $R=1000$. SiO$_2$(s) (crystalline) radiative properties are computed from refractive indices \citep{palik1998handbook,zeidler2013optical}. See \citet[][Sec 3.1]{grant_miritransit_2023} and \cite{Mullens2024_arxiv} for a more complete description of the aerosol implementation in \texttt{POSEIDON}. We additionally fit for two dataset offsets, one for SOSS order 1 and order 2 together, and one for JWST MIRI LRS, keeping the archival HST and {\it Spitzer} data fixed. The priors for our \texttt{POSEIDON} retrievals are summarized in Appendix Table \ref{tab:retrieval_model_params}. 

Our \texttt{POSEIDON} retrievals sample the parameter space using MultiNest, here with 2,000 live points (with the exception of Model D which was sampled with 500 live points for expediency). Model transmission spectra are computed at a spectral resolution of R = 20,000 from 0.2-13.0 \textmu m. The model transmission spectra are then convolved with the point spread function of each instrument and binned down to the data resolution and compared to data during the retrieval likelihood evaluation.

In Figure \ref{fig:RetrievalResults}, we compare our combined JWST+HST+Spitzer retrieval results with those without NIRISS SOSS from \citet{grant_miritransit_2023}. In particular, we display the results of Model B, the model highlighted in \citet[][Figure 5]{grant_miritransit_2023}. Our retrievals find that the SOSS data is offset down $\sim$ 400 ppm to agree with the continuum level of the archival HST data. The JWST MIRI LRS data is pushed down $\sim$ 170 ppm, consistent within 1 sigma of the offset found in \citet{grant_miritransit_2023}. The inclusion of the SOSS data with the full dataset allows us to better constrain water and potassium. The retrieved H$_2$O abundance ($\log$ H$_2$O = -2.96$^{+0.31}_{-0.24}$) is two orders of magnitude larger than results found in \citet{grant_miritransit_2023}, solving the sub-solar water abundance mystery discussed in \citet{grant_miritransit_2023} and the bimodal water distribution found in \citet{Alderson2022}.

The largest deviation between the archival HST data and the SOSS data occurs in the 0.6 to 0.8 \textmu m region. The SOSS data has a much flatter spectrum in those wavelengths. The flat spectrum from 0.6 to 0.8 \textmu m is reminiscent of `gray' cloud deck opacity, which cuts off the spectral features of potassium and water. The flat spectrum at these wavelengths is seemingly at odds with the amplified scattering slope found in the archival HST data. This is notable as the SiO$_2$ retrieved particle size is sensitive to the scattering slope found in STIS wavelengths, while the aerosol mixing ratio and pressure extent are sensitive to the amplitude of the 8.6 \textmu m SiO$_2$ feature in the infrared. We found that retrievals including the full data set had a difficult time simultaneously fitting an SiO$_2$ cloud that could explain the HST-STIS scattering slope, the flatter SOSS order 2 spectrum, and the JWST MIRI LRS SiO$_2$ feature. In particular, Model B, as the only model without H$^-$, finds a slightly bimodal SiO$_2$ abundance with an unconstrained cloud vertical extent that can fit the flatter SOSS data, but not the STIS slope (Figure \ref{fig:RetrievalResults}, top panel, red). SiO$_2$ in this model retrieves a shallower STIS slope, smaller particles, and larger cloud vertical extent than what was found in \citet{grant_miritransit_2023}. The inclusion of H$^-$ continuum opacity, with the addition of SiO$_2$ Mie scattering, provides an optimal fit to both the SOSS and HST data. For example, a nested Bayesian model comparison using Model B with and without H$^-$ favors its inclusion at 5.1$\sigma$ (Appendix Table \ref{table:transmission_stats}). Model B with H$^-$ is highlighted in Figure \ref{fig:RetrievalResults} (magenta) with a retrieved $\log$ H$^-$ abundance of -10.19$^{+0.30}_{-0.23}$. (See Appendix Figure \ref{fig:spectral_contribution}, which displays the spectral contribution of each species included in the retrieval). The complete corner plot for this retrieval is found in Appendix Figure \ref{fig:ModelBwH-Corner}. Water, potassium, and H$^-$ show a slight degeneracy in the corner plot, given that the wings of potassium and water absorption can contribute to the flat opacity in SOSS order 2.

We also tested Model B with a gray infinite opacity cloud deck in lieu of H$^-$ opacity and found that Model B with H$^-$ was preferred by 3.1$\sigma$ (Appendix Table \ref{table:transmission_stats}). The transit depth floor found from 0.6 to 0.8 \textmu m extends out to about 1.5 \textmu m, while the dip in the spectrum between 1.6 and 1.7 \textmu m extends below this `floor' (see H$^-$ contribution in Appendix Figure \ref{fig:spectral_contribution}). H$^-$ is preferred over a gray opacity cloud solely due to a dip in the spectrum after 1.5 \textmu m (see Appendix Figure \ref{fig:gray_vs_H-}), since including an additional flat-opacity to fit SOSS Order 2 allows the STIS slope to be fit by SiO$_2$ in both retrievals. \citet{Lewis2020} found a similar pattern in the hot Jupiter HAT-P-41b's spectrum and found H$^-$ continuum opacity to provide the best explanation. For completeness, we also tested inhomogeneous `patchy' SiO$_2$ clouds and found that the retrieved cloud fraction was consistent with 100\% terminator cloud coverage ($82^{+13}_{-20}\%$), and that models without patchy clouds were preferred by 2.0$\sigma$ (Appendix Table \ref{table:transmission_stats}). 

\begin{figure*}
    \includegraphics[width=0.95\textwidth]{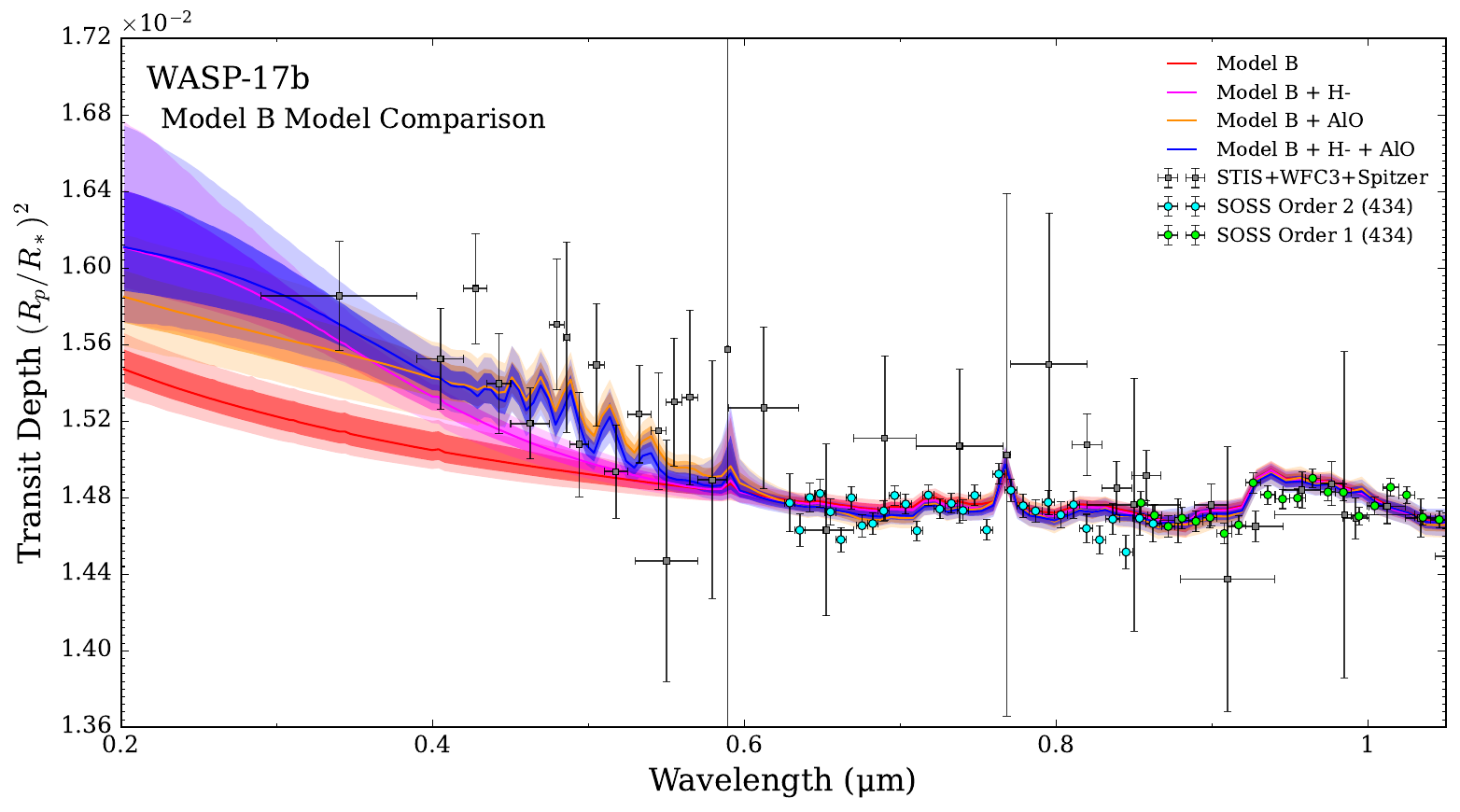}
    
    \caption{Comparison of nested \texttt{POSEIDON} Model B retrievals with and without H$^-$ and AlO, zooming on the 0.2 to 1.05 \textmu m region. We display median retrieved spectra (solid lines) and 1$\sigma$ confidence regions (shaded regions). Model B with both H$^-$ and AlO (blue) is the best fit model, but only due to imparting fine structure to fit the STIS slope. Future observations of the 0.2-0.8 \textmu m region will provide more reliable inferences of metal species.}    
    \label{fig:model-b-comparison}
\end{figure*}

As mentioned above, we also ran retrievals with the chemical inventories of Models A and C, and an exploratory retrieval, D. Statistics for each model are shown in Appendix Table \ref{table:transmission_stats}. In our exploratory retrieval (Model D) we found evidence for aluminum oxide (AlO). We ran nested retrievals with the chemical inventories of Model B with AlO and with H$^-$ and AlO. We find that Model B with both H$^-$ and AlO is preferred over Model B with H$^-$ by 4.7$\sigma$. We display the effect of including AlO in Figure \ref{fig:model-b-comparison}, zooming in on the 0.2 to 1.05 \textmu m region. From the plot, we glean that AlO is preferred as it adds additional opacity to the slope in the STIS data which results in a better fit. Additionally, while the retrieved spectra for Model B with AlO (orange) is within the 1 sigma shaded region of the Model B with AlO and H$^-$ (blue) retrieved spectra in the displayed wavelength region, Model B with AlO and H$^-$ is still preferred by 3.1$\sigma$ due to being able to fit the SOSS Order 1 dip between 1.6-1.7 \textmu m (not shown). We note that the current STIS data are not particularly precise due to transit observations not fully capturing egress or any post-transit baseline. We conclude that the inference of AlO due to the fine structure it imparts to fit the STIS data is not particularly reliable. Future observations of WASP-17b via the HUSTLE program (Hubble Ultraviolet-optical Survey of Transiting Legacy Exoplanets, GO 17183, PI H. R. Wakeford) will take place over the 0.2--0.8 \textmu m region at higher resolution with the WFC3 UVIS G280 mode, and therefore be able to confidently detect the fine structure that metals impart at these wavelengths, as well as provide a more robust slope shape.

We ran an additional test where we held NIRISS-SOSS data constant, and retrieved three data offsets: one for HST data, one for \textit{Spitzer} data, and one for JWST MIRI-LRS data. We find that our retrieved H$^{-}$, H$_2$O, K abundances, as well as SiO$_2$ properties, are all within one sigma of the results shown in Figure \ref{fig:RetrievalResults} and Appendix Figure \ref{fig:ModelBwH-Corner}, showing that our decision to hold archival HST data constant was arbitrary. We find that the retrieved offset for \textit{Spitzer} data is less constrained than the retrieved offset for HST and JWST MIRI-LRS data, causing the retrieved CO$_2$ abundance to become unconstrained. With the addition of NIRSpec/G395H data (3--5\,$\mu$m, Lewis et al. in prep), the CO$_2$ abundance will be explored in more detail.

In order to quantify what information can be gleaned from NIRISS SOSS, we ran retrievals with the chemical inventory of Model B and Model B + H$^-$ on the SOSS data alone. Our results are displayed in Figure \ref{fig:RetrievalResults_SOSSOnly}. With the SOSS data, we find that our H$_2$O, H$^{-}$, and K abundances are within one sigma of the retrieved abundances found in retrievals including the full data set displayed in Figure \ref{fig:RetrievalResults}. In particular, we find that H$^{-}$ is still preferred by 4$\sigma$ (Appendix Table \ref{table:soss_alone_stats}).

\subsection{\texttt{petitRADTRANS (pRT)} Retrieval Configuration and Results}\label{subsec:pRTconfig}

To complement and validate our {\tt POSEIDON} results, we also use {\tt pRT} to perform free retrievals on our \texttt{Ahsoka} NIRISS SOSS transmission spectrum. We perform four retrievals with different model configurations for the NIRISS SOSS data alone: A, B, B with H$^{-}$, C, and D$^{*}$, which are shown in Appendix Table \ref{tab:model_set_up}, using MultiNest \citep{Feroz2008,Buchner2014}. We also perform retrievals of model B and model B with H$^-$ for the combined JWST+HST+Spitzer dataset described in the Section \ref{sec:Retrievals} introduction. All retrievals include H$_{2}$ and He as the background gas and include the effects of Rayleigh scattering and H$_{2}$-H$_{2}$ and H$_{2}$-He collision-induced absorption. All models also include SiO$_{2}$ clouds, following the \citet{Ackerman-Marley2001} cloud model built into {\tt pRT}. Unlike how clouds are modeled in {\tt POSEIDON}, the clouds in {\tt pRT} are described by the width of the particle size distribution, $\sigma_\mathrm{lnorm}$; their vertical mixing parameter, K$_{zz}$; and sedimentation parameter, f$_\mathrm{sed}$. We fix the radius of the host star and allow the planetary radius, surface gravity (log(g)), and temperature to vary freely. The pressure-temperature profile is assumed to be isothermal and the reference pressure is set at 100 bar. Modeled pressures are evenly distributed in 100 layers in log-pressure space. The priors and posteriors for our retrieval of model B with SiO$_{2}$ clouds are summarized in Appendix Table \ref{tab:retrieval_model_params}. Our initial tests used 4000 live points, but we found comparable results using fewer live points. For models B and B with H$^-$, our full data set retrievals are run with 500 live points and our SOSS-alone retrievals are run with 2000 live points. We chose to utilize 500 live points for our full data retrievals in order to validate the 2000 live point full data retrievals from \texttt{POSEIDON}. For model D$^{*}$, only 400 live points were used to expedite the validation.

Our results for the full data set retrievals are shown in Figure \ref{fig:RetrievalResults}, compared with the results from \texttt{POSEIDON}. We find good agreement with the results from \texttt{POSEIDON} for the temperature and chemical abundances (for Model B + H$^{-}$: log H$_2$O = -3.00$^{+0.17}_{-0.16}$, log K = -7.97$^{+0.41}_{-0.43}$, log H$^{-}$ = -10.3$^{+0.18}_{-0.16}$) where the inclusion of H$^{-}$ in Model B is preferred by 6.5 $\sigma $ (Appendix Table \ref{table:transmission_stats}). However, the differing cloud implementations in \texttt{pRT} and \texttt{POSEIDON} lead to discrepancies in the retrieved cloud parameters. In particular, without the inclusion of H$^{-}$ (Model B + SiO$_2$, dark blue), the SiO$_2$ feature in the MIRI LRS bandpass is not well fit. Additionally, like \texttt{POSEIDON}, SiO$_2$ scattering fits the flatter SOSS data but not the STIS slope. When H$^-$ is not included, $f_{\text{sed}}$ is driven to lower values against the edge of the prior. The small sedimentation efficiency represents smaller cloud particles that are vertically extended \citep{Gao2018}. With the inclusion of  H$^{-}$, our retrieval fits the STIS slope, the MIRI feature, and the dip in opacity after 1.5 \textmu m in SOSS Order 1. Our results for the SOSS data alone retrievals are shown in Figure \ref{fig:RetrievalResults_SOSSOnly}. Similar to the results from \texttt{POSEIDON}, we find that our H$_2$O, H$^{-}$, and K abundances are within one sigma of the retrieved abundances found in retrievals including the full data set and that H$^{-}$ is still preferred by 3.3 $\sigma$ for the SOSS data alone (Appendix Table \ref{table:soss_alone_stats}).

\section{Discussion} \label{sec:discuss}

\begin{figure*}
    \includegraphics[width=0.95\textwidth]{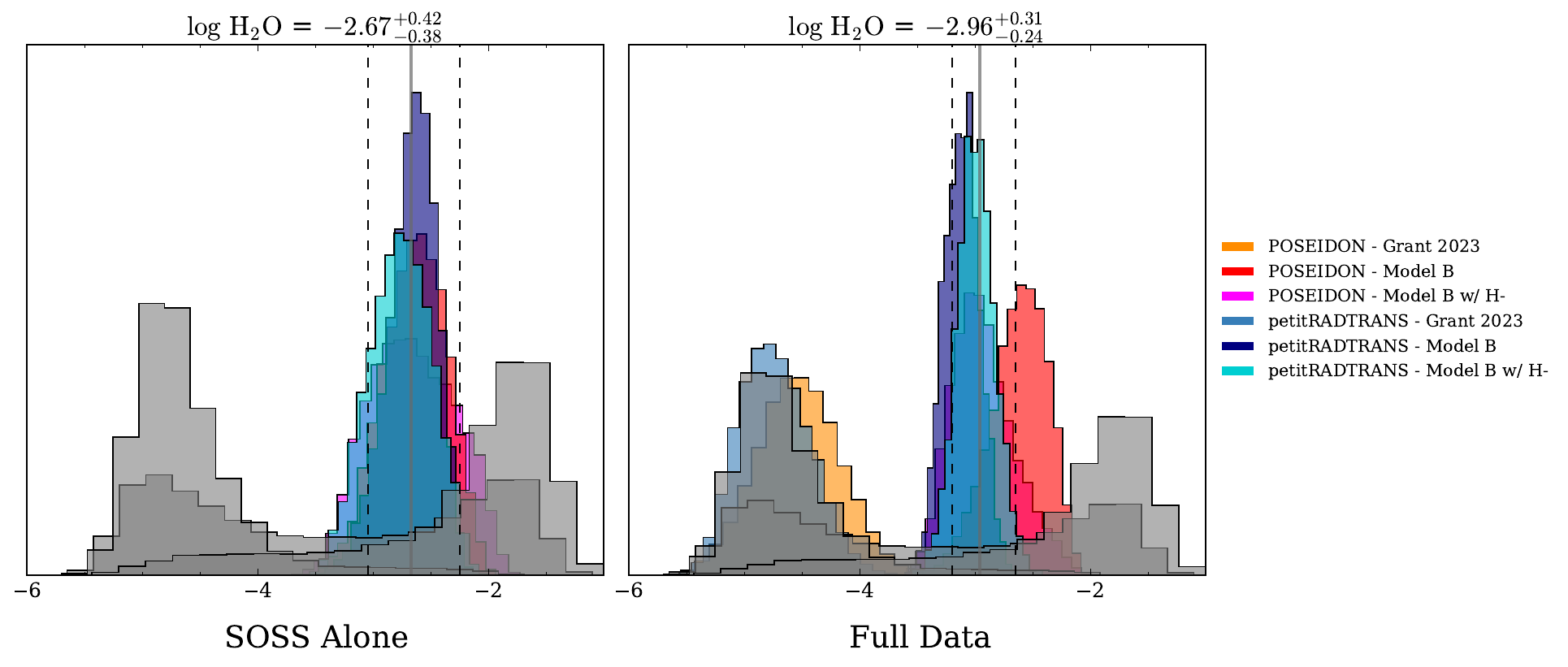}
    
    \caption{Comparison of WASP-17b retrieved H$_2$O abundances from \cite{Alderson2022} (gray posteriors) to those shown in Figures \ref{fig:RetrievalResults} and \ref{fig:RetrievalResults_SOSSOnly} (posteriors correspond to the same color scheme) and retrievals performed in \citet{grant_miritransit_2023}. \cite{Alderson2022}'s analysis of HST+Spitzer data found that sub-solar (Model B), super-solar and bimodal posterior distributions were all statistically valid, depending upon the model configuration. Retrievals from \cite{grant_miritransit_2023} (dark orange and light blue posteriors) included MIRI LRS data and yielded similar results using model B. Adding NIRISS SOSS data, we find tight constraints on a super-solar water abundance from retrievals on HST+SOSS+Spitzer+MIRI, and retrievals on SOSS alone. Clearly, NIRISS SOSS is driving the abundance constraints, since the SOSS Alone retrievals with both \texttt{POSEIDON} and \texttt{pRT} yield similarly constrained abundances.
    }
    \label{fig:H2OAbundanceComparison}
\end{figure*}

Our NIRISS SOSS observations of WASP-17b present the opportunity to explore the NIR transmission spectrum in more detail than previously achievable, due to the increased resolution, precision and wavelength range compared to previous HST WFC3/NIR observations \citep{Alderson2022}. The retrievals presented in Section \ref{sec:Retrievals} achieve tighter constraints on the H$_2$O abundance, breaking the degeneracy seen in \texttt{POSEIDON} retrievals based on the HST+Spitzer data presented in \citet{Alderson2022}, and also confirm the presence of H$^{-}$, a species also hinted at in \citet{Alderson2022}. Furthermore, our POSEIDON retrievals allow us to constrain the potassium abundance ($\log$ K = -8.07$^{+0.58}_{-0.52}$), a species which was not identified in previous space-based observations \citep{sing2016,Alderson2022}.

Though the presence of H$_2$O in WASP-17b's transmission spectrum is not a new detection, retrievals run on the HST+Spitzer data could not constrain its abundance, finding that based on the model configuration, bimodal, sub-solar and super-solar posterior distributions were all statistically valid (\citealt{Alderson2022}, labeled as models A, B, and C), while retrievals run with the inclusion of MIRI LRS data based on the model B configuration in \citet{Alderson2022} obtained similar values \citep{grant_miritransit_2023}.  As shown in Figure \ref{fig:H2OAbundanceComparison}, \texttt{POSEIDON} and \texttt{pRT} retrievals run on the HST+SOSS+Spitzer+MIRI data using the model B configuration now obtain a well-constrained super-solar H$_2$O abundance, albeit one which is lower than found from the HST+Spitzer data alone. It is clear that the NIRISS SOSS data is driving this abundance constraint as the SOSS alone retrievals, which consider the NIRISS SOSS data only, obtain similar abundances and constraints to retrievals performed on the full data set. This result is unsurprising given the wealth of H$_2$O absorption features within the NIRISS SOSS bandpass, and demonstrates the improvement which can be gained over comparable HST observations. It is important to note, however, that the current HST+SOSS+Spitzer+MIRI configuration contains limited information as to the carbon-bearing species in WASP-17b's atmosphere, and that the CO$_2$ abundance constraints shown in Figure \ref{fig:RetrievalResults} are more consistent with values from the sub-solar H$_2$O mode found in \citet{Alderson2022}. This may indicate that the overall atmospheric metallicity may shift once more should carbon-bearing species be more robustly detected, a hypothesis that will be better explored with the addition of NIRSpec/G395H data (3--5\,$\mu$m, Lewis et al. in prep). 

The retrievals presented in Section \ref{sec:Retrievals} also highlight that opacity from bound-free H$^-$ absorption, where a high-energy photon liberates an electron from the hydrogen anion, is required to best fit the NIRISS SOSS transmission spectrum. H$^-$ was considered as a potential opacity source in \citet{Alderson2022}, indeed Model 10 in \citet{Alderson2022} uses the same configuration as the neon pink \texttt{POSEIDON} retrieval in Figure \ref{fig:RetrievalResults} but without the inclusion of SiO$_2$ clouds (i.e., Model 10 = Model B + H$^-$). However, given the information available within the HST+Spitzer transmission spectrum, Model 10 and Model B were statistically comparable. The NIRISS SOSS transmission spectrum presented in this work is able to confirm the presence of H$^-$, given the flattening of the 0.6--0.8\,\micron\, region compared to the STIS data, and the downward shapes of the wings of the H$_2$O absorption features which were hinted at with WFC3/G141. Our HST+SOSS+Spitzer+MIRI \texttt{POSEIDON} retrievals find similar abundances of H$^-$ as the HST+Spitzer retrievals presented in \citet{Alderson2022}, but obtain significantly tighter constraints (-10.19$^{+0.30}_{-0.23}$ compared to the previous -9.65$^{+1.25}_{-1.68}$). As discussed in \citet{Alderson2022}, such abundances of H$^-$ would only be expected under equilibrium conditions at much hotter temperatures than WASP-17b \citep{Kitzmann2018}. However, disequilibrium processes such as H$_2$ dissociative electron attachment and destruction by atomic H collisional detachment could produce the required mixing ratios \citep{Lavvas2014, Lewis2020}. This detection of H$^-$ is yet more observational evidence that disequilibrium chemistry is an important consideration for hot Jupiter exoplanet atmospheres \citep[e.g,][]{Tsai2023}, and demonstrates the power of JWST in exploring new chemistry paradigms.

Bound-free H$^{-}$ opacity has been explored in other work. \citet{Lewis2020} found that H$^{-}$ continuum opacity provided a good fit to HAT-P-41b's transmission spectrum. \citet{Welbanks2023} further explored the significance of the H$^{-}$ detection utilizing a Bayesian leave-one-out approach and found that the detection was driven slightly by WFC3 points after 1.5 \textmu m and strongly by a 3.6 \textmu m \textit{Spitzer} point. They found that the H$^{-}$ claim depended on how reliable the \textit{Spitzer} point is, but additionally confirmed the \citet{Lewis2020} detection with a similar significance when they ran a retrieval with the complete HAT-P-41b dataset. From our gray vs H$^{-}$ retrievals (Appendix Figure \ref{fig:gray_vs_H-}), we find that H$^{-}$ is preferred by 3.1$\sigma$. This is due to the resolution provided by NIRISS SOSS.

Our work demonstrates that a combination of SiO$_2$ Mie scattering aerosols and a continuum opacity that falls off at 1.5 \textmu m is needed to fit the full dataset. Other work has shown that opacity sources other than H$^{-}$ can produce a similar continuum opacity with a wavelength-dependent fall-off. \citet{ConstantinouMadhu2024} formulated a sigmoid cloud parameterization that captures the general trend of aerosols having near-constant opacity near 1 \textmu m that falls off at longer wavelengths. This sigmoid opacity mimics the general shape of the H$^{-}$ opacity. We ran an additional \texttt{POSEIDON} retrieval with the updated crystalline SiO$_2$ refractive indices found in \citet{Herbin2023} that have more complete wavelength coverage to ensure that Mie scattering SiO$_2$ could not account for both sources of opacity. We find that $\log$ H$^{-}$ $\sim$ -10 is still needed to fit the spectra, even with updated refractive indices. Given that SiO$_2$ cannot mimic an H$^{-}$ opacity, we conclude that H$^{-}$ remains the most probable source of opacity we are detecting. 

Regardless of the new inferences gained from the NIRISS SOSS data, Figure \ref{fig:RetrievalResults} shows that the inclusion of SOSS in the \texttt{POSEIDON} retrieval does not change the conclusions of \citet{grant_miritransit_2023} that WASP-17b's atmosphere contains small particle SiO$_2$ clouds, with the SiO$_2$ particle size and abundance parameters retrieving similar values and constraints. Our addition of the NIRISS SOSS data does however provide additional constraints over the HST STIS data, with the flatter Order 2 region resulting in a change in the cloud pressure (log\,P$_\mathrm{top,slab}$ SiO$_2$), and tighter constraints on the extent ($\Delta$log P SiO$_2$). \texttt{pRT} results for  K$_{\text{zz}}$, $f_{\rm{sed}}$, and $P_{\rm{base}}$\,SiO$_2$ similarly agree with the results of \citet{grant_miritransit_2023}. The SiO$_{2\,\rm{base}}$ value does differ slightly from our prior results as the inclusion of H$^-$ likely reduces the amount of SiO$_2$ needed for a good fit. \texttt{POSEIDON} similarly found less SiO$_2$ was needed compared with the earlier \citet{grant_miritransit_2023} results. Regardless of the differences in aerosol modeling between \texttt{pRT} and \texttt{POSEIDON}, retrieval modeling using the full dataset presented in this work finds that \texttt{pRT} and \texttt{POSEIDON} agree on the nature of the cloud opacity. Overall, we find that despite different abundances of the gaseous species, we obtain the same cloud inferences as \citet{grant_miritransit_2023} when considering the complete available data, and that the previous detection of SiO$_2$ clouds is robust to the inclusion of the additional data presented here. 

When only the SOSS data are considered, both \texttt{POSEIDON} and \texttt{pRT} retrievals have unconstrained and inconsistent cloud properties with \citet{grant_miritransit_2023}. This result is not surprising, given that SOSS data alone does not have the constraining information provided by the cloud feature seen in the MIRI LRS data and the blueward optical to NUV slope suggested by the full HST STIS wavelength range.

\section{Conclusions} \label{sec:conclusions}

Here we have presented NIRISS SOSS transmission observations of the hot Jupiter WASP-17b as part of the JWST TST-DREAMS GTO program. We analyzed the data using three independent data reduction pipelines: \texttt{Ahsoka}, \texttt{transitspectroscopy}, and \texttt{supreme-SPOON}. Our resultant transmission spectra show excellent agreement between pipelines, with minor differences  traceable to specific variations in our analyses. Our R=100 \texttt{Ahsoka} spectra achieve average (median) uncertainties across  order 1 of 78 (55) ppm, and across order 2 of 67 (60) ppm. 

We investigated the effects of different limb darkening treatments on the resultant R=100 \texttt{Ahsoka} transmission spectrum. We compared \textit{fixing} limb darkening coefficients (LDCs) to theoretical values, versus \textit{fitting} LDCs during light curve fits, and we also examined the effects of \textit{prebinning} (binning and then fitting) LDCs versus \textit{postbinning} (fitting then binning) LDCs. When comparing \textit{fixing} versus \textit{fitting} postbinned LDCs, we found that resultant transit depths diverged beyond the 1$\sigma$ error bars at wavelengths beyond 2\,$\mu$m, with transit depth differences reaching 600 ppm at the longest wavelengths. We found similar disagreements when comparing \textit{prebinned} to \textit{postbinned} \textit{fit} LDCs. Our findings call for careful consideration in the application of limb darkening treatments to analysis of JWST high-precision NIR continuous time-series data. 

Our transmission spectrum shows multiple H$_2$O absorption features and a somewhat flatter slope towards the optical than seen in previous HST STIS observations. Using \texttt{POSEIDON} and \texttt{pRT} retrievals, we obtain a well-constrained super-solar H$_2$O abundance, breaking the degeneracy seen in the HST data. We also constrain the H$^-$ abundance, finding that our model including H$^-$ is preferred over our model without H$^-$ to 5.1$\sigma$. Finally, we constrain the potassium abundance on WASP-17b for the first time using space-based observations. 

Our H$_2$O and H$^-$ detections are driven by the multiple H$_2$O absorption features and the flatter optical slope combined with a dip in the NIRISS SOSS spectrum beyond 1.5 \textmu m. The presence of H$^-$ in WASP-17b's atmosphere is not expected under equilibrium chemistry conditions, suggesting that disequilibrium processes are at play. We also find that the addition of our NIRISS SOSS data continues to support the previous MIRI LRS driven detection of SiO$_2$ clouds, demonstrating that the detection is robust.

The current HST+SOSS+Spitzer+MIRI configuration contains limited information as to the carbon-bearing species in WASP-17b's atmosphere, and the overall atmospheric metallicity may shift once more should carbon-bearing species be more robustly detected, a hypothesis that will be better explored with the addition of NIRSpec/G395H data (Lewis et al. in prep).

\vspace{0.5cm}

We thank the anonymous referee for insightful recommendations and comments, which have increased the quality of this work. 

This paper reports work carried out in the context of the JWST Telescope Scientist Team (\url{https://www.stsci.edu/~marel/jwsttelsciteam.html}, PI: M. Mountain). Funding is provided to the team by NASA through grant 80NSSC20K0586. Based on observations with the NASA/ESA/CSA JWST, associated with program GTO-1353 (PI: N.K. Lewis), obtained at the Space Telescope Science Institute, which is operated by AURA, Inc., under NASA contract NAS 5-03127. 

DRL and CIC acknowledge research support by an appointment to the NASA Postdoctoral Program at the NASA Goddard Space Flight center (GSFC), administered by Oak Ridge Associated Universities (ORAU) under contract with NASA. Additionally, DRL and KDC acknowledge support from the GSFC Sellers Exoplanet Environments Collaboration (SEEC), which is supported by NASA's Planetary Science Division's Research Program. DRL also acknowledges support by NASA under award number 80GSFC21M0002. 

EM acknowledges that this material is based upon work supported by the National Science Foundation Graduate Research Fellowship under Grant No. 2139899. 

LA acknowledges funding from UKRI STFC Grant ST/W507337/1, UKRI STFC Consolidated Grant ST/V000454/1 and from the University of Bristol School of Physics PhD Scholarship Fund, and is currently supported by the Klarman Fellowship at Cornell University. 

HRW and DG are funded by UK Research and Innovation (UKRI) framework under the UK government’s Horizon Europe funding guarantee for an ERC Starter Grant [grant number EP/Y006313/1]. 

Resources supporting this work were provided by the NASA High-End Computing (HEC) Program through the NASA Advanced Supercomputing (NAS) Division at Ames Research Center. NEB acknowledges support from NASA's Interdisciplinary Consortia for Astrobiology Research (grant No. NNH19ZDA001N-ICAR) under award number 19-ICAR19\_2-0041. 

We also acknowledge the MIT SuperCloud and Lincoln Laboratory Supercomputing Center for providing high performance computing resources that have contributed to the research results reported within this paper.

MR acknowledges financial support from the Natural Sciences and Engineering Research Council of Canada, the Fonds de recherche du Québec, and the Trottier Institute for Research on Exoplanets. 

RJM is supported by NASA through the NASA Hubble Fellowship grant HST-HF2-51513.001, awarded by the Space Telescope Science Institute, which is operated by the Association of Universities for Research in Astronomy, Inc., for NASA, under contract NAS 5-26555.


\vspace{5mm}
\textit{Data availability:} The specific observations analyzed can be accessed via the STScI MAST archive \dataset[DOI:10.17909/580k-bb85]{http://dx.doi.org/10.17909/580k-bb85}. All data products and models are available at \url{https://doi.org/10.5281/zenodo.14193061}.

\facilities{JWST (NIRISS SOSS)}

\software{\texttt{astropy} \citep{astropy2013,astropy2018}, \texttt{batman} \citep{kreidberg2015}, \texttt{celerite} \citep{Foreman_Mackey_2017}, \texttt{dynesty} \citep{Speagle_2020}, \texttt{emcee} \citep{Foreman-Mackey2013}, \texttt{Eureka!} \citep{bell2022}, \texttt{ExoTiC-LD} \citep{david_grant_2022_7437681}, \texttt{juliet} \citep{espinoza2019}, \texttt{JWST Science Calibration Pipeline} \citep{bushouse2023}, \texttt{MultiNest} \citep{Feroz2008,Buchner2014}, \texttt{nirHiss} \citep{feinstein2023}, \texttt{PASTASOSS}  \citep{baines2023traces,baines2023wavelength}, \texttt{petitRADTRANS} v2.7.6 \citep{Molliere2019}, \texttt{PICASO} v3.1 \citep{batalha2019exoplanet,mukherjee2023picaso}, \texttt{POSEIDON} v1.2 \citep[][]{MacDonaldMadhusudhan2017,MacDonald2023,Mullens2024_arxiv}, \texttt{supreme-SPOON} (\texttt{exoTEDRF}) \citep{radica2023, Radica2024_exotedrf}, \texttt{Virga} \citep{Batalha2020,Rooney2022}, \texttt{Ultranest} \citep{Ultranest}. } 



\appendix

\section{Appendix Information}

Here, we provide Appendix Figures 15, 16, and 17, as well as Appendix Tables 3, 4, 5, 6, and 7, which were mentioned in the main text. Figure 15 shows the spectral contribution of the \texttt{POSEIDON} model B + H$^{-}$ + SiO$_2$ median retrieved spectrum (black), which was presented in Figure 11. Figure 16 shows the complete \texttt{POSEIDON} retrieval corner plot for the same model. Figure 17 compares nested \texttt{POSEIDON} retrievals of model B + H$^{-}$ + SiO$_2$ to a model with an infinite opacity gray cloud deck. Table 3 provides the light curve fitting parameters for our \texttt{transitspectroscopy/juliet} analysis. Table 4 shows chemical inventories and opacity references for molecules in our retrieval analyses. Table 5 summarizes retrieved parameters for model B+SiO$_2$ with H$^{-}$ for both \texttt{POSEIDON} and \texttt{pRT}. Tables 6 and 7 summarize our retrieval statistics for both the full data set and SOSS data alone, respectively.

\begin{table*}[htpb]
  	\caption{White Light Curve Fitting Parameter Information for \texttt{transitspectroscopy/juliet} Analysis. }        
  	\label{table:wlc_fitting_transitspectroscopy}
        \begin{flushleft}
  	\begin{tabular}{ l c c c }        
  	\hline\hline
   &\multicolumn{3}{c}{\hspace{28 mm} Fixed Values and \texttt{juliet} Fit Values} \\ 
    \cmidrule(l){3-4}
    Parameters\tablenotemark{a} & Prior &  Order 1 & Order 2 \\
    \hline
    $P$ (days) & fixed \tablenotemark{b}  & 3.73548546 & 3.73548546 \\
    $t_0$ (BMJD$_{\rm TDB}$) & $\mathcal{N}(60023.6973372, 0.2)$ & $60023.69740955^{+3.844 \times 10^{-5}}_{-3.725 \times 10^{-5}}$ & $60023.697295757^{+4.812\times 10^{-5}}_{-4.760 \times 10^{-5}}$ \\
    $a/R_{\star}$ & $\mathcal{N}(7.025, 0.5)$ \tablenotemark{c} & $7.1236^{+0.0300}_{-0.0288}$ & $7.1298^{+0.0433}_{-0.0426}$ \\
    $b$  & $\mathcal{N}({0.361, 0.1})$ \tablenotemark{c} & ${0.3411}^{+0.0123}_{-0.0130}$ & ${0.3375}^{+0.0175}_{-0.0194}$ \\
    $R_{\rm p}/R_{\star}$ & $\mathcal{U}(0,0.2)$  &  $0.123891^{+0.000190}_{-0.000199}$ & $0.122935^{+0.000312}_{-0.000317}$ \\
    $q_{1}$ & $\mathcal{U}(0, 1)$   & $0.0966^{+0.0169}_{-0.0146}$ & $0.2117^{+0.0309}_{-0.0284}$ \\
    $q_{2}$ & $\mathcal{U}(0, 1)$  & $0.2179^{+0.0624}_{-0.0561}$ & $ 0.2607^{+0.0529}_{-0.0476}$ \\
    \\ [-2.5 ex]
    \hline \\ [-2.5 ex]
    $M_{SOSS}$ & $\mathcal{N}(0.0, 0.1)$  & $-0.0000016532^{+11.503 \times 10^{-6} }_{-11.469 \times 10^{-6} }$ & $-0.0000037002^{+13.872 \times 10^{-6} }_{-13.911\times 10^{-6} }$ \\
    $\sigma_{w,SOSS}$ & log$\mathcal{U}(10, 1000)$  & $116.371^{+5.191}_{-6.082}$ & $224.621^{+9.221}_{-12.306}$ \\
    $\sigma_{\mathrm{GP}_\mathrm{SOSS}}$ & log$\mathcal{U}(10^{-5}, 1000.0)$  & $0.0000656336^{+1.0226 \times 10^{-5} }_{-9.4831 \times 10^{-6} }$ & $0.0000460152^{+3.9930 \times 10^{-5} }_{-2.7656 \times 10^{-5} }$ \\
    $\rho_{\mathrm{GP}_\mathrm{SOSS}}$ & log$\mathcal{U}(10^{-3}, 0.5)$  & $0.01911^{+0.01167}_{-0.00782}$ & $0.01020^{+0.03172}_{-0.00756}$ \\
    \\ [-2.5 ex]
    \hline
    \end{tabular}
    \tablenotetext{a}{Parameter definitions: time of transit center, $t_0$, where BMJD$_{\rm TDB} =$ BJD$_{\rm TDB}$ - 2400000.5; semi-major axis in units of stellar radii, $a/R_{\star}$; impact parameter, $b$; planet radius in units of stellar radii, $R_{\rm p}/R_{\star}$; quadratic limb darkening coefficients  $q_{1}$ and $q_{2}$ using the \cite{kipping2013} parameterization; mean-out-of-transit offset, $M_{SOSS}$; jitter parameter, $\sigma_{w,SOSS}$; systematics Gaussian Process parameters $\sigma_{\mathrm{GP}_\mathrm{SOSS}}$ (amplitude) and $\rho_{\mathrm{GP}_\mathrm{SOSS}}$ (length-scale). }
    \tablenotetext{b}{\cite{Alderson2022}}
    \tablenotetext{c} {\cite{sedaghati2016}}
    \end{flushleft}
\end{table*}

\begin{figure}
    \includegraphics[width=0.95\textwidth]{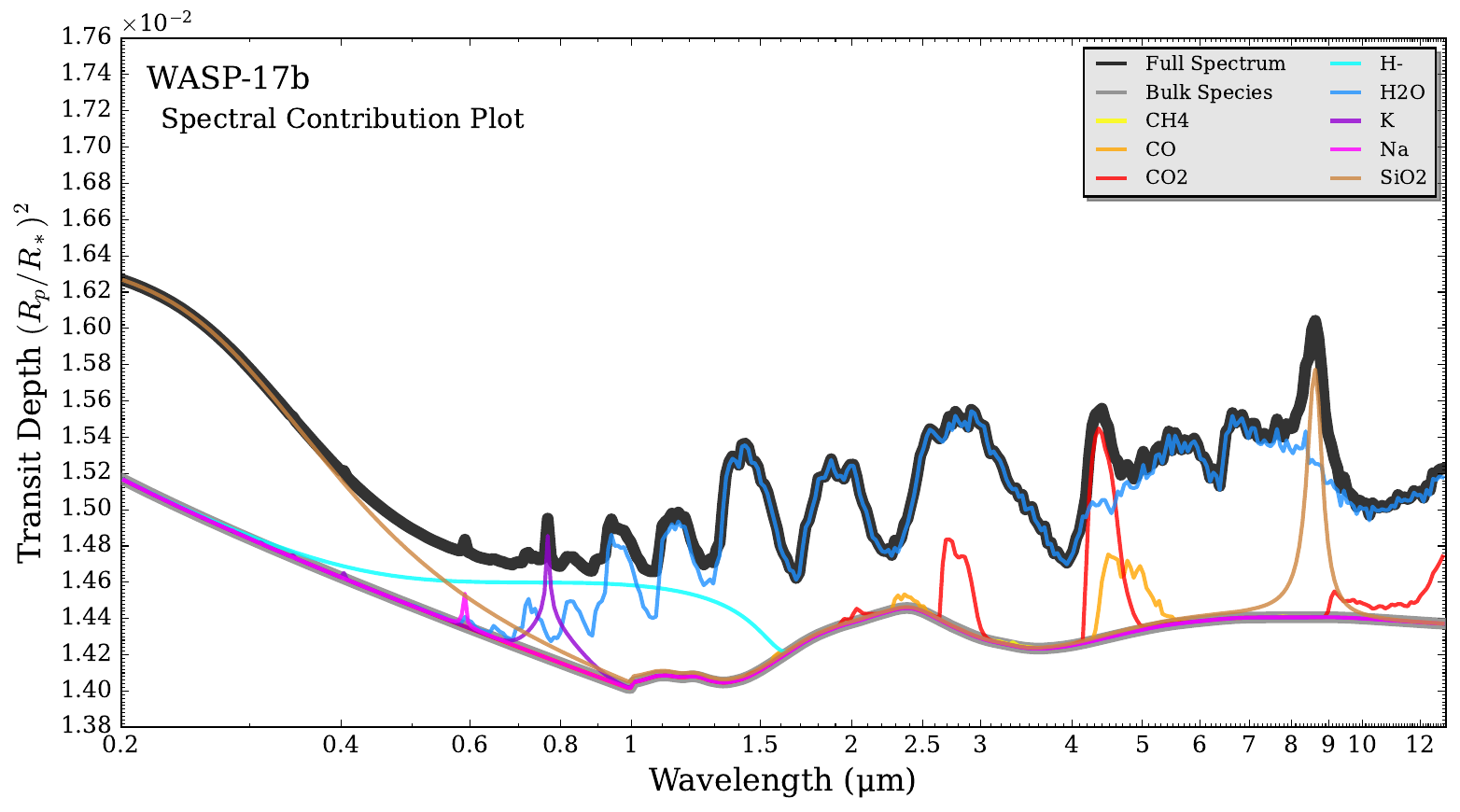}
    
    \caption{Spectral contribution of \texttt{POSEIDON} - Model B + H$^-$ + SiO$_2$ spectrum (black) displayed in Figure \ref{fig:RetrievalResults}. H$_2$ + He continuum opacity (gray) provides a baseline to the spectrum. SiO$_2$ (brown) dominates the opacity 0.2--0.4 \textmu m via Mie scattering. Na (magenta) has a weak absorption feature at 0.60 \textmu m. K (purple) has an absorption feature at 0.76 \textmu m. H$_2$O (blue) dominates the spectrum in a series of absorption features in the infrared. H$^-$ (cyan) has a flat continuum opacity extending from 0.4 to 1.5--1.6 \textmu m, giving rise to the flat SOSS Order 2 data and the deeper dip in transit depth in SOSS Order 1 after 1.5 \textmu m. CO$_2$ (red) has an absorption feature at 4.5 \textmu m that is broadened slightly by CO (orange). SiO$_2$ (brown) has an absorption feature at 8.6 \textmu m that is broadened slightly by H$_2$O. CH$_4$ (yellow) does not contribute to the spectrum.}    
    \label{fig:spectral_contribution}
\end{figure}

\begin{deluxetable*}{llll}
    \renewcommand{\arraystretch}{1.1}
    \tabletypesize{\footnotesize}
    \tablecolumns{4} 
    \tablecaption{Retrieval Chemical Inventory}

    \tablehead{Molecule \phantom{space} & Model\phantom{space} & POSEIDON Opacity References\phantom{space} & pRT Opacity References\phantom{space}}
    \startdata
    \hline
    AlO & D & \citet{Patrascu2015} & \citet{Patrascu2015}\\
    CaO & D & \citet{Yurchenko2016} & N/A\\
    TiO & A, C, D & \citet{McKemmish2019} & \citet{McKemmish2019}\\
    VO & C, D & \citet{McKemmish2016} & \citet{McKemmish2016}\\
    CrH & D & \citet{Burrows2002} & \citet{Burrows2002}\\
    FeH & D & \citet{Wende2010} & \citet{Wende2010}\\
    ScH & D & \citet{Lodi2015} & N/A\\
    TiH & D & \citet{Burrows2005} & N/A\\
    Na & B, C, D & \citet{Barklem2016} & \citet{Piskunov1995}\\
    K & B, C, D & \citet{Barklem2016}  & \citet{Piskunov1995}\\
    Li & D & \citet{Barklem2016}  & \citet{Kurucz1993}\\
    H$_2$O & A, B, C, D & \citet{Polyansky2018} & \citet{Rothman2010}\\
    CO & B, C, D & \citet{Li2015} & \citet{Rothman2010}\\
    CO$_2$ & A, B, C, D & \citet{Tashkun2011} & \citet{Yurchenko2020}\\
    CH$_4$ & A, B, C, D & \citet{Yurchenko2017} & \citet{Yurchenko2017} \\
    HCN & D & \citet{Barber2014} & \citet{Barber2014}\\
    NH$_3$ & D & \citet{Coles2019} & \citet{Coles2019} \\
    SO$_2$ & D & \citet{Underwood2016} & \citet{Underwood2016, Chubb2020}\\
    H$_2$S & D & \citet{Azzam2016} & N/A\\
    H$^-$ & A, B$^{*}$, C, D & \citet{John1988} & \citet{Gray2005}\\
    SiO$_2$ & A, B, C, D & \citet{palik1998handbook,andersen2006infrared} & \citet{Kitzmann2018}\\
    \enddata
    \label{tab:model_set_up}
    \tablecomments{Chemical inventory used in the retrieval analysis. Models refer to model A, B, C from \citet{Alderson2022}, Table 8. Model D represents an expanded, exploratory chemical inventory. Model D for {\tt pRT} is missing CaO, ScH, TiH, and H$_{2}$S so we refer to it as D$^{*}$ for clarity. B$^{*}$ refers to Model B with H$^-$. H$^-$ refers to continuum opacity due to bound-free absorption. SiO$_2$ refers to the refractive index reference.}
\end{deluxetable*}

\clearpage
\movetabledown = 2in
\begin{rotatetable}
\begin{deluxetable*}{lcccccc}
    \renewcommand{\arraystretch}{1.1}
    \tabletypesize{\tiny}
    \tablecolumns{7} 
    \tablecaption{Summary of Atmospheric Retrievals for Model B + SiO$_{2}$ with H$^-$.}
    \tablehead{ & \multicolumn{3}{c}{\textsc{POSEIDON}} & \multicolumn{3}{c}{\textsc{petitRADTRANS}}}
    \startdata
    Parameter \phantom{space} & \phantom{space} Prior\phantom{space} & \phantom{space}Posterior (SOSS)\phantom{space} & \phantom{space}Posterior (Full Data)\phantom{space} & \phantom{space}Prior\phantom{space} & \phantom{space}Posterior (SOSS)\phantom{space} &\phantom{space}Posterior (Full Data)\phantom{space} \\
    \hline
    $\log$\,g [cm/s$^2$] & $\cdots$ & $\cdots$ & $\cdots$ &                                                                   $\mathcal{U}$(2.2, 3.0)   &  $2.80^{+0.016}_{-0.019}$    & $2.80_{-0.016}^{+0.015}$ \\ 
    $T$ [K] & $\mathcal{U}$(400, 2300) &  $1244^{+44}_{-44}$& $1271^{+37}_{-36}$ &                                            $\mathcal{U}$(400, 2300) &  $1184^{+38}_{-51}$           & $1196_{-37}^{+26}$ \\ 
    $R_{\rm{p, ref}}$ [R$_{\rm{J}}$] & $\mathcal{U}$(1.5895, 2.1505) & $1.71^{+0.01}_{-0.01}$  & $1.69^{+0.01}_{-0.01}$ &     $\mathcal{U}$(1.8, 2.1)   &  $1.84\pm0.01$               & $1.82\pm0.01$ \\ 
    $\log$\,H$_2$O & $\mathcal{U}$(-12, -0.3) & $-2.67^{+0.42}_{-0.37}$   & $-2.96^{+0.31}_{-0.24}$ &                         $\mathcal{U}$(-14, -2)    &  $-2.76^{+0.28}_{-0.29}$     & $-3.00_{-0.16}^{+0.17}$ \\ 
    $\log$\,CH$_4$ & $\mathcal{U}$(-12, -0.3) & $-9.06^{+2.04}_{-1.91}$   & $-9.03^{+2.00}_{-1.93}$ &                         $\mathcal{U}$(-14, -2)    &  $-10.02^{+2.46}_{-2.43}$    & $-10.42_{-2.13}^{+2.24}$ \\ 
    $\log$\,CO$_2$ & $\mathcal{U}$(-12, -0.3) & $-9.00^{+2.10}_{-1.97}$   & $-5.80^{+0.62}_{-1.31}$ &                         $\mathcal{U}$(-14, -2)    &  $-10.15^{+2.70}_{-2.67}$    & $-6.23_{-2.74}^{+0.75}$ \\ 
    $\log$\,CO & $\mathcal{U}$(-12, -0.3) & $-4.48^{+1.58}_{-4.96}$   &  $-3.97^{+1.00}_{-4.59}$ &                            $\mathcal{U}$(-14, -2)    &  $-6.56^{+3.29}_{-4.93}$    &  $-3.63_{-4.64}^{+0.69}$ \\ 
    $\log$\,Na & $\mathcal{U}$(-12, -0.3) & $-5.28^{+0.93}_{-2.31}$   & $-9.32^{+1.75}_{-1.76}$ &                             $\mathcal{U}$(-14, -2)    &  $-5.02^{+0.74}_{-2.98}$    &  $-9.97_{-2.28}^{+2.23}$ \\  
    $\log$\,K & $\mathcal{U}$(-12, -0.3) & $-7.13^{+0.73}_{-0.71}$   & $-8.07^{+0.58}_{-0.52}$ &                              $\mathcal{U}$(-14, -2)    &  $-7.00^{+0.63}_{-0.69}$    &  $-7.97_{-0.43}^{+0.41}$ \\ 
    $\log$\,H$^-$ & $\mathcal{U}$(-12, -0.3) & $-9.75^{+0.42}_{-0.37}$   & $-10.19^{+0.30}_{-0.23}$ &                         $\mathcal{U}$(-14.6, 0.4)  &  $-9.75^{+0.42}_{-0.37}$    & $-10.3_{-0.15}^{+0.16}$ \\ 
    $\log$\,SiO$_2$ & $\mathcal{U}$(-30, -1) & $-20.01^{+10.38}_{-6.52}$   & $-11.23^{+0.60}_{-0.54}$ &                       $\cdots$                  & $\cdots$ & $\cdots$ \\
    $\log$\,$r_{\rm{m}}$\,SiO$_2$ [\textmu m] & $\mathcal{U}$(-3, -1) & $-1.19^{+1.42}_{-1.20}$   & $-1.85^{+0.10}_{-0.10}$ & $\cdots$                  & $\cdots$ & $\cdots$ \\
    $\log$\,$P_{\rm{top,slab}}$\,SiO$_2$ [bars] & $\mathcal{U}$(-8, 2) & $-1.40^{+2.23}_{-4.18}$  &$-6.60^{+1.09}_{-0.94}$ &  $\cdots$                  & $\cdots$ & $\cdots$ \\
    $\Delta\log$\,$P$\,SiO$_2$ [bars] & $\mathcal{U}$(0, 10) & $4.90^{+3.28}_{-3.23}$  & $1.96^{+1.27}_{-1.02}$ &             $\cdots$                  & $\cdots$  & $\cdots$ \\
    $\sigma_{\rm{lnorm}}$ [cm] & $\cdots$ & $\cdots$ & $\cdots$ &                                                             $\mathcal{U}$(1.05, 3.0)  & $2.02^{+0.63}_{-0.61}$   & $1.37_{-0.21}^{+0.34}$ \\ 
    $\log$\,$K_{zz}$ [cm$^2$/s] & $\cdots$ & $\cdots$ & $\cdots$ &                                                            $\mathcal{U}$(5.0, 13.0)  & $9.08^{+2.64}_{-2.75}$   & $7.51_{-0.57}^{+0.66}$ \\ 
    $f_{\rm{sed}}$ & $\cdots$ & $\cdots$ & $\cdots$ &                                                                         $\mathcal{U}$(0.1, 10.1)  & $5.46^{+2.95}_{-3.26}$   & $5.86_{-2.29}^{+2.33}$ \\ 
    $\log$\,SiO$_{2\,\rm{base}}$ & $\cdots$ & $\cdots$ & $\cdots$ &                                                           $\mathcal{U}$(-11, -1)    & $-7.43^{+3.08}_{-2.56}$  & $-3.90_{-0.43}^{0.44}$ \\ 
    $\log$\,$P_{\rm{base}}$\,SiO$_2$ [bars] & $\cdots$ & $\cdots$ & $\cdots$ &                                                $\mathcal{U}$(-6, 3.0)    & $-0.87^{+2.54}_{-3.48}$  & $-4.24_{-0.49}^{+0.39}$ \\ 
    $\delta_{\rm{rel},SOSS}$ [ppm] & $\mathcal{U}$(-500, 500) & $\cdots$ &  $434^{+17}_{-17}$ &                               $\cdots$ &  $\cdots$ \\
    $\delta_{\rm{rel},MIRI}$ [ppm] &  $\mathcal{U}$(-500, 500) & $\cdots$ & $169^{+42}_{-43}$ &                               $\cdots$ &  $\cdots$ \\
    \enddata
    \label{tab:retrieval_model_params}
    \tablecomments{Retrieved parameters for model B with H$^-$ and SiO$_{2}$ clouds considering SOSS-alone and the full dataset. Parameter definitions: gravity, $g$; isothermal atmospheric temperature, $T$; reference radius, $R_{\rm{p, ref}}$; gas-phase volume mixing ratios, $\log$\,X; constant aerosol mixing ratio, $\log$\,SiO$_2$; mean particle size, $\log$\,$r_{\rm{m}}$\,SiO$_2$; cloud-top pressure, $\log$\,$P_{\rm{top,slab}}$\,SiO$_2$; width of the cloud, $\Delta\log$\,$P$\,SiO$_2$; width of the particle-size distribution, $\sigma_{\rm{lnorm}}$; vertical mixing, $\log$\,$K_{zz}$; sedimentation efficiency, $f_{\rm{sed}}$; aerosol mixing ratio at the base, $\log$\,SiO$_{2\,\rm{base}}$; cloud-base pressure, $\log$\,$P_{\rm{base}}$\,SiO$_2$; SOSS dataset offset, $\delta_{\rm{rel,SOSS}}$; MIRI dataset offset, $\delta_{\rm{rel,MIRI}}$. Units for logarithmic parameters refer to the argument.}
\end{deluxetable*}
\end{rotatetable}
\clearpage

\begin{figure*}
    \includegraphics[width=\textwidth]{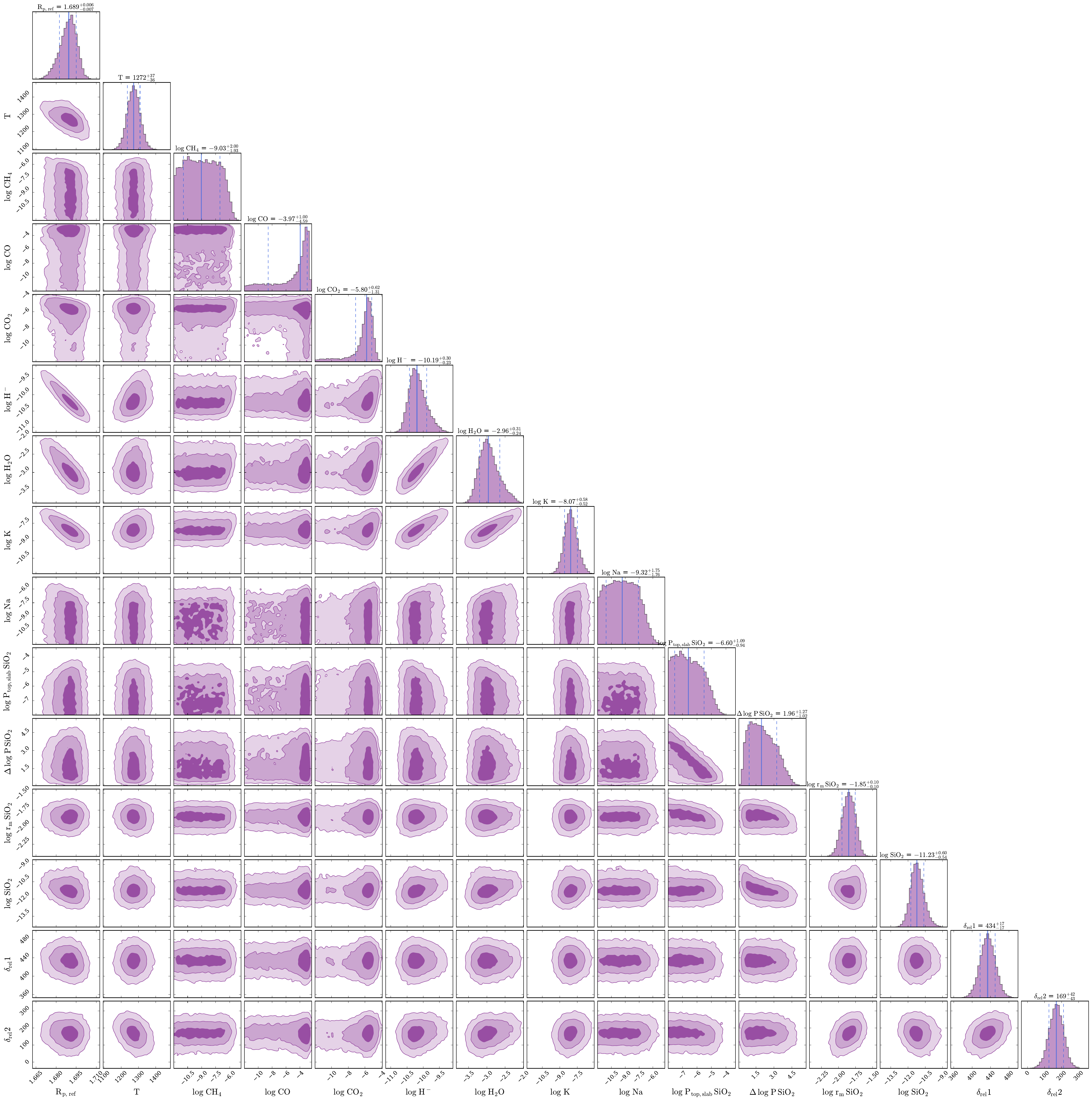}
    
    \caption{ POSEIDON Retrieval Complete Corner Plot for Model B with H$^-$ 
    }
    \label{fig:ModelBwH-Corner}
\end{figure*}

\begin{figure*}
    \includegraphics[width=0.95\textwidth]{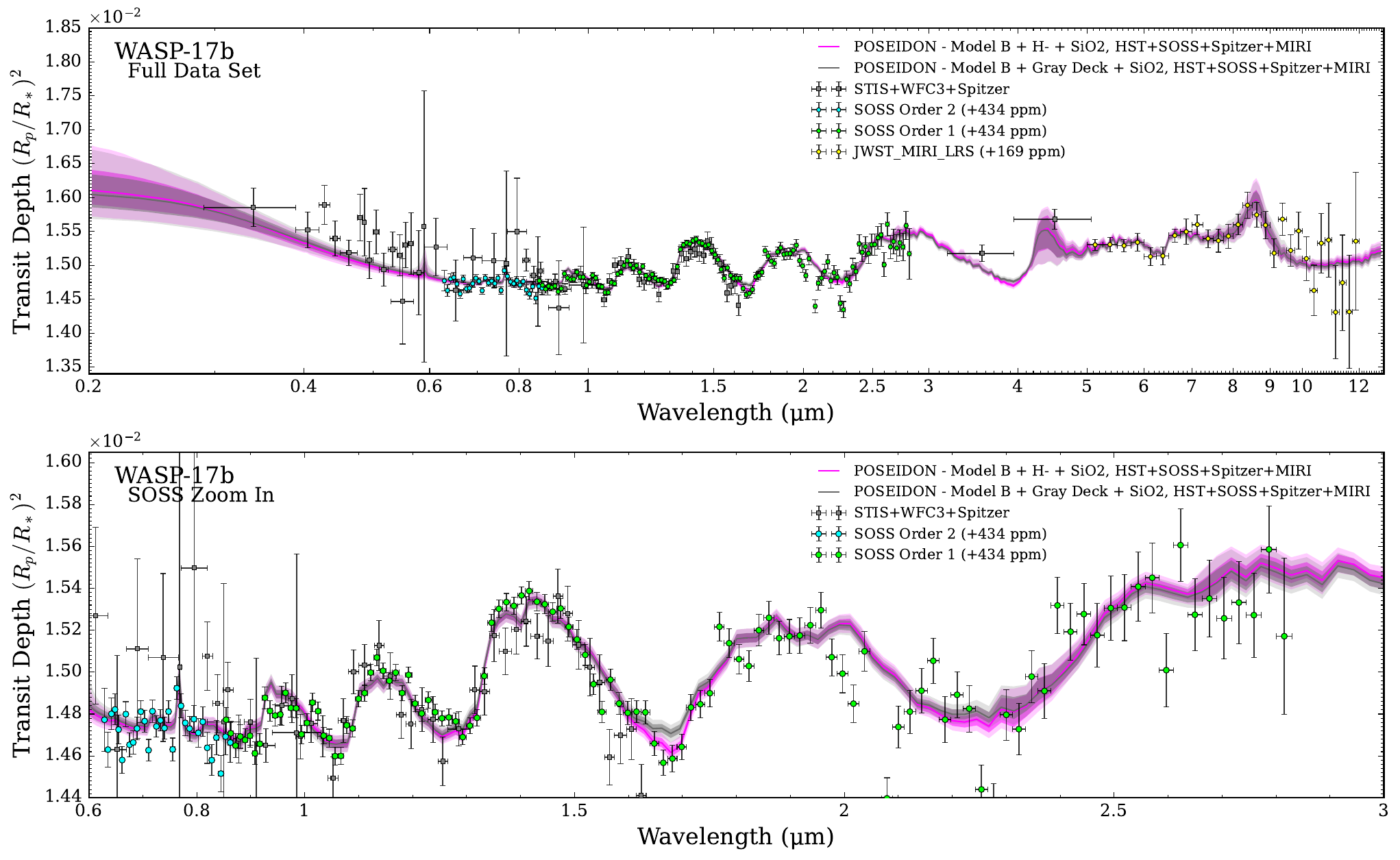}
    \caption{Comparison of nested \texttt{POSEIDON} retrievals with H$^{-}$ opacity and an infinite opacity gray cloud deck. The full data set (top panel) as well as a zoom-in on the SOSS data (bottom panel) are shown. Though an infinite opacity deck can fit the flat SOSS Order 2 data, H$^{-}$ is still preferred by 3.1 $\sigma$ due to a decrease in opacity past 1.5 \textmu m. } 
    \label{fig:gray_vs_H-}
\end{figure*}

\begin{deluxetable*}{llllll}
\tablecolumns{4}
\tablewidth{0pt}
\tabletypesize{\footnotesize}
\tablecaption{Transmission Statistics (Full Data) \label{table:transmission_stats}}
\tablehead{Model & lnZ & DoF & $\chi^2$ & $\chi^2_{red}$ & $\sigma$ Significance}
\startdata
 & & POSEIDON & & & \\ 
\hline
Model A & $1757.66^{+0.13}_{-0.13}$ & 233 & 479.92 & 2.06 & 3.9 \\
Model B & $1756.54^{+0.12}_{-0.12}$ & 232 & 479.98 & 2.07 & 4.2 \\
Model C{$^\dagger$} & $1763.72^{+0.13}_{-0.13}$ & 229 & 457.78 & 2.00 & --- \\
Model D & $1763.61^{+0.30}_{-0.30}$ & 218 & 433.00 & 1.99 & 1.2 \\
\hline
Model B & $1756.54^{+0.12}_{-0.12}$ & 232 & 479.98 & 2.07 & 5.1\\
Model B + Gray Cloud & $1764.28^{+0.12}_{-0.12}$ & 231 & 464.64 & 2.01 & 3.1 \\
Model B + H$^-${$^\dagger$} & $1767.75^{+0.13}_{-0.13}$ & 231 & 456.92 & 1.98 & ---  \\
Model B + H$^-$ + Patchy SiO$_2$ Clouds & $1766.73^{+0.13}_{-0.13}$ & 230 & 456.73 & 1.99 & 2.0 \\
\hline
Model B & $1756.54^{+0.12}_{-0.12}$ & 232 & 479.98 & 2.07 &  6.8\\
Model B + H$^-$ & $1767.75^{+0.13}_{-0.13}$ & 231 & 456.92 & 1.98 & 4.7 \\
Model B + AlO & $1773.91^{+0.13}_{-0.13}$ & 231 & 444.83 & 1.93 & 3.1 \\
Model B + H$^-$ + AlO{$^\dagger$} & $1777.27^{+0.13}_{-0.13}$ & 230 & 433.55 & 1.89 & --- \\
\hline
\hline
 & & pRT & & & \\ 
\hline
Model B & $1745.5\pm0.1$ & 230 & 494.97 & 2.15 &  6.5\\
Model B + H$^-${$^\dagger$} & $1764.5\pm0.1$ & 229  & 449.29 & 1.96 &  --- \\
\enddata
\tablecomments{{$^\dagger$} denotes base model for $\sigma$ significance. Model comparison and statistics for retrievals run on the complete dataset. Model C was the most preferred out of Model A, B, C, and D due to inclusion of K and H$^-$. We then ran Model B with H$^-$ included, which was preferred over Model B (without H$^-$) by 5.1 $\sigma$. A gray cloud and patchy SiO$_2$ clouds were not preferred. In our exploratory model, Model D, strong evidence for AlO was found. We ran nested Model B retrievals with H$^-$ and AlO and found that Model B with AlO and H$^-$ were preferred, but the detection of AlO is driven by archival data (Figure \ref{fig:model-b-comparison}). All models include SiO$_{2}$, even the gray cloud model.}
\end{deluxetable*}

\begin{deluxetable*}{llllll}
\tablecolumns{4}
\tablewidth{0pt}
\tabletypesize{\footnotesize}
\tablecaption{Transmission Statistics (SOSS) \label{table:soss_alone_stats}}
\tablehead{Model & lnZ & DoF & $\chi^2$ & $\chi^2_{red}$ & $\sigma$ Significance}
\startdata
 & & POSEIDON & & & \\ 
\hline
Model B & $1116.25^{+0.10}_{-0.10}$ & 138 & 323.61 & 2.35 & 4.0\\
Model B + H$^-${$^\dagger$} & $1122.48^{+0.09}_{-0.09}$ & 137 & 320.46 & 2.34 & ---  \\
\hline
\hline
 & & pRT & & & \\ 
\hline
Model A & $1105.8\pm0.0$ & 150 & 341.07 & 2.27 & 4.8 \\
Model B & $1111.8\pm0.0$ & 136 & 313.69 & 2.31 & 3.3 \\ 
Model B with H$^{-}${$^\dagger$} & $1115.7\pm0.1$   & 135 & 322.24 & 2.39 & --- \\
Model C & $1110.8\pm0.0$ & 150 & 309.60 & 2.06 & 3.6 \\
Model D$^{*}$ & $1106.9\pm0.0$ & 127 & 307.93 & 2.4 & 4.6 \\ 
\enddata
\tablecomments{{$^\dagger$} denotes base model for $\sigma$ significance. Model comparison and statistics for retrievals run on the SOSS data alone. All models include SiO$_{2}$. Molecules included for D$^{*}$ are listed in Table \ref{tab:retrieval_model_params}.}
\end{deluxetable*}



\bibliography{WASP17_NIRISS_transit}{}

\begin{thebibliography}{}
\expandafter\ifx\csname natexlab\endcsname\relax\def\natexlab#1{#1}\fi
\providecommand{\url}[1]{\href{#1}{#1}}
\providecommand{\dodoi}[1]{doi:~\href{http://doi.org/#1}{\nolinkurl{#1}}}
\providecommand{\doeprint}[1]{\href{http://ascl.net/#1}{\nolinkurl{http://ascl.net/#1}}}
\providecommand{\doarXiv}[1]{\href{https://arxiv.org/abs/#1}{\nolinkurl{https://arxiv.org/abs/#1}}}

\bibitem[{{Ackerman} \& {Marley}(2001)}]{Ackerman-Marley2001}
{Ackerman}, A.~S., \& {Marley}, M.~S. 2001, \apj, 556, 872,
  \dodoi{10.1086/321540}

\bibitem[{{Albert} {et~al.}(2023){Albert}, {Lafreni{\`e}re}, {Ren{\'e}},
  {Artigau}, {Volk}, {Goudfrooij}, {Martel}, {Radica}, {Rowe}, {Espinoza},
  {Roy}, {Filippazzo}, {Darveau-Bernier}, {Talens}, {Sivaramakrishnan},
  {Willott}, {Fullerton}, {LaMassa}, {Hutchings}, {Rowlands}, {Vila}, {Zhou},
  {Aldridge}, {Maszkiewicz}, {Beaulieu}, {Cook}, {Piaulet}, {Roy},
  {Lamontagne}, {Morel}, {Frost}, {Salhi}, {Coulombe}, {Benneke}, {MacDonald},
  {Johnstone}, {Turner}, {Fournier-Tondreau}, {Allart}, \&
  {Kaltenegger}}]{albert2023}
{Albert}, L., {Lafreni{\`e}re}, D., {Ren{\'e}}, D., {et~al.} 2023, \pasp, 135,
  075001, \dodoi{10.1088/1538-3873/acd7a3}

\bibitem[{{Alderson} {et~al.}(2022){Alderson}, {Wakeford}, {MacDonald},
  {Lewis}, {May}, {Grant}, {Sing}, {Stevenson}, {Fowler}, {Goyal}, {Batalha},
  \& {Kataria}}]{Alderson2022}
{Alderson}, L., {Wakeford}, H.~R., {MacDonald}, R.~J., {et~al.} 2022, \mnras,
  512, 4185, \dodoi{10.1093/mnras/stac661}

\bibitem[{Andersen {et~al.}(2006)Andersen, Mutschke, Posch, Min, \&
  Tamanai}]{andersen2006infrared}
Andersen, A.~C., Mutschke, H., Posch, T., Min, M., \& Tamanai, A. 2006, Journal
  of Quantitative Spectroscopy and Radiative Transfer, 100, 4

\bibitem[{{Anderson} {et~al.}(2010){Anderson}, {Hellier}, {Gillon}, {Triaud},
  {Smalley}, {Hebb}, {Collier Cameron}, {Maxted}, {Queloz}, {West}, {Bentley},
  {Enoch}, {Horne}, {Lister}, {Mayor}, {Parley}, {Pepe}, {Pollacco},
  {S{\'e}gransan}, {Udry}, \& {Wilson}}]{anderson2010}
{Anderson}, D.~R., {Hellier}, C., {Gillon}, M., {et~al.} 2010, \apj, 709, 159,
  \dodoi{10.1088/0004-637X/709/1/159}

\bibitem[{{Anderson} {et~al.}(2011){Anderson}, {Smith}, {Lanotte}, {Barman},
  {Collier Cameron}, {Campo}, {Gillon}, {Harrington}, {Hellier}, {Maxted},
  {Queloz}, {Triaud}, \& {Wheatley}}]{anderson2011}
{Anderson}, D.~R., {Smith}, A.~M.~S., {Lanotte}, A.~A., {et~al.} 2011, \mnras,
  416, 2108, \dodoi{10.1111/j.1365-2966.2011.19182.x}

\bibitem[{{Astropy Collaboration} {et~al.}(2013){Astropy Collaboration},
  {Robitaille}, {Tollerud}, {Greenfield}, {Droettboom}, {Bray}, {Aldcroft},
  {Davis}, {Ginsburg}, {Price-Whelan}, {Kerzendorf}, {Conley}, {Crighton},
  {Barbary}, {Muna}, {Ferguson}, {Grollier}, {Parikh}, {Nair}, {Unther},
  {Deil}, {Woillez}, {Conseil}, {Kramer}, {Turner}, {Singer}, {Fox}, {Weaver},
  {Zabalza}, {Edwards}, {Azalee Bostroem}, {Burke}, {Casey}, {Crawford},
  {Dencheva}, {Ely}, {Jenness}, {Labrie}, {Lim}, {Pierfederici}, {Pontzen},
  {Ptak}, {Refsdal}, {Servillat}, \& {Streicher}}]{astropy2013}
{Astropy Collaboration}, {Robitaille}, T.~P., {Tollerud}, E.~J., {et~al.} 2013,
  \aap, 558, A33, \dodoi{10.1051/0004-6361/201322068}

\bibitem[{{Astropy Collaboration} {et~al.}(2018){Astropy Collaboration},
  {Price-Whelan}, {Sip{\H{o}}cz}, {G{\"u}nther}, {Lim}, {Crawford}, {Conseil},
  {Shupe}, {Craig}, {Dencheva}, {Ginsburg}, {VanderPlas}, {Bradley},
  {P{\'e}rez-Su{\'a}rez}, {de Val-Borro}, {Aldcroft}, {Cruz}, {Robitaille},
  {Tollerud}, {Ardelean}, {Babej}, {Bach}, {Bachetti}, {Bakanov}, {Bamford},
  {Barentsen}, {Barmby}, {Baumbach}, {Berry}, {Biscani}, {Boquien}, {Bostroem},
  {Bouma}, {Brammer}, {Bray}, {Breytenbach}, {Buddelmeijer}, {Burke},
  {Calderone}, {Cano Rodr{\'\i}guez}, {Cara}, {Cardoso}, {Cheedella}, {Copin},
  {Corrales}, {Crichton}, {D'Avella}, {Deil}, {Depagne}, {Dietrich}, {Donath},
  {Droettboom}, {Earl}, {Erben}, {Fabbro}, {Ferreira}, {Finethy}, {Fox},
  {Garrison}, {Gibbons}, {Goldstein}, {Gommers}, {Greco}, {Greenfield},
  {Groener}, {Grollier}, {Hagen}, {Hirst}, {Homeier}, {Horton}, {Hosseinzadeh},
  {Hu}, {Hunkeler}, {Ivezi{\'c}}, {Jain}, {Jenness}, {Kanarek}, {Kendrew},
  {Kern}, {Kerzendorf}, {Khvalko}, {King}, {Kirkby}, {Kulkarni}, {Kumar},
  {Lee}, {Lenz}, {Littlefair}, {Ma}, {Macleod}, {Mastropietro}, {McCully},
  {Montagnac}, {Morris}, {Mueller}, {Mumford}, {Muna}, {Murphy}, {Nelson},
  {Nguyen}, {Ninan}, {N{\"o}the}, {Ogaz}, {Oh}, {Parejko}, {Parley}, {Pascual},
  {Patil}, {Patil}, {Plunkett}, {Prochaska}, {Rastogi}, {Reddy Janga},
  {Sabater}, {Sakurikar}, {Seifert}, {Sherbert}, {Sherwood-Taylor}, {Shih},
  {Sick}, {Silbiger}, {Singanamalla}, {Singer}, {Sladen}, {Sooley},
  {Sornarajah}, {Streicher}, {Teuben}, {Thomas}, {Tremblay}, {Turner},
  {Terr{\'o}n}, {van Kerkwijk}, {de la Vega}, {Watkins}, {Weaver}, {Whitmore},
  {Woillez}, {Zabalza}, \& {Astropy Contributors}}]{astropy2018}
{Astropy Collaboration}, {Price-Whelan}, A.~M., {Sip{\H{o}}cz}, B.~M., {et~al.}
  2018, \aj, 156, 123, \dodoi{10.3847/1538-3881/aabc4f}

\bibitem[{{Azzam} {et~al.}(2016){Azzam}, {Tennyson}, {Yurchenko}, \&
  {Naumenko}}]{Azzam2016}
{Azzam}, A. A.~A., {Tennyson}, J., {Yurchenko}, S.~N., \& {Naumenko}, O.~V.
  2016, \mnras, 460, 4063, \dodoi{10.1093/mnras/stw1133}

\bibitem[{{Baines} {et~al.}(2023{\natexlab{a}}){Baines}, {Espinoza},
  {Filippazzo}, \& {Volk}}]{baines2023wavelength}
{Baines}, T., {Espinoza}, N., {Filippazzo}, J., \& {Volk}, K.
  2023{\natexlab{a}}, {Characterization of the visit-to-visit Stability of the
  GR700XD Wavelength Calibration for NIRISS/SOSS Observations}, Technical
  Report JWST-STScI-008571, 12 pages

\bibitem[{{Baines} {et~al.}(2023{\natexlab{b}}){Baines}, {Espinoza},
  {Filippazzo}, \& {Volk}}]{baines2023traces}
---. 2023{\natexlab{b}}, arXiv e-prints, arXiv:2311.07769,
  \dodoi{10.48550/arXiv.2311.07769}

\bibitem[{{Barber} {et~al.}(2014){Barber}, {Strange}, {Hill}, {Polyansky},
  {Mellau}, {Yurchenko}, \& {Tennyson}}]{Barber2014}
{Barber}, R.~J., {Strange}, J.~K., {Hill}, C., {et~al.} 2014, \mnras, 437,
  1828, \dodoi{10.1093/mnras/stt2011}

\bibitem[{{Barklem} \& {Collet}(2016)}]{Barklem2016}
{Barklem}, P.~S., \& {Collet}, R. 2016, \aap, 588, A96,
  \dodoi{10.1051/0004-6361/201526961}

\bibitem[{{Barstow} {et~al.}(2017){Barstow}, {Aigrain}, {Irwin}, \&
  {Sing}}]{barstow2017}
{Barstow}, J.~K., {Aigrain}, S., {Irwin}, P.~G.~J., \& {Sing}, D.~K. 2017,
  \apj, 834, 50, \dodoi{10.3847/1538-4357/834/1/50}

\bibitem[{{Batalha} {et~al.}(2020){Batalha}, {Caoimherooney11}, \&
  {Sagnickm}}]{Batalha2020}
{Batalha}, N., {Caoimherooney11}, \& {Sagnickm}. 2020, {natashabatalha/virga:
  Initial Release}, v0.0, Zenodo,  Zenodo, \dodoi{10.5281/zenodo.3759888}

\bibitem[{{Batalha} \& {Line}(2017)}]{Batalha2017}
{Batalha}, N.~E., \& {Line}, M.~R. 2017, \aj, 153, 151,
  \dodoi{10.3847/1538-3881/aa5faa}

\bibitem[{Batalha {et~al.}(2019)Batalha, Marley, Lewis, \&
  Fortney}]{batalha2019exoplanet}
Batalha, N.~E., Marley, M.~S., Lewis, N.~K., \& Fortney, J.~J. 2019, The
  Astrophysical Journal, 878, 70

\bibitem[{{Bean} {et~al.}(2018){Bean}, {Stevenson}, {Batalha},
  {Berta-Thompson}, {Kreidberg}, {Crouzet}, {Benneke}, {Line}, {Sing},
  {Wakeford}, {Knutson}, {Kempton}, {D{\'e}sert}, {Crossfield}, {Batalha}, {de
  Wit}, {Parmentier}, {Harrington}, {Moses}, {Lopez-Morales}, {Alam}, {Blecic},
  {Bruno}, {Carter}, {Chapman}, {Decin}, {Dragomir}, {Evans}, {Fortney},
  {Fraine}, {Gao}, {Garc{\'\i}a Mu{\~n}oz}, {Gibson}, {Goyal}, {Heng}, {Hu},
  {Kendrew}, {Kilpatrick}, {Krick}, {Lagage}, {Lendl}, {Louden}, {Madhusudhan},
  {Mandell}, {Mansfield}, {May}, {Morello}, {Morley}, {Nikolov}, {Redfield},
  {Roberts}, {Schlawin}, {Spake}, {Todorov}, {Tsiaras}, {Venot}, {Waalkes},
  {Wheatley}, {Zellem}, {Angerhausen}, {Barrado}, {Carone}, {Casewell},
  {Cubillos}, {Damiano}, {de Val-Borro}, {Drummond}, {Edwards}, {Endl},
  {Espinoza}, {France}, {Gizis}, {Greene}, {Henning}, {Hong}, {Ingalls}, {Iro},
  {Irwin}, {Kataria}, {Lahuis}, {Leconte}, {Lillo-Box}, {Lines}, {Lothringer},
  {Mancini}, {Marchis}, {Mayne}, {Palle}, {Rauscher}, {Roudier}, {Shkolnik},
  {Southworth}, {Swain}, {Taylor}, {Teske}, {Tinetti}, {Tremblin}, {Tucker},
  {van Boekel}, {Waldmann}, {Weaver}, \& {Zingales}}]{Bean2018}
{Bean}, J.~L., {Stevenson}, K.~B., {Batalha}, N.~M., {et~al.} 2018, \pasp, 130,
  114402, \dodoi{10.1088/1538-3873/aadbf3}

\bibitem[{{Beichman} {et~al.}(2014){Beichman}, {Benneke}, {Knutson}, {Smith},
  {Lagage}, {Dressing}, {Latham}, {Lunine}, {Birkmann}, {Ferruit}, {Giardino},
  {Kempton}, {Carey}, {Krick}, {Deroo}, {Mandell}, {Ressler}, {Shporer},
  {Swain}, {Vasisht}, {Ricker}, {Bouwman}, {Crossfield}, {Greene}, {Howell},
  {Christiansen}, {Ciardi}, {Clampin}, {Greenhouse}, {Sozzetti}, {Goudfrooij},
  {Hines}, {Keyes}, {Lee}, {McCullough}, {Robberto}, {Stansberry}, {Valenti},
  {Rieke}, {Rieke}, {Fortney}, {Bean}, {Kreidberg}, {Ehrenreich}, {Deming},
  {Albert}, {Doyon}, \& {Sing}}]{Beichman2014}
{Beichman}, C., {Benneke}, B., {Knutson}, H., {et~al.} 2014, \pasp, 126, 1134,
  \dodoi{10.1086/679566}

\bibitem[{Bell {et~al.}(2022)Bell, Ahrer, Brande, Carter, Feinstein, {Guzman
  Caloca}, Mansfield, Zieba, Piaulet, Benneke, Filippazzo, May, Roy, Kreidberg,
  \& Stevenson}]{bell2022}
Bell, T.~J., Ahrer, E.-M., Brande, J., {et~al.} 2022, Journal of Open Source
  Software, 7, 4503, \dodoi{10.21105/joss.04503}

\bibitem[{{Benneke} {et~al.}(2024){Benneke}, {Roy}, {Coulombe}, {Radica},
  {Piaulet}, {Ahrer}, {Pierrehumbert}, {Krissansen-Totton}, {Schlichting},
  {Hu}, {Yang}, {Christie}, {Thorngren}, {Young}, {Pelletier}, {Knutson},
  {Miguel}, {Evans-Soma}, {Dorn}, {Gagnebin}, {Fortney}, {Komacek},
  {MacDonald}, {Raul}, {Cloutier}, {Acuna}, {Lafreni{\`e}re}, {Cadieux},
  {Doyon}, {Welbanks}, \& {Allart}}]{Benneke2024}
{Benneke}, B., {Roy}, P.-A., {Coulombe}, L.-P., {et~al.} 2024, arXiv e-prints,
  arXiv:2403.03325, \dodoi{10.48550/arXiv.2403.03325}

\bibitem[{{Brown}(2001)}]{Brown2001}
{Brown}, T.~M. 2001, \apj, 553, 1006, \dodoi{10.1086/320950}

\bibitem[{{Buchner}(2021)}]{Ultranest}
{Buchner}, J. 2021, The Journal of Open Source Software, 6, 3001,
  \dodoi{10.21105/joss.03001}

\bibitem[{{Buchner} {et~al.}(2014){Buchner}, {Georgakakis}, {Nandra}, {Hsu},
  {Rangel}, {Brightman}, {Merloni}, {Salvato}, {Donley}, \&
  {Kocevski}}]{Buchner2014}
{Buchner}, J., {Georgakakis}, A., {Nandra}, K., {et~al.} 2014, \aap, 564, A125,
  \dodoi{10.1051/0004-6361/201322971}

\bibitem[{{Burningham} {et~al.}(2021){Burningham}, {Faherty}, {Gonzales},
  {Marley}, {Visscher}, {Lupu}, {Gaarn}, {Fabienne Bieger}, {Freedman}, \&
  {Saumon}}]{Burningham2021}
{Burningham}, B., {Faherty}, J.~K., {Gonzales}, E.~C., {et~al.} 2021, \mnras,
  506, 1944, \dodoi{10.1093/mnras/stab1361}

\bibitem[{{Burrows} {et~al.}(2005){Burrows}, {Dulick}, {Bauschlicher},
  {Bernath}, {Ram}, {Sharp}, \& {Milsom}}]{Burrows2005}
{Burrows}, A., {Dulick}, M., {Bauschlicher}, C.~W., J., {et~al.} 2005, \apj,
  624, 988, \dodoi{10.1086/429366}

\bibitem[{{Burrows} {et~al.}(2002){Burrows}, {Ram}, {Bernath}, {Sharp}, \&
  {Milsom}}]{Burrows2002}
{Burrows}, A., {Ram}, R.~S., {Bernath}, P., {Sharp}, C.~M., \& {Milsom}, J.~A.
  2002, \apj, 577, 986, \dodoi{10.1086/342242}

\bibitem[{{Bushouse} {et~al.}(2023){Bushouse}, {Eisenhamer}, {Dencheva},
  {Davies}, {Greenfield}, {Morrison}, {Hodge}, {Simon}, {Grumm}, {Droettboom},
  {Slavich}, {Sosey}, {Pauly}, {Miller}, {Jedrzejewski}, {Hack}, {Davis},
  {Crawford}, {Law}, {Gordon}, {Regan}, {Cara}, {MacDonald}, {Bradley},
  {Shanahan}, {Jamieson}, {Teodoro}, {Williams}, \&
  {Pena-Guerrero}}]{bushouse2023}
{Bushouse}, H., {Eisenhamer}, J., {Dencheva}, N., {et~al.} 2023, {JWST
  Calibration Pipeline}, 1.12.0, Zenodo,  Zenodo,
  \dodoi{10.5281/zenodo.6984365}

\bibitem[{{Chubb} {et~al.}(2021){Chubb}, {Rocchetto, Marco}, {Yurchenko, Sergei
  N.}, {Min, Michiel}, {Waldmann, Ingo}, {Barstow, Joanna K.}, {Mollière,
  Paul}, {Al-Refaie, Ahmed F.}, {Phillips, Mark W.}, \& {Tennyson,
  Jonathan}}]{Chubb2020}
{Chubb}, K.~L., {Rocchetto, Marco}, {Yurchenko, Sergei N.}, {et~al.} 2021,
  A\&A, 646, A21, \dodoi{10.1051/0004-6361/202038350}

\bibitem[{{Coles} {et~al.}(2019){Coles}, {Yurchenko}, \&
  {Tennyson}}]{Coles2019}
{Coles}, P.~A., {Yurchenko}, S.~N., \& {Tennyson}, J. 2019, \mnras, 490, 4638,
  \dodoi{10.1093/mnras/stz2778}

\bibitem[{{Constantinou} \& {Madhusudhan}(2024)}]{ConstantinouMadhu2024}
{Constantinou}, S., \& {Madhusudhan}, N. 2024, \mnras, 530, 3252,
  \dodoi{10.1093/mnras/stae633}

\bibitem[{{Coulombe} {et~al.}(2023){Coulombe}, {Benneke}, {Challener},
  {Piette}, {Wiser}, {Mansfield}, {MacDonald}, {Beltz}, {Feinstein}, {Radica},
  {Savel}, {Dos Santos}, {Bean}, {Parmentier}, {Wong}, {Rauscher}, {Komacek},
  {Kempton}, {Tan}, {Hammond}, {Lewis}, {Line}, {Lee}, {Shivkumar},
  {Crossfield}, {Nixon}, {Rackham}, {Wakeford}, {Welbanks}, {Zhang}, {Batalha},
  {Berta-Thompson}, {Changeat}, {D{\'e}sert}, {Espinoza}, {Goyal},
  {Harrington}, {Knutson}, {Kreidberg}, {L{\'o}pez-Morales}, {Shporer}, {Sing},
  {Stevenson}, {Aggarwal}, {Ahrer}, {Alam}, {Bell}, {Blecic}, {Caceres},
  {Carter}, {Casewell}, {Crouzet}, {Cubillos}, {Decin}, {Fortney}, {Gibson},
  {Heng}, {Henning}, {Iro}, {Kendrew}, {Lagage}, {Leconte}, {Lendl},
  {Lothringer}, {Mancini}, {Mikal-Evans}, {Molaverdikhani}, {Nikolov}, {Ohno},
  {Palle}, {Piaulet}, {Redfield}, {Roy}, {Tsai}, {Venot}, \&
  {Wheatley}}]{coulombe2023}
{Coulombe}, L.-P., {Benneke}, B., {Challener}, R., {et~al.} 2023, \nat, 620,
  292, \dodoi{10.1038/s41586-023-06230-1}

\bibitem[{{Darveau-Bernier} {et~al.}(2022){Darveau-Bernier}, {Albert},
  {Talens}, {Lafreni{\`e}re}, {Radica}, {Doyon}, {Cook}, {Rowe}, {Allart},
  {Artigau}, {Benneke}, {Cowan}, {Dang}, {Espinoza}, {Johnstone},
  {Kaltenegger}, {Lim}, {Pauly}, {Pelletier}, {Piaulet}, {Roy}, {Roy},
  {Splinter}, {Taylor}, \& {Turner}}]{darveau-bernier2022}
{Darveau-Bernier}, A., {Albert}, L., {Talens}, G.~J., {et~al.} 2022, \pasp,
  134, 094502, \dodoi{10.1088/1538-3873/ac8a77}

\bibitem[{{Doyon} {et~al.}(2023){Doyon}, {Willott}, {Hutchings},
  {Sivaramakrishnan}, {Albert}, {Lafreni{\`e}re}, {Rowlands}, {Bego{\~n}a
  Vila}, {Martel}, {LaMassa}, {Aldridge}, {Artigau}, {Cameron}, {Chayer},
  {Cook}, {Cooper}, {Darveau-Bernier}, {Dupuis}, {Earnshaw}, {Espinoza},
  {Filippazzo}, {Fullerton}, {Gaudreau}, {Gawlik}, {Goudfrooij}, {Haley},
  {Kammerer}, {Kendall}, {Lambros}, {Ignat}, {Maszkiewicz}, {McColgan},
  {Morishita}, {Ouellette}, {Pacifici}, {Philippi}, {Radica}, {Ravindranath},
  {Rowe}, {Roy}, {Roy}, {Saad}, {Sohn}, {Talens}, {Touahri}, {Thatte},
  {Taylor}, {Vandal}, {Volk}, {Wander}, {Warner}, {Zheng}, {Zhou}, {Abraham},
  {Beaulieu}, {Benneke}, {Ferrarese}, {Jayawardhana}, {Johnstone},
  {Kaltenegger}, {Meyer}, {Pipher}, {Rameau}, {Rieke}, {Salhi}, \&
  {Sawicki}}]{doyon2023}
{Doyon}, R., {Willott}, C.~J., {Hutchings}, J.~B., {et~al.} 2023, \pasp, 135,
  098001, \dodoi{10.1088/1538-3873/acd41b}

\bibitem[{Espinoza(2022)}]{espinoza_nestor_2022}
Espinoza, N. 2022, \dodoi{10.5281/zenodo.6960924}

\bibitem[{{Espinoza} \& {Jord{\'a}n}(2015)}]{espinoza2015}
{Espinoza}, N., \& {Jord{\'a}n}, A. 2015, \mnras, 450, 1879,
  \dodoi{10.1093/mnras/stv744}

\bibitem[{{Espinoza} \& {Jord{\'a}n}(2016)}]{espinoza2016}
---. 2016, \mnras, 457, 3573, \dodoi{10.1093/mnras/stw224}

\bibitem[{Espinoza {et~al.}(2019)Espinoza, Kossakowski, \&
  Brahm}]{Espinoza_2019}
Espinoza, N., Kossakowski, D., \& Brahm, R. 2019, Monthly Notices of the Royal
  Astronomical Society, 490, 2262, \dodoi{10.1093/mnras/stz2688}

\bibitem[{{Espinoza} {et~al.}(2019){Espinoza}, {Kossakowski}, \&
  {Brahm}}]{espinoza2019}
{Espinoza}, N., {Kossakowski}, D., \& {Brahm}, R. 2019, \mnras, 490, 2262,
  \dodoi{10.1093/mnras/stz2688}

\bibitem[{{Espinoza} {et~al.}(2023){Espinoza}, {{\'U}beda}, {Birkmann},
  {Ferruit}, {Valenti}, {Sing}, {Rustamkulov}, {Regan}, {Kendrew}, {Sabbi},
  {Schlawin}, {Beatty}, {Albert}, {Greene}, {Nikolov}, {Karakla}, {Keyes},
  {Alves de Oliveira}, {B{\"o}ker}, {Pena-Guerrero}, {Giardino}, {Kumari},
  {Manjavacas}, {Proffitt}, \& {Rawle}}]{espinoza2023}
{Espinoza}, N., {{\'U}beda}, L., {Birkmann}, S.~M., {et~al.} 2023, \pasp, 135,
  018002, \dodoi{10.1088/1538-3873/aca3d3}

\bibitem[{{Fegley} \& {Lodders}(1994)}]{fegleylodders1994}
{Fegley}, Bruce, J., \& {Lodders}, K. 1994, \icarus, 110, 117,
  \dodoi{10.1006/icar.1994.1111}

\bibitem[{{Feinstein} {et~al.}(2023){Feinstein}, {Radica}, {Welbanks},
  {Murray}, {Ohno}, {Coulombe}, {Espinoza}, {Bean}, {Teske}, {Benneke}, {Line},
  {Rustamkulov}, {Saba}, {Tsiaras}, {Barstow}, {Fortney}, {Gao}, {Knutson},
  {MacDonald}, {Mikal-Evans}, {Rackham}, {Taylor}, {Parmentier}, {Batalha},
  {Berta-Thompson}, {Carter}, {Changeat}, {dos Santos}, {Gibson}, {Goyal},
  {Kreidberg}, {L{\'o}pez-Morales}, {Lothringer}, {Miguel}, {Molaverdikhani},
  {Moran}, {Morello}, {Mukherjee}, {Sing}, {Stevenson}, {Wakeford}, {Ahrer},
  {Alam}, {Alderson}, {Allen}, {Batalha}, {Bell}, {Blecic}, {Brande},
  {Caceres}, {Casewell}, {Chubb}, {Crossfield}, {Crouzet}, {Cubillos}, {Decin},
  {D{\'e}sert}, {Harrington}, {Heng}, {Henning}, {Iro}, {Kempton}, {Kendrew},
  {Kirk}, {Krick}, {Lagage}, {Lendl}, {Mancini}, {Mansfield}, {May}, {Mayne},
  {Nikolov}, {Palle}, {Petit dit de la Roche}, {Piaulet}, {Powell}, {Redfield},
  {Rogers}, {Roman}, {Roy}, {Nixon}, {Schlawin}, {Tan}, {Tremblin}, {Turner},
  {Venot}, {Waalkes}, {Wheatley}, \& {Zhang}}]{feinstein2023}
{Feinstein}, A.~D., {Radica}, M., {Welbanks}, L., {et~al.} 2023, \nat, 614,
  670, \dodoi{10.1038/s41586-022-05674-1}

\bibitem[{{Feroz} \& {Hobson}(2008)}]{Feroz2008}
{Feroz}, F., \& {Hobson}, M.~P. 2008, \mnras, 384, 449,
  \dodoi{10.1111/j.1365-2966.2007.12353.x}

\bibitem[{{Fisher} \& {Heng}(2018)}]{fisherHeng2018}
{Fisher}, C., \& {Heng}, K. 2018, \mnras, 481, 4698,
  \dodoi{10.1093/mnras/sty2550}

\bibitem[{Foreman-Mackey {et~al.}(2017)Foreman-Mackey, Agol, Ambikasaran, \&
  Angus}]{Foreman_Mackey_2017}
Foreman-Mackey, D., Agol, E., Ambikasaran, S., \& Angus, R. 2017, The
  Astronomical Journal, 154, 220, \dodoi{10.3847/1538-3881/aa9332}

\bibitem[{{Foreman-Mackey} {et~al.}(2013){Foreman-Mackey}, {Hogg}, {Lang}, \&
  {Goodman}}]{Foreman-Mackey2013}
{Foreman-Mackey}, D., {Hogg}, D.~W., {Lang}, D., \& {Goodman}, J. 2013, \pasp,
  125, 306, \dodoi{10.1086/670067}

\bibitem[{{Fournier-Tondreau} {et~al.}(2024){Fournier-Tondreau}, {MacDonald},
  {Radica}, {Lafreni{\`e}re}, {Welbanks}, {Piaulet}, {Coulombe}, {Allart},
  {Morel}, {Artigau}, {Albert}, {Lim}, {Doyon}, {Benneke}, {Rowe},
  {Darveau-Bernier}, {Cowan}, {Lewis}, {Cook}, {Flagg}, {Genest}, {Pelletier},
  {Johnstone}, {Dang}, {Kaltenegger}, {Taylor}, \&
  {Turner}}]{Fournier-Tondreau2024}
{Fournier-Tondreau}, M., {MacDonald}, R.~J., {Radica}, M., {et~al.} 2024,
  \mnras, 528, 3354, \dodoi{10.1093/mnras/stad3813}

\bibitem[{{Fu} {et~al.}(2022){Fu}, {Espinoza}, {Sing}, {Lothringer}, {Dos
  Santos}, {Rustamkulov}, {Deming}, {Kempton}, {Komacek}, {Knutson}, {Albert},
  {Pontoppidan}, {Volk}, \& {Filippazzo}}]{fu2022}
{Fu}, G., {Espinoza}, N., {Sing}, D.~K., {et~al.} 2022, \apjl, 940, L35,
  \dodoi{10.3847/2041-8213/ac9977}

\bibitem[{{Gao} {et~al.}(2018){Gao}, {Marley}, \& {Ackerman}}]{Gao2018}
{Gao}, P., {Marley}, M.~S., \& {Ackerman}, A.~S. 2018, \apj, 855, 86,
  \dodoi{10.3847/1538-4357/aab0a1}

\bibitem[{{Goodman} \& {Weare}(2010)}]{GoodmanWeare2010}
{Goodman}, J., \& {Weare}, J. 2010, Communications in Applied Mathematics and
  Computational Science, 5, 65, \dodoi{10.2140/camcos.2010.5.65}

\bibitem[{Gordon \& McBride(1994)}]{gordon1994computer}
Gordon, S., \& McBride, B.~J. 1994, Computer program for calculation of complex
  chemical equilibrium compositions and applications. Part 1: Analysis, Tech.
  rep.

\bibitem[{{Grant} \& {Wakeford}(2024)}]{Grant_2024JOSS}
{Grant}, D., \& {Wakeford}, H. 2024, The Journal of Open Source Software, 9,
  6816, \dodoi{10.21105/joss.06816}

\bibitem[{Grant \& Wakeford(2022)}]{david_grant_2022_7437681}
Grant, D., \& Wakeford, H.~R. 2022, Exo-TiC/ExoTiC-LD: ExoTiC-LD v3.0.0,
  v3.0.0,  Zenodo, \dodoi{10.5281/zenodo.7437681}

\bibitem[{{Grant} {et~al.}(2023){Grant}, {Lewis}, {Wakeford}, {Batalha},
  {Glidden}, {Goyal}, {Mullens}, {MacDonald}, {May}, {Seager}, {Stevenson},
  {Valenti}, {Visscher}, {Alderson}, {Allen}, {Ca{\~n}as}, {Col{\'o}n},
  {Clampin}, {Espinoza}, {Gressier}, {Huang}, {Lin}, {Long}, {Louie},
  {Pe{\~n}a-Guerrero}, {Ranjan}, {Sotzen}, {Valentine}, {Anderson}, {Balmer},
  {Bellini}, {Hoch}, {Kammerer}, {Libralato}, {Mountain}, {Perrin}, {Pueyo},
  {Rickman}, {Rebollido}, {Sohn}, {van der Marel}, \&
  {Watkins}}]{grant_miritransit_2023}
{Grant}, D., {Lewis}, N.~K., {Wakeford}, H.~R., {et~al.} 2023, \apjl, 956, L29,
  \dodoi{10.3847/2041-8213/acfc3b}

\bibitem[{{Gray}(2005)}]{Gray2005}
{Gray}, D.~F. 2005, {The Observation and Analysis of Stellar Photospheres},
  \dodoi{10.1017/CBO9781316036570}

\bibitem[{{Greene} {et~al.}(2016){Greene}, {Line}, {Montero}, {Fortney},
  {Lustig-Yaeger}, \& {Luther}}]{Greene2016}
{Greene}, T.~P., {Line}, M.~R., {Montero}, C., {et~al.} 2016, \apj, 817, 17,
  \dodoi{10.3847/0004-637X/817/1/17}

\bibitem[{{Gressier} {et~al.}(2024){Gressier}, {MacDonald}, {Espinoza},
  {Wakeford}, {Lewis}, {Goyal}, {Louie}, {Radica}, {Batalha}, {Long}, {May},
  {Mullens}, {Seager}, {Stevenson}, {Valenti}, {Alderson}, {Allen},
  {Ca{\~n}as}, {Challener}, {Col{\`o}n}, {Glidden}, {Grant}, {Huang}, {Lin},
  {Valentine}, {Mountain}, {Pueyo}, {Perrin}, \& {van der
  Marel}}]{Gressier2024_arxiv}
{Gressier}, A., {MacDonald}, R.~J., {Espinoza}, N., {et~al.} 2024, arXiv
  e-prints, arXiv:2410.08149, \dodoi{10.48550/arXiv.2410.08149}

\bibitem[{Hervé~Herbin \& Petitprez(2023)}]{Herbin2023}
Hervé~Herbin, Lise~Deschutter, A.~D., \& Petitprez, D. 2023, Aerosol Science
  and Technology, 57, 255, \dodoi{10.1080/02786826.2023.2165899}

\bibitem[{{Hohm}(1994)}]{Hohm1994}
{Hohm}, U. 1994, Chemical Physics, 179, 533,
  \dodoi{10.1016/0301-0104(94)87028-4}

\bibitem[{{Holmberg} \& {Madhusudhan}(2023)}]{holmberg2023}
{Holmberg}, M., \& {Madhusudhan}, N. 2023, \mnras, 524, 377,
  \dodoi{10.1093/mnras/stad1580}

\bibitem[{{Howe} {et~al.}(2017){Howe}, {Burrows}, \& {Deming}}]{Howe2017}
{Howe}, A.~R., {Burrows}, A., \& {Deming}, D. 2017, \apj, 835, 96,
  \dodoi{10.3847/1538-4357/835/1/96}

\bibitem[{{John}(1988)}]{John1988}
{John}, T.~L. 1988, \aap, 193, 189

\bibitem[{{Karman} {et~al.}(2019){Karman}, {Gordon}, {van der Avoird},
  {Baranov}, {Boulet}, {Drouin}, {Groenenboom}, {Gustafsson}, {Hartmann},
  {Kurucz}, {Rothman}, {Sun}, {Sung}, {Thalman}, {Tran}, {Wishnow},
  {Wordsworth}, {Vigasin}, {Volkamer}, \& {van der Zande}}]{Karman2019}
{Karman}, T., {Gordon}, I.~E., {van der Avoird}, A., {et~al.} 2019, \icarus,
  328, 160, \dodoi{10.1016/j.icarus.2019.02.034}

\bibitem[{{Kataria} {et~al.}(2016){Kataria}, {Sing}, {Lewis}, {Visscher},
  {Showman}, {Fortney}, \& {Marley}}]{kataria2016}
{Kataria}, T., {Sing}, D.~K., {Lewis}, N.~K., {et~al.} 2016, \apj, 821, 9,
  \dodoi{10.3847/0004-637X/821/1/9}

\bibitem[{{Khalafinejad} {et~al.}(2018){Khalafinejad}, {Salz}, {Cubillos},
  {Zhou}, {von Essen}, {Husser}, {Bayliss}, {L{\'o}pez-Morales}, {Dreizler},
  {Schmitt}, \& {L{\"u}ftinger}}]{Khalafinejad2018}
{Khalafinejad}, S., {Salz}, M., {Cubillos}, P.~E., {et~al.} 2018, \aap, 618,
  A98, \dodoi{10.1051/0004-6361/201732029}

\bibitem[{{Kipping}(2013)}]{kipping2013}
{Kipping}, D.~M. 2013, \mnras, 435, 2152, \dodoi{10.1093/mnras/stt1435}

\bibitem[{{Kitzmann} \& {Heng}(2018)}]{Kitzmann2018}
{Kitzmann}, D., \& {Heng}, K. 2018, \mnras, 475, 94,
  \dodoi{10.1093/mnras/stx3141}

\bibitem[{{Kreidberg}(2015)}]{kreidberg2015}
{Kreidberg}, L. 2015, \pasp, 127, 1161, \dodoi{10.1086/683602}

\bibitem[{{Kurucz}(1993)}]{Kurucz1993}
{Kurucz}, R. 1993, Robert Kurucz CD-ROM, 18

\bibitem[{{Lavvas} {et~al.}(2014){Lavvas}, {Koskinen}, \& {Yelle}}]{Lavvas2014}
{Lavvas}, P., {Koskinen}, T., \& {Yelle}, R.~V. 2014, \apj, 796, 15,
  \dodoi{10.1088/0004-637X/796/1/15}

\bibitem[{{Lewis} {et~al.}(2020){Lewis}, {Wakeford}, {MacDonald}, {Goyal},
  {Sing}, {Barstow}, {Powell}, {Kataria}, {Mishra}, {Marley}, {Batalha},
  {Moses}, {Gao}, {Wilson}, {Chubb}, {Mikal-Evans}, {Nikolov}, {Pirzkal},
  {Spake}, {Stevenson}, {Valenti}, \& {Zhang}}]{Lewis2020}
{Lewis}, N.~K., {Wakeford}, H.~R., {MacDonald}, R.~J., {et~al.} 2020, \apjl,
  902, L19, \dodoi{10.3847/2041-8213/abb77f}

\bibitem[{{Li} {et~al.}(2015){Li}, {Gordon}, {Rothman}, {Tan}, {Hu}, {Kassi},
  {Campargue}, \& {Medvedev}}]{Li2015}
{Li}, G., {Gordon}, I.~E., {Rothman}, L.~S., {et~al.} 2015, \apjs, 216, 15,
  \dodoi{10.1088/0067-0049/216/1/15}

\bibitem[{{Lim} {et~al.}(2023){Lim}, {Benneke}, {Doyon}, {MacDonald},
  {Piaulet}, {Artigau}, {Coulombe}, {Radica}, {L'Heureux}, {Albert}, {Rackham},
  {de Wit}, {Salhi}, {Roy}, {Flagg}, {Fournier-Tondreau}, {Taylor}, {Cook},
  {Lafreni{\`e}re}, {Cowan}, {Kaltenegger}, {Rowe}, {Espinoza}, {Dang}, \&
  {Darveau-Bernier}}]{lim2023}
{Lim}, O., {Benneke}, B., {Doyon}, R., {et~al.} 2023, \apjl, 955, L22,
  \dodoi{10.3847/2041-8213/acf7c4}

\bibitem[{{Lodders}(1999)}]{lodders99}
{Lodders}, K. 1999, \apj, 519, 793, \dodoi{10.1086/307387}

\bibitem[{Lodders(2002)}]{lodders02}
Lodders, K. 2002, The Astrophysical Journal, 577, 974, \dodoi{10.1086/342241}

\bibitem[{{Lodders}(2010)}]{Lodders2010}
{Lodders}, K. 2010, in Astrophysics and Space Science Proceedings, Vol.~16,
  Principles and Perspectives in Cosmochemistry, 379,
  \dodoi{10.1007/978-3-642-10352-0_8}

\bibitem[{{Lodders} \& {Fegley}(2002)}]{LoddersFegley2002}
{Lodders}, K., \& {Fegley}, B. 2002, \icarus, 155, 393,
  \dodoi{10.1006/icar.2001.6740}

\bibitem[{{Lodi} {et~al.}(2015){Lodi}, {Yurchenko}, \& {Tennyson}}]{Lodi2015}
{Lodi}, L., {Yurchenko}, S.~N., \& {Tennyson}, J. 2015, Molecular Physics, 113,
  1998, \dodoi{10.1080/00268976.2015.1029996}

\bibitem[{{Louie} {et~al.}(2018){Louie}, {Deming}, {Albert}, {Bouma}, {Bean},
  \& {Lopez-Morales}}]{Louie2018}
{Louie}, D.~R., {Deming}, D., {Albert}, L., {et~al.} 2018, \pasp, 130, 044401,
  \dodoi{10.1088/1538-3873/aaa87b}

\bibitem[{{MacDonald}(2023)}]{MacDonald2023}
{MacDonald}, R.~J. 2023, The Journal of Open Source Software, 8, 4873,
  \dodoi{10.21105/joss.04873}

\bibitem[{MacDonald \& Lewis(2022)}]{macdonald2022trident}
MacDonald, R.~J., \& Lewis, N.~K. 2022, The Astrophysical Journal, 929, 20

\bibitem[{{MacDonald} \& {Madhusudhan}(2017)}]{MacDonaldMadhusudhan2017}
{MacDonald}, R.~J., \& {Madhusudhan}, N. 2017, \mnras, 469, 1979,
  \dodoi{10.1093/mnras/stx804}

\bibitem[{{Mandell} {et~al.}(2013){Mandell}, {Haynes}, {Sinukoff},
  {Madhusudhan}, {Burrows}, \& {Deming}}]{mandell2013}
{Mandell}, A.~M., {Haynes}, K., {Sinukoff}, E., {et~al.} 2013, \apj, 779, 128,
  \dodoi{10.1088/0004-637X/779/2/128}

\bibitem[{{May} {et~al.}(2023){May}, {MacDonald}, {Bennett}, {Moran},
  {Wakeford}, {Peacock}, {Lustig-Yaeger}, {Highland}, {Stevenson}, {Sing},
  {Mayorga}, {Batalha}, {Kirk}, {L{\'o}pez-Morales}, {Valenti}, {Alam},
  {Alderson}, {Fu}, {Gonzalez-Quiles}, {Lothringer}, {Rustamkulov}, \&
  {Sotzen}}]{May2023}
{May}, E.~M., {MacDonald}, R.~J., {Bennett}, K.~A., {et~al.} 2023, \apjl, 959,
  L9, \dodoi{10.3847/2041-8213/ad054f}

\bibitem[{{McKemmish} {et~al.}(2019){McKemmish}, {Masseron}, {Hoeijmakers},
  {P{\'e}rez-Mesa}, {Grimm}, {Yurchenko}, \& {Tennyson}}]{McKemmish2019}
{McKemmish}, L.~K., {Masseron}, T., {Hoeijmakers}, H.~J., {et~al.} 2019,
  \mnras, 488, 2836, \dodoi{10.1093/mnras/stz1818}

\bibitem[{{McKemmish} {et~al.}(2016){McKemmish}, {Yurchenko}, \&
  {Tennyson}}]{McKemmish2016}
{McKemmish}, L.~K., {Yurchenko}, S.~N., \& {Tennyson}, J. 2016, \mnras, 463,
  771, \dodoi{10.1093/mnras/stw1969}

\bibitem[{{Molli{\`e}re} {et~al.}(2019){Molli{\`e}re}, {Wardenier}, {van
  Boekel}, {Henning}, {Molaverdikhani}, \& {Snellen}}]{Molliere2019}
{Molli{\`e}re}, P., {Wardenier}, J.~P., {van Boekel}, R., {et~al.} 2019, \aap,
  627, A67, \dodoi{10.1051/0004-6361/201935470}

\bibitem[{Morello {et~al.}(2020)Morello, Claret, Martin-Lagarde, Cossou,
  Tsiaras, \& Lagage}]{Morello_2020}
Morello, G., Claret, A., Martin-Lagarde, M., {et~al.} 2020, The Astronomical
  Journal, 159, 75, \dodoi{10.3847/1538-3881/ab63dc}

\bibitem[{{Morley} {et~al.}(2024){Morley}, {Mukherjee}, {Marley}, {Fortney},
  {Visscher}, {Lupu}, {Gharib-Nezhad}, {Thorngren}, {Freedman}, \& {Batalha
  7}}]{Diamondback}
{Morley}, C.~V., {Mukherjee}, S., {Marley}, M.~S., {et~al.} 2024, arXiv
  e-prints, arXiv:2402.00758, \dodoi{10.48550/arXiv.2402.00758}

\bibitem[{Mukherjee {et~al.}(2023)Mukherjee, Batalha, Fortney, \&
  Marley}]{mukherjee2023picaso}
Mukherjee, S., Batalha, N.~E., Fortney, J.~J., \& Marley, M.~S. 2023, The
  Astrophysical Journal, 942, 71

\bibitem[{{Mullens} {et~al.}(2024){Mullens}, {Lewis}, \&
  {MacDonald}}]{Mullens2024_arxiv}
{Mullens}, E., {Lewis}, N.~K., \& {MacDonald}, R.~J. 2024, arXiv e-prints,
  arXiv:2410.19253, \dodoi{10.48550/arXiv.2410.19253}

\bibitem[{Palik(1998)}]{palik1998handbook}
Palik, E.~D. 1998, Handbook of optical constants of solids, Vol.~3 (Academic
  press)

\bibitem[{{Patel} \& {Espinoza}(2022)}]{patelEspinoza2022}
{Patel}, J.~A., \& {Espinoza}, N. 2022, \aj, 163, 228,
  \dodoi{10.3847/1538-3881/ac5f55}

\bibitem[{{Patrascu} {et~al.}(2015){Patrascu}, {Yurchenko}, \&
  {Tennyson}}]{Patrascu2015}
{Patrascu}, A.~T., {Yurchenko}, S.~N., \& {Tennyson}, J. 2015, \mnras, 449,
  3613, \dodoi{10.1093/mnras/stv507}

\bibitem[{{Piaulet-Ghorayeb} {et~al.}(2024){Piaulet-Ghorayeb}, {Benneke},
  {Radica}, {Raul}, {Coulombe}, {Ahrer}, {Kubyshkina}, {Howard},
  {Krissansen-Totton}, {MacDonald}, {Roy}, {Louca}, {Christie},
  {Fournier-Tondreau}, {Allart}, {Miguel}, {Schlichting}, {Welbanks},
  {Cadieux}, {Dorn}, {Evans-Soma}, {Fortney}, {Pierrehumbert},
  {Lafreni{\`e}re}, {Acu{\~n}a}, {Komacek}, {Innes}, {Beatty}, {Cloutier},
  {Doyon}, {Gagnebin}, {Gapp}, \& {Knutson}}]{Piaulet2024}
{Piaulet-Ghorayeb}, C., {Benneke}, B., {Radica}, M., {et~al.} 2024, The
  Astrophysical Journal Letters, 974, L10, \dodoi{10.3847/2041-8213/ad6f00}

\bibitem[{{Pinhas} {et~al.}(2019){Pinhas}, {Madhusudhan}, {Gandhi}, \&
  {MacDonald}}]{pinhas2019}
{Pinhas}, A., {Madhusudhan}, N., {Gandhi}, S., \& {MacDonald}, R. 2019, \mnras,
  482, 1485, \dodoi{10.1093/mnras/sty2544}

\bibitem[{Piskunov {et~al.}(1995)Piskunov, Kupka, Ryabchikova, Weiss, \&
  Jeffery}]{Piskunov1995}
Piskunov, N., Kupka, F., Ryabchikova, T., Weiss, W., \& Jeffery, S. 1995,
  Astronomy and Astrophysics Supplement Series, 112, 525

\bibitem[{{Polyansky} {et~al.}(2018){Polyansky}, {Kyuberis}, {Zobov},
  {Tennyson}, {Yurchenko}, \& {Lodi}}]{Polyansky2018}
{Polyansky}, O.~L., {Kyuberis}, A.~A., {Zobov}, N.~F., {et~al.} 2018, \mnras,
  480, 2597, \dodoi{10.1093/mnras/sty1877}

\bibitem[{{Pontoppidan} {et~al.}(2022){Pontoppidan}, {Barrientes}, {Blome},
  {Braun}, {Brown}, {Carruthers}, {Coe}, {DePasquale}, {Espinoza}, {Marin},
  {Gordon}, {Henry}, {Hustak}, {James}, {Jenkins}, {Koekemoer}, {LaMassa},
  {Law}, {Lockwood}, {Moro-Martin}, {Mullally}, {Pagan}, {Player}, {Proffitt},
  {Pulliam}, {Ramsay}, {Ravindranath}, {Reid}, {Robberto}, {Sabbi}, {Ubeda},
  {Balogh}, {Flanagan}, {Gardner}, {Hasan}, {Meinke}, \&
  {Nota}}]{Pontoppidan2022}
{Pontoppidan}, K.~M., {Barrientes}, J., {Blome}, C., {et~al.} 2022, \apjl, 936,
  L14, \dodoi{10.3847/2041-8213/ac8a4e}

\bibitem[{{Radica}(2024)}]{Radica2024_exotedrf}
{Radica}, M. 2024, The Journal of Open Source Software, 9, 6898,
  \dodoi{10.21105/joss.06898}

\bibitem[{{Radica} {et~al.}(2022){Radica}, {Albert}, {Taylor},
  {Lafreni{\`e}re}, {Coulombe}, {Darveau-Bernier}, {Doyon}, {Cook}, {Cowan},
  {Espinoza}, {Johnstone}, {Kaltenegger}, {Piaulet}, {Roy}, \&
  {Talens}}]{radica2022}
{Radica}, M., {Albert}, L., {Taylor}, J., {et~al.} 2022, \pasp, 134, 104502,
  \dodoi{10.1088/1538-3873/ac9430}

\bibitem[{{Radica} {et~al.}(2023){Radica}, {Welbanks}, {Espinoza}, {Taylor},
  {Coulombe}, {Feinstein}, {Goyal}, {Scarsdale}, {Albert}, {Baghel}, {Bean},
  {Blecic}, {Lafreni{\`e}re}, {MacDonald}, {Zamyatina}, {Allart1}, {Artigau},
  {Batalha}, {Cook}, {Cowan}, {Dang}, {Doyon}, {Fournier-Tondreau},
  {Johnstone}, {Line}, {Moran}, {Mukherjee}, {Pelletier}, {Roy}, {Talens},
  {Filippazzo}, {Pontoppidan}, \& {Volk}}]{radica2023}
{Radica}, M., {Welbanks}, L., {Espinoza}, N., {et~al.} 2023, \mnras, 524, 835,
  \dodoi{10.1093/mnras/stad1762}

\bibitem[{{Radica} {et~al.}(2024{\natexlab{a}}){Radica}, {Coulombe}, {Taylor},
  {Albert}, {Allart}, {Benneke}, {Cowan}, {Dang}, {Lafreni{\`e}re},
  {Thorngren}, {Artigau}, {Doyon}, {Flagg}, {Johnstone}, {Pelletier}, \&
  {Roy}}]{Radica2024}
{Radica}, M., {Coulombe}, L.-P., {Taylor}, J., {et~al.} 2024{\natexlab{a}},
  \apjl, 962, L20, \dodoi{10.3847/2041-8213/ad20e4}

\bibitem[{{Radica} {et~al.}(2024{\natexlab{b}}){Radica}, {Piaulet-Ghorayeb},
  {Taylor}, {Coulombe}, {Albert}, {Artigau}, {Benneke}, {Cowan}, {Doyon},
  {Lafreni{\`e}re}, {L'Heureux}, \& {Lim}}]{Radica2024b}
{Radica}, M., {Piaulet-Ghorayeb}, C., {Taylor}, J., {et~al.}
  2024{\natexlab{b}}, arXiv e-prints, arXiv:2409.19333,
  \dodoi{10.48550/arXiv.2409.19333}

\bibitem[{{Rauscher} {et~al.}(2014){Rauscher}, {Boehm}, {Cagiano}, {Delo},
  {Foltz}, {Greenhouse}, {Hickey}, {Hill}, {Kan}, {Lindler}, {Mott},
  {Waczynski}, \& {Wen}}]{rauscher2014}
{Rauscher}, B.~J., {Boehm}, N., {Cagiano}, S., {et~al.} 2014, \pasp, 126, 739,
  \dodoi{10.1086/677681}

\bibitem[{{Rooney} {et~al.}(2022){Rooney}, {Batalha}, {Gao}, \&
  {Marley}}]{Rooney2022}
{Rooney}, C.~M., {Batalha}, N.~E., {Gao}, P., \& {Marley}, M.~S. 2022, \apj,
  925, 33, \dodoi{10.3847/1538-4357/ac307a}

\bibitem[{Rothman {et~al.}(2010)Rothman, Gordon, Barber, Dothe, Gamache,
  Goldman, Perevalov, Tashkun, \& Tennyson}]{Rothman2010}
Rothman, L., Gordon, I., Barber, R., {et~al.} 2010, Journal of Quantitative
  Spectroscopy and Radiative Transfer, 111, 2139,
  \dodoi{https://doi.org/10.1016/j.jqsrt.2010.05.001}

\bibitem[{{Schlawin} {et~al.}(2020){Schlawin}, {Leisenring}, {Misselt},
  {Greene}, {McElwain}, {Beatty}, \& {Rieke}}]{schlawin2020}
{Schlawin}, E., {Leisenring}, J., {Misselt}, K., {et~al.} 2020, \aj, 160, 231,
  \dodoi{10.3847/1538-3881/abb811}

\bibitem[{{Seager} \& {Sasselov}(2000)}]{Seager2000}
{Seager}, S., \& {Sasselov}, D.~D. 2000, \apj, 537, 916, \dodoi{10.1086/309088}

\bibitem[{{Sedaghati} {et~al.}(2016){Sedaghati}, {Boffin},
  {Je{\v{r}}abkov{\'a}}, {Garc{\'\i}a Mu{\~n}oz}, {Grenfell}, {Smette},
  {Ivanov}, {Csizmadia}, {Cabrera}, {Kabath}, {Rocchetto}, \&
  {Rauer}}]{sedaghati2016}
{Sedaghati}, E., {Boffin}, H.~M.~J., {Je{\v{r}}abkov{\'a}}, T., {et~al.} 2016,
  \aap, 596, A47, \dodoi{10.1051/0004-6361/201629090}

\bibitem[{{Sing} {et~al.}(2016){Sing}, {Fortney}, {Nikolov}, {Wakeford},
  {Kataria}, {Evans}, {Aigrain}, {Ballester}, {Burrows}, {Deming},
  {D{\'e}sert}, {Gibson}, {Henry}, {Huitson}, {Knutson}, {Lecavelier Des
  Etangs}, {Pont}, {Showman}, {Vidal-Madjar}, {Williamson}, \&
  {Wilson}}]{sing2016}
{Sing}, D.~K., {Fortney}, J.~J., {Nikolov}, N., {et~al.} 2016, \nat, 529, 59,
  \dodoi{10.1038/nature16068}

\bibitem[{{Southworth} {et~al.}(2012){Southworth}, {Hinse}, {Dominik}, {Fang},
  {Harps{\o}e}, {J{\o}rgensen}, {Kerins}, {Liebig}, {Mancini}, {Skottfelt},
  {Anderson}, {Smalley}, {Tregloan-Reed}, {Wertz}, {Alsubai}, {Bozza}, {Calchi
  Novati}, {Dreizler}, {Gu}, {Hundertmark}, {Jessen-Hansen}, {Kains},
  {Kjeldsen}, {Lund}, {Lundkvist}, {Mathiasen}, {Penny}, {Rahvar}, {Ricci},
  {Scarpetta}, {Snodgrass}, \& {Surdej}}]{southworth2012}
{Southworth}, J., {Hinse}, T.~C., {Dominik}, M., {et~al.} 2012, \mnras, 426,
  1338, \dodoi{10.1111/j.1365-2966.2012.21781.x}

\bibitem[{Speagle(2020)}]{Speagle_2020}
Speagle, J.~S. 2020, Monthly Notices of the Royal Astronomical Society, 493,
  3132, \dodoi{10.1093/mnras/staa278}

\bibitem[{{Stevenson} {et~al.}(2016){Stevenson}, {Lewis}, {Bean}, {Beichman},
  {Fraine}, {Kilpatrick}, {Krick}, {Lothringer}, {Mandell}, {Valenti}, {Agol},
  {Angerhausen}, {Barstow}, {Birkmann}, {Burrows}, {Charbonneau}, {Cowan},
  {Crouzet}, {Cubillos}, {Curry}, {Dalba}, {de Wit}, {Deming}, {D{\'e}sert},
  {Doyon}, {Dragomir}, {Ehrenreich}, {Fortney}, {Garc{\'\i}a Mu{\~n}oz},
  {Gibson}, {Gizis}, {Greene}, {Harrington}, {Heng}, {Kataria}, {Kempton},
  {Knutson}, {Kreidberg}, {Lafreni{\`e}re}, {Lagage}, {Line}, {Lopez-Morales},
  {Madhusudhan}, {Morley}, {Rocchetto}, {Schlawin}, {Shkolnik}, {Shporer},
  {Sing}, {Todorov}, {Tucker}, \& {Wakeford}}]{Stevenson2016}
{Stevenson}, K.~B., {Lewis}, N.~K., {Bean}, J.~L., {et~al.} 2016, \pasp, 128,
  094401, \dodoi{10.1088/1538-3873/128/967/094401}

\bibitem[{{Tashkun} \& {Perevalov}(2011)}]{Tashkun2011}
{Tashkun}, S.~A., \& {Perevalov}, V.~I. 2011, \jqsrt, 112, 1403,
  \dodoi{10.1016/j.jqsrt.2011.03.005}

\bibitem[{{Taylor} {et~al.}(2023){Taylor}, {Radica}, {Welbanks}, {MacDonald},
  {Blecic}, {Zamyatina}, {Roth}, {Bean}, {Parmentier}, {Coulombe}, {Feinstein},
  {Espinoza}, {Benneke}, {Lafreni{\`e}re}, {Doyon}, \& {Ahrer}}]{taylor2023}
{Taylor}, J., {Radica}, M., {Welbanks}, L., {et~al.} 2023, \mnras, 524, 817,
  \dodoi{10.1093/mnras/stad1547}

\bibitem[{{Tsai} {et~al.}(2023){Tsai}, {Lee}, {Powell}, {Gao}, {Zhang},
  {Moses}, {H{\'e}brard}, {Venot}, {Parmentier}, {Jordan}, {Hu}, {Alam},
  {Alderson}, {Batalha}, {Bean}, {Benneke}, {Bierson}, {Brady}, {Carone},
  {Carter}, {Chubb}, {Inglis}, {Leconte}, {Line}, {L{\'o}pez-Morales},
  {Miguel}, {Molaverdikhani}, {Rustamkulov}, {Sing}, {Stevenson}, {Wakeford},
  {Yang}, {Aggarwal}, {Baeyens}, {Barat}, {de Val-Borro}, {Daylan}, {Fortney},
  {France}, {Goyal}, {Grant}, {Kirk}, {Kreidberg}, {Louca}, {Moran},
  {Mukherjee}, {Nasedkin}, {Ohno}, {Rackham}, {Redfield}, {Taylor}, {Tremblin},
  {Visscher}, {Wallack}, {Welbanks}, {Youngblood}, {Ahrer}, {Batalha}, {Behr},
  {Berta-Thompson}, {Blecic}, {Casewell}, {Crossfield}, {Crouzet}, {Cubillos},
  {Decin}, {D{\'e}sert}, {Feinstein}, {Gibson}, {Harrington}, {Heng},
  {Henning}, {Kempton}, {Krick}, {Lagage}, {Lendl}, {Lothringer}, {Mansfield},
  {Mayne}, {Mikal-Evans}, {Palle}, {Schlawin}, {Shorttle}, {Wheatley}, \&
  {Yurchenko}}]{Tsai2023}
{Tsai}, S.-M., {Lee}, E. K.~H., {Powell}, D., {et~al.} 2023, \nat, 617, 483,
  \dodoi{10.1038/s41586-023-05902-2}

\bibitem[{{Underwood} {et~al.}(2016){Underwood}, {Tennyson}, {Yurchenko},
  {Huang}, {Schwenke}, {Lee}, {Clausen}, \& {Fateev}}]{Underwood2016}
{Underwood}, D.~S., {Tennyson}, J., {Yurchenko}, S.~N., {et~al.} 2016, \mnras,
  459, 3890, \dodoi{10.1093/mnras/stw849}

\bibitem[{{Valentine} {et~al.}(2024){Valentine}, {Wakeford}, {Challener},
  {Batalha}, {Lewis}, {Grant}, {Mullens}, {Alderson}, {Goyal}, {MacDonald},
  {May}, {Seager}, {Stevenson}, {Valenti}, {Allen}, {Espinoza}, {Glidden},
  {Gressier}, {Huang}, {Lin}, {Long}, {Louie}, {Clampin}, {Perrin}, {van der
  Marel}, \& {Mountain}}]{Valentine2024}
{Valentine}, D., {Wakeford}, H.~R., {Challener}, R.~C., {et~al.} 2024, \aj,
  168, 123, \dodoi{10.3847/1538-3881/ad5c61}

\bibitem[{Visscher {et~al.}(2010)Visscher, Lodders, \& Fegley}]{channon10}
Visscher, C., Lodders, K., \& Fegley, B. 2010, The Astrophysical Journal, 716,
  1060, \dodoi{10.1088/0004-637x/716/2/1060}

\bibitem[{{Visscher} {et~al.}(2006){Visscher}, {Lodders}, \&
  {Fegley}}]{visscher06}
{Visscher}, C., {Lodders}, K., \& {Fegley}, Bruce, J. 2006, \apj, 648, 1181,
  \dodoi{10.1086/506245}

\bibitem[{{Wakeford} {et~al.}(2016){Wakeford}, {Sing}, {Evans}, {Deming}, \&
  {Mandell}}]{wakeford2016}
{Wakeford}, H.~R., {Sing}, D.~K., {Evans}, T., {Deming}, D., \& {Mandell}, A.
  2016, \apj, 819, 10, \dodoi{10.3847/0004-637X/819/1/10}

\bibitem[{{Wakeford} {et~al.}(2017){Wakeford}, {Visscher}, {Lewis}, {Kataria},
  {Marley}, {Fortney}, \& {Mandell}}]{wakeford2016high}
{Wakeford}, H.~R., {Visscher}, C., {Lewis}, N.~K., {et~al.} 2017, \mnras, 464,
  4247, \dodoi{10.1093/mnras/stw2639}

\bibitem[{{Welbanks} {et~al.}(2019){Welbanks}, {Madhusudhan}, {Allard},
  {Hubeny}, {Spiegelman}, \& {Leininger}}]{welbanks2019}
{Welbanks}, L., {Madhusudhan}, N., {Allard}, N.~F., {et~al.} 2019, \apjl, 887,
  L20, \dodoi{10.3847/2041-8213/ab5a89}

\bibitem[{{Welbanks} {et~al.}(2023){Welbanks}, {McGill}, {Line}, \&
  {Madhusudhan}}]{Welbanks2023}
{Welbanks}, L., {McGill}, P., {Line}, M., \& {Madhusudhan}, N. 2023, \aj, 165,
  112, \dodoi{10.3847/1538-3881/acab67}

\bibitem[{{Wende} {et~al.}(2010){Wende}, {Reiners}, {Seifahrt}, \&
  {Bernath}}]{Wende2010}
{Wende}, S., {Reiners}, A., {Seifahrt}, A., \& {Bernath}, P.~F. 2010, \aap,
  523, A58, \dodoi{10.1051/0004-6361/201015220}

\bibitem[{{Wood} {et~al.}(2011){Wood}, {Maxted}, {Smalley}, \&
  {Iro}}]{Wood2011}
{Wood}, P.~L., {Maxted}, P.~F.~L., {Smalley}, B., \& {Iro}, N. 2011, \mnras,
  412, 2376, \dodoi{10.1111/j.1365-2966.2010.18061.x}

\bibitem[{{Yurchenko} {et~al.}(2017){Yurchenko}, {Amundsen}, {Tennyson}, \&
  {Waldmann}}]{Yurchenko2017}
{Yurchenko}, S.~N., {Amundsen}, D.~S., {Tennyson}, J., \& {Waldmann}, I.~P.
  2017, \aap, 605, A95, \dodoi{10.1051/0004-6361/201731026}

\bibitem[{{Yurchenko} {et~al.}(2016){Yurchenko}, {Blissett}, {Asari},
  {Vasilios}, {Hill}, \& {Tennyson}}]{Yurchenko2016}
{Yurchenko}, S.~N., {Blissett}, A., {Asari}, U., {et~al.} 2016, \mnras, 456,
  4524, \dodoi{10.1093/mnras/stv2858}

\bibitem[{Yurchenko {et~al.}(2020)Yurchenko, Mellor, Freedman, \&
  Tennyson}]{Yurchenko2020}
Yurchenko, S.~N., Mellor, T.~M., Freedman, R.~S., \& Tennyson, J. 2020, Monthly
  Notices of the Royal Astronomical Society, 496, 5282,
  \dodoi{10.1093/mnras/staa1874}

\bibitem[{{Zamyatina} {et~al.}(2023){Zamyatina}, {H{\'e}brard}, {Drummond},
  {Mayne}, {Manners}, {Christie}, {Tremblin}, {Sing}, \&
  {Kohary}}]{Zamyatina2023}
{Zamyatina}, M., {H{\'e}brard}, E., {Drummond}, B., {et~al.} 2023, \mnras, 519,
  3129, \dodoi{10.1093/mnras/stac3432}

\bibitem[{Zeidler {et~al.}(2013)Zeidler, Posch, \&
  Mutschke}]{zeidler2013optical}
Zeidler, S., Posch, T., \& Mutschke, H. 2013, Astronomy \& Astrophysics, 553,
  A81

\bibitem[{{Zhou} \& {Bayliss}(2012)}]{Zhou2012}
{Zhou}, G., \& {Bayliss}, D.~D.~R. 2012, \mnras, 426, 2483,
  \dodoi{10.1111/j.1365-2966.2012.21817.x}

\end{thebibliography}
\bibliographystyle{aasjournal}



\end{document}